\documentclass[12pt,final]{article}
\usepackage{jheppub}
\usepackage{graphicx}
\usepackage{amssymb}
\usepackage{amsmath}
\usepackage{cancel}
\usepackage{epstopdf}
\usepackage{mathtools}
\usepackage[export]{adjustbox}
\usepackage{subcaption}
\usepackage[usenames,dvipsnames]{xcolor}
\usepackage{tikz}
\usepackage{tikz-feynman}
\usepackage{relsize}

\usepackage[markifdraft]{gitinfo2}

\renewcommand{\v}[1]{\boldsymbol{{#1}}}

\usepackage[obeyDraft]{todonotes}
\usetikzlibrary{calc}

\let\emptyset\varnothing

\tikzset{invclip/.style={clip,insert path={{[reset cm]
      (-16383.99999pt,-16383.99999pt) rectangle (16383.99999pt,16383.99999pt)
    }}}}

\def\dd{\hat d}
\def\del{\hat \delta}
\def\delp{\hat \delta^{(+)}}

\newcommand{\df}{d \Phi}
\def\Del{\del_\Phi}
\def\wn{\bar}
\definecolor{allOrderBlue}{rgb}{0.4,0.5,1}
\definecolor{patternBlue}{rgb}{0,0,1}
\definecolor{photonRed}{rgb}{1,0.2,0.2}

\def\sect#1{sect.~\ref{#1}}
\def\sects#1#2{sects.~\ref{#1} and~\ref{#2}}
\def\pb{\bar p}
\def\qb{\bar q}
\def\Ord{\mathcal{O}}
\def\Lexp{\biggl\langle\!\!\!\biggl\langle}
\def\Rexp{\biggr\rangle\!\!\!\biggr\rangle}
\def\backdentA{\hspace*{-6mm}}

\newbox\charbox
\newbox\slabox
\def\s#1{{      
        \setbox\charbox=\hbox{$#1$}
        \setbox\slabox=\hbox{$/$}
        \dimen\charbox=\ht\slabox
        \advance\dimen\charbox by -\dp\slabox
        \advance\dimen\charbox by -\ht\charbox
        \advance\dimen\charbox by \dp\charbox
        \divide\dimen\charbox by 2
        \raise-\dimen\charbox\hbox to \wd\charbox{\hss/\hss}
        \llap{$#1$}
}}
\def\cut#1{{      
        \setbox\charbox=\hbox{$#1$}
        \setbox\slabox=\hbox{$|$}
        \dimen\charbox=\ht\slabox
        \advance\dimen\charbox by -\dp\slabox
        \advance\dimen\charbox by -\ht\charbox
        \advance\dimen\charbox by \dp\charbox
        \divide\dimen\charbox by 2
        \raise-\dimen\charbox\hbox to \wd\charbox{\hss$|$\hss}
        \llap{$#1$}
}}

\def\finalk{r}
\def\initialk{p}
\def\initialkc{p'}

\newcommand{\DeltaPlo}{\Delta p_{1}^{\mu, (0)}}
\newcommand{\DeltaPloTwo}{\Delta p_{2}^{\mu, (0)}}
\newcommand{\DeltaPnlo}{\Delta p_{1}^{\mu, (1)}}
\newcommand{\DeltaPnloTwo}{\Delta p_{2}^{\mu, (1)}}
\newcommand{\impKer}{\mathcal{I}^\mu}
\newcommand{\impKerCl}{\mathcal{I}_\class^{\mu, (1)}}
\newcommand{\RadKer}{\mathcal{R}}
\newcommand{\RadKerCl}{\mathcal{R}^{(0)}}
\def\Rad{R}

\def\impKerTerm#1{\mathcal{I}^\mu_{#1}}

\DeclareMathOperator{\csch}{csch}

\title{Amplitudes, Observables, and Classical Scattering}
\author[a]{David A. Kosower,}
\author[b]{Ben Maybee,}
\author[b]{Donal O'Connell}
\affiliation[a]{
Institut de Physique Th\'eorique, CEA, CNRS, Universit\'e Paris-Saclay, F-91191 Gif-sur-Yvette cedex, France
}
\affiliation[b]{
Higgs Centre for Theoretical Physics, School of Physics and Astronomy, The University of Edinburgh, Edinburgh EH9 3JZ, Scotland, UK%
}
\date\today

\abstract{%
We present a formalism for computing classically measurable quantities
directly from on-shell quantum scattering amplitudes.  We discuss the ingredients
needed for obtaining the classical result, and show how to set up
the calculation to derive the result efficiently.
We do this without specializing to a specific theory.
We study in detail two examples in electrodynamics: the momentum transfer
in spinless scattering to next-to-leading order, and the momentum
radiated to leading order.
}

\begin{document}
\maketitle

\def\LIGO{LIGO}
\def\eqn#1{eq.~(\ref{#1})}
\def\Eqn#1{Equation~(\ref{#1})}
\def\eqns#1#2{eqs.~(\ref{#1}) and~(\ref{#2})}
\newcommand{\Ampl}{\mathcal{A}}
\newcommand{\AmplB}{\mathcal{\bar A}}

\section{Introduction}

The dawn of gravitational-wave astronomy, heralded by the binary black-hole
and neutron-star mergers detected by the \LIGO{} collaboration~\cite{LIGO}, 
has spawned
interest in new techniques for solving
the two-body problem in gravity and
generating the theoretical waveforms
required~\cite{Buonanno:2014aza} for event detection as well as parameter extraction from 
observed mergers.  Such techniques would complement methods based
on the `traditional' Arnowitt--Deser--Misner Hamiltonian 
formalism~\cite{ADM,Schafer:2018kuf},
direct post-Newtonian solutions in harmonic gauge~\cite{Blanchet:2013haa},
long-established effective-one-body (EOB) methods~\cite{EOB1,EOB2}, numerical-relativity
approaches~\cite{NumericalRelativity},
and the effective-field theory approach
pioneered by Goldberger and Rothstein~\cite{EFT1,EFT2,EFT3,EFT4,EFT5,EFT6}.

Our broader interest is in exploring the
application of modern scattering-amplitudes techniques to this question.
Indeed, amplitudes have already been applied successfully to understand aspects
of the general relativistic two-body problem, notably to computing the
potential between two masses~\cite{Neill:2013wsa,Bjerrum-Bohr:2013bxa,Bjerrum-Bohr:2014lea,Bjerrum-Bohr:2014zsa,%
Bjerrum-Bohr:2016hpa,Cachazo:2017jef,Guevara:2017csg,Bjerrum-Bohr:2018xdl}.
The relevance of
a scattering amplitude---in particular, a loop amplitude---to the classical
potential was understood in
earlier work on gravity as an effective field theory~\cite{Donoghue:1993eb,%
Donoghue:1994dn,Donoghue:1996mt,%
Donoghue:2001qc,BjerrumBohr:2002ks,BjerrumBohr:2002kt}, and 
was emphasised by Donoghue and Holstein~\cite{Holstein:2004dn}.
This connection will be important for us below. More recently, Damour
has emphasised that methods based on scattering amplitudes are
relevant to the EOB formalism~\cite{Damour:2016gwp,Damour:2017zjx}.

An important insight arising from the study of scattering amplitudes
is that gravitational amplitudes are simpler than one would expect,
and in particular are closely connected to the amplitudes of Yang--Mills
theory. This connection is called the double copy, because gravitational
amplitudes are obtained as a product of two Yang--Mills quantities. One can
implement this double copy in a variety of ways: the original statement,
by Kawai, Lewellen and Tye~\cite{Kawai:1985xq} presents a tree-level
 gravitational amplitude
as a sum over terms, each of which is a product of two tree-level
color-ordered
Yang--Mills amplitudes (multiplied
by appropriate Mandelstam invariants). More recently, Bern, Carrasco and 
Johansson~\cite{DoubleCopy} demonstrated that the double copy can be 
understood very simply in terms of a diagrammatic expansion of a scattering 
amplitude: the gravitational numerators are simply the square of the kinematic numerators
in Yang--Mills theory, once a property known as colour-kinematics duality
is imposed on the numerators. These advances were particularly exciting
as they lead to a clear generalisation to loop level.
The work of BCJ suggests that gravity may be simpler than it seems,
and also more closely connected to Yang--Mills theory than one would guess
after inspecting their Lagrangians. We may hope that these insights
will be relevant to the real-world physics of gravitational waves.

The double copy
indeed connects classical solutions of Yang--Mills theory and gravity. In
particular, point charges in Yang--Mills theory map to point sources in 
gravity~\cite{Monteiro:2014cda,Luna:2016due,Goldberger:2016iau,Luna:2016hge}.
This holds true to all orders, even for accelerating particles~\cite{Luna:2016due,Luna:2018dpt},
and we know that the classical radiation emitted by accelerating particles
does indeed double copy from Yang--Mills theory to gravity~\cite{Luna:2016due,Goldberger:2016iau,Goldberger:2017frp,Shen:2018ebu},
even for a particle moving in an arbitrary manner~\cite{Goldberger:2016iau,Shen:2018ebu}, 
at least to the first two orders of perturbation theory. 
There are also indications that the double copy can encompass bound 
states~\cite{Goldberger:2017vcg,Plefka:2018dpa} and perhaps
spinning particles~\cite{Goldberger:2017ogt,Li:2018qap}.
These observations suggest that scattering-amplitudes methods,
which naturally incorporate spin,
should apply to the classical gravitational physics of 
spinning matter~\cite{Holstein:2008sx}. 

In addition to offering us
the double copy, the techniques of scattering amplitudes~\cite{Cheung:2018wkq}
or an analysis of soft limits~\cite{Laddha:2018rle}
may help to simplify the computation of physical waveforms relevant
for gravitational wave observatories.
First, though, we must understand systematically how to extract the classical result using
on-shell quantum-mechanical scattering amplitudes 
in order to take full advantage of amplitude methods in the gravitational-wave problem.  

The present article is a step in this direction.  We focus directly
on physical observables, extracting the classical values from
a fully relativistic quantum-mechanical computation.  We examine
two particular observables.  The first is the change in momentum during a scattering
event, both with and without accompanying radiation.  
The second is the radiated momentum during the event.
We shall use them as a laboratory to explore certain
conceptual and practical issues in approaching the classical limit.  
Our formalism applies to both electrodynamics and gravity.
We will
work out in detail explicit examples in electrodynamics, 
but many of the issues we explore also arise in 
the gravitational case. For simplicity,
we restrict to spinless scattering in this article.

Our two observables are not completely independent. Indeed the relation between
them goes to the heart of one of the difficulties in traditional
approaches to classical field theory with point sources. 
In two-particle scattering in
classical electrodynamics, for example, momentum is transferred from one particle to the
other via the electromagnetic field, as described by the Lorentz force. 
But the energy-momentum lost by point particles to radiation is not accounted for by the 
Lorentz force.
Conservation of momentum is restored by taking into account an additional force,
the Abraham--Lorentz--Dirac (ALD)
force~\cite{Lorentz,Abraham,Dirac:1938nz,LandauLifshitz}, see e.g.  refs.~\cite{Higuchi:2002qc,Galley:2006gs,%
Galley:2010es,Birnholtz:2013nta,Birnholtz:2014fwa,Birnholtz:2014gna} for more recent treatments. Inclusion of this radiation reaction force is not without cost: 
rather, it leads to the celebrated issues of runaway solutions or causality
violations in the classical electrodynamics of point sources.

The quantum-mechanical description of charged-particle scattering should
cure these ills.  Indeed we will see explicitly that a quantum-mechanical
description will conserve energy and momentum in particle scattering 
automatically.

In the next section,
we begin by describing where the factors of $\hbar$ appear in scattering 
amplitudes.  As we
will see it is straightforward to make these factors explicit, but 
nevertheless there are aspects of extracting the classical result that remain
obscure. 
This motivates the following section, where we give a formal definition of 
the momentum transfer to a particle in quantum field 
theory and of the expectation value of the momentum emitted in radiation 
during a scattering event.  We also give expressions for these observables
in terms of on-shell scattering amplitudes.
In \sect{sec:PointParticleLimit}, we
discuss the quantum wavefunctions
suitable for studying classical point particles,
and derive
simplified formul\ae{} for our observables.
In \sect{sec:examples} we 
apply our formalism in electrodynamics
to compute the two classical observables,
the momentum transfer and the radiated momentum, from scattering amplitudes.
In \sect{ClassicalCalculations}, we perform the corresponding classical calculations, 
and compare the results with those obtained from quantum field theory.
Section~\ref{sec:discussion} contains a discussion of our
results and our conclusions. 
In the appendices, we provide some details on our conventions and some
of the integrals we used, as well as dwelling in more detail on radiation in
the classical theory.

\section{Restoring $\hbar$}
\label{RestoringHBar}

A straightforward and pragmatic approach to restoring all factors
of $\hbar$ in
an expression is dimensional analysis\footnote{Paraphrasing the late 
	Sidney Coleman, natural units are natural to use because one can always
    put the units, expressed through $\hbar$s and $c$s, back through dimensional
    analysis.}.  We will continue to use relativistically natural units, with
$c=1$. We denote the dimensions of mass and length by $[M]$ and $[L]$,
respectively.

We may choose the dimensions of an $n$-point
scattering amplitude in four dimensions to be $[M]^{4-n}$ even when 
$\hbar \neq 1$. This is consistent with choosing 
the dimensions of creation and annihilation 
operators so that,
\begin{equation}
[a_p^{\vphantom{\dagger}}, a^\dagger_{p'}] = (2\pi)^3 \delta^{(3)}(\v{p} - \v{p}')\,,
\end{equation}
where bold symbols (here $\v{p}$ and $\v{p}'$) denote spatial three-vectors.
We define single-particle states by,
\begin{equation}
|p \rangle = \sqrt{2 E_p} \, a^\dagger_p |0\rangle\,.
\end{equation}
The dimension of $|p\rangle$ is thus $[M]^{-1}$. (The vacuum state is taken to be
dimensionless.)
We further define $n$-particle asymptotic states as tensor products of these
normalised single particle states. In order to avoid an unsightly splatter
of factors of $2\pi$, it is convenient to define,
\begin{equation}
\del^{(n)}(p) \equiv (2\pi)^n \delta^{(n)}(p)\,,
\label{delDefinition}
\end{equation}
for the $n$-fold Dirac $\delta$ distribution.
The scattering matrix $S$ and the transition matrix $T$ are both,
of course, dimensionless.  We define
the amplitudes in four dimensions as usual by
\begin{equation}
\langle p'_1 \cdots p'_m | T | p_1 \cdots p_n \rangle = 
\Ampl(p_1 \cdots p_n \rightarrow p'_1 \cdots p'_m) 
\del^{(4)}(p_1 + \cdots p_n - p'_1 - \cdots - p'_m),
\end{equation}
leading to the advertised dimensions for amplitudes.
\def\lcomp{\ell_c}
\def\lpack{\ell_w}
\def\lscatt{\ell_s}

When restoring powers of $\hbar$, we must distinguish between the 
momentum $p^\mu$ of a particle and its wavenumber, which has dimensions of
$[L]^{-1}$. This distinction will be important for us later, so we introduce 
a notation for the wavenumber $\pb$ associated with a momentum $p$:
\begin{align}
\wn p \equiv p / \hbar.
\label{eq:notationWavenumber}
\end{align}
In the course of restoring powers of $\hbar$ by dimensional analysis, we first 
treat the momenta of all particles as genuine momenta. We also treat 
any mass as a mass rather than the associated Compton wavelength.

When is a point-particle description appropriate?  We will consider the
scattering of two point-like objects, with momenta $p_{1,2}$,
initially separated by a transverse
impact parameter $b^\mu$.  (The impact parameter is transverse in the sense that
$p_i \cdot b = 0$ for $i = 1, 2$.) At the quantum level, the particles are
described by wavefunctions.  We will discuss these wavefunctions in more detail 
in~\sect{sec:PointParticleLimit}. 
We expect the point-particle description to be valid when the separation
of the two scattering particles is always very large compared to their 
(reduced) Compton
wavelengths $\lcomp^{(i)}\equiv \hbar/m_i$,
so the point-particle description will be accurate provided
that
\begin{equation}
\sqrt{-b^2} \gg \lcomp^{(1,2)}\,.
\end{equation}
The impact parameter and the Compton wavelengths are not the only scales
we must consider, however.  The wavefunctions have another intrinsic scale,
given by the spread of the wavepackets, $\lpack$.  The quantum-mechanical
expectation values of observables, as we will discuss, are well-approximated by
the corresponding classical ones, when the packet spreads are in the
`Goldilocks' zone, $\lcomp\ll \lpack\ll \sqrt{-b^2}$.

Let us now imagine restoring the $\hbar$s in a given amplitude. When $\hbar = 1$,
the amplitude has dimensions of $[M]^{4-n}$.  When $\hbar \neq 1$, the dimensions
of the momenta and masses in the amplitude are unchanged. Similarly there is no
change to the dimensions of polarisation vectors or tensors or of any 
Lie-algebraic factors in Yang--Mills theories. However, we must remember that the
dimensionless coupling in electrodynamics is $e/\sqrt{\hbar}$. Similarly, in
gravity a factor of $1/\sqrt{\hbar}$ appears as the appropriate coupling with
dimensions of inverse mass is $\kappa = \sqrt{32 \pi G/ \hbar}$. The algorithm to
restore the dimensions of any amplitude in scalar electrodynamics or scalar gravity is thus simple: each factor of a coupling is multiplied by an additional
factor of $1/\sqrt{\hbar}$. For example, an $n$-point, $L$-loop amplitude in
scalar QED is proportional to $\hbar^{1-n/2-L}$.

This conclusion, though well-known, may be surprising in the present context
because it seems naively that as $\hbar \rightarrow 0$, higher multiplicities
and higher loop orders are \textit{more\/} important.  As we will see, 
however, the approach to the classical limits --- 
for observables that make sense classically ---
effectively forces certain momenta to scale with $\hbar$.  These momenta have the classical 
interpretation of wavenumbers. Examples include the momenta of massless 
particles, such as photons or gravitons.  In putting the factors of $\hbar$
back into the couplings, we have therefore not yet made manifest all of 
the physically relevant factors of $\hbar$. This provides one motivation for the 
remainder of our paper: we wish to construct on-shell observables which are both 
classically and quantum-mechanically sensible. We will then carefully analyse the 
small-$\hbar$ region to understand how scattering amplitudes encode classical 
physics. We will see that the appropriate treatment is one
where point particles have 
momenta which are fixed as we take $\hbar$ to zero, whereas for massless particles
and momentum transfers between massive particles,
it is the wavenumber which we should treat as fixed in the limit.

\section{Impulse and radiated momentum in quantum field theory}
\label{QFTsetup}

We examine scattering events in which two
widely separated particles are prepared at $t \rightarrow -\infty$, 
and then shot at each other 
with impact parameter $b^\mu$.  We begin with a discussion of the 
appropriate incoming state, setting up convenient notation.  We then
describe the observables of interest.

Our formalism is quite general; for simplicity, we will nonetheless focus on 
scattering of two stable quanta of different scalar fields with different 
masses. We will also restrict our attention to scattering processes in 
which quanta of
fields 1 and 2 are both present in the final state.
This will happen, for example, if 
the particles have separately conserved quantum numbers.  We also assume 
that no new quanta of fields 1 and 2 can be produced during the collision, for 
example because the centre-of-mass energy is too small.

\def\in{\textbf{in}}
\subsection{The incoming state}

As we prepare the particles in the far past, the appropriate states are
incoming states  $| \psi \rangle_\textrm{in}$.
 We describe the incoming particles 
by wavefunctions $\phi_i(p_i)$. The main application we have in mind is to 
the scattering of point-like classical particles, and so we take our 
wave functions to have reasonably well-defined positions and
momenta. We will discuss the requirements on the wavepackets
in considerably more detail in \sect{sec:PointParticleLimit}. In this
section we focus on a general discussion of on-shell observables
associated with the scattering of localised particles, without
specialising to the kinds of wavefunctions which are most
appropriate for approaching a classical limit.

The initial state is then, 
\begin{equation}
| \psi \rangle_\textrm{in} = \int \! \dd^4 p_1 \dd^4 p_2 \, \delp(p_1^2 - m_1^2) 
\delp(p_2^2 - m_2^2) \, \phi_1(p_1) \phi_2(p_2) \, e^{i b \cdot p_1/\hbar} | p_1 p_2 
\rangle_\textrm{in}\,,
\label{InitialState}
\end{equation}
where $\dd p$ absorbs a factor of $2 \pi$; more generally $\dd^n p$ is defined by
\begin{equation}
\dd^n p \equiv \frac{d^n p}{(2 \pi)^n}\,.
\label{eq:ddxDefinition}
\end{equation}
We restrict the integration to positive-energy solutions of the
delta functions of $p_i^2-m_i^2$, as indicated by the $(+)$ superscript
in $\delp$, as well as absorbing a factor of $2\pi$ just as for
$\del(p)$,
\begin{equation}
\delp(p^2-m^2) \equiv 2\pi\Theta(p^0)\delta(p^2-m^2)\,.
\label{delpDefinition}
\end{equation}
In \eqn{InitialState} we have translated the wavepacket 
of particle 1 relative to particle 2 by the impact parameter $b$.
In the following, we will often omit the subscript ``in'', any unlabeled state being understood to be an ``in'' state.

We will find it convenient to further abbreviate the notation for
on-shell integrals (over Lorentz-invariant phase space),
\begin{equation}
\df(p_i) \equiv \dd^4 p_i \, \delp(p_i^2-m_i^2)\,.
\label{eq:dfDefinition}
\end{equation}
We will generally leave the mass implicit, along with the designation
of the integration variable as the first summand when the argument is a sum.

As noted in \sect{RestoringHBar}, we follow the standard convention
for normalizing states, so that
\begin{equation}
\langle p' | p \rangle = 2 E_p \del^{(3)} (\v{p}-\v{p}').
\label{MomentumStateNormalization}
\end{equation}
As the expression on the right-hand side
is the appropriately normalized delta function for the on-shell 
measure, 
\begin{equation}
\label{eq:norm1}
\int \df(p_1') \, 2 E_{p'_1} \, \del^{(3)} (\v{p}_1-\v{p}_1') f(p_1') = f(p_1)\,,
\end{equation}
for any function $f(p_1')$, we define,
\begin{equation}
\Del(p_1-p_1') \equiv  2 E_{p'_1} \del^{(3)} (\v{p}_1-\v{p}_1')\,.
\end{equation}
The argument on the left-hand side is understood as a function of four-vectors.
This 
leads to a notationally clearer version of \eqn{eq:norm1}:
\begin{equation}
\int \df(p'_1) \, \Del(p_1 - p_1') f(p_1') = f(p_1)\,,
\end{equation}
and of \eqn{MomentumStateNormalization}:
\begin{equation}
\langle p' | p \rangle = \Del(p-p')\,.
\end{equation}
We can also rewrite \eqn{InitialState},
\begin{equation}
| \psi \rangle_\textrm{in} = \int \! \df(p_1) \df(p_2)\;
  \phi_1(p_1) \phi_2(p_2) \, e^{i b \cdot p_1/\hbar} | p_1 p_2 \rangle_\textrm{in}\,.
\label{InitialStateSimple}
\end{equation}
Using this simplified notation, the normalisation condition is
\begin{equation}
\begin{aligned}
1 &= \langle \psi | \psi \rangle \\
&= \int \! \df(p_1) \df(p_2) \df(p'_1) \df(p'_2) e^{i b\cdot(p_1-p_1')/\hbar}
\\&\hspace*{15mm}\times \phi_1(p_1) \phi_1^*(p_1') \, \phi_2(p_2) \phi_2^*(p_2') 
\, \Del(p_1 - p_1') \, \Del (p_2 - p_2') \\
&= \int \! \df(p_1)\df(p_2)\; |\phi_1(p_1)|^2 |\phi_2(p_2)|^2\,.
\end{aligned}
\end{equation}
We can obtain this normalization by requiring both
wavefunctions $\phi_i$ to be normalized to unity:
\begin{equation}
\int \! \df(p_1)\; |\phi_1(p_1)|^2 = 1\,.
\label{WavefunctionNormalization}
\end{equation}

\subsection{The impulse on a particle}

At a gravitational wave observatory, we are of course interested in the gravitational radiation emitted by the source of interest.
We will discuss the radiated momentum in~\sect{sec:radiatedmomentum}.
Gravitational waves also carry information about the potential experienced by, for example, a black hole binary system. This observation motivates our interest in an on-shell observable related to the potential. We choose to explore the \textit{impulse\/} on a particle during a scattering event: at the classical level, this is simply the total change in the momentum of one of the particles --- say particle~1 --- during the collision. 

\def\outp#1{p_{\textrm{out},#1}}
\def\operator{\mathbb}
To define the observable, we place detectors at asymptotically large distances
pointing at the collision region.
The detectors measure only the momentum of particle 1. We assume that these detectors cover all possible scattering angles. Let $\operator P_i^\mu$ be the momentum operator for particle~$i$;
the expectation of the
first particle's outgoing momentum $\outp1^\mu$ is then
\begin{equation}
\begin{aligned}
\langle \outp1^\mu \rangle &= 
{}_\textrm{out}{\langle}\psi|\operator P_1^\mu |\psi\rangle_\textrm{out} \\
&= {}_\textrm{out}{\langle}\psi|\operator P_1^\mu U(\infty,-\infty)\,
                            |\psi\rangle_\textrm{in} \\
&= {}_\textrm{in}{\langle} \psi | \, U(\infty, -\infty)^\dagger \operator P_1^\mu U(\infty, -\infty) \, | \psi \rangle_\textrm{in},
\end{aligned}
\end{equation}
where $U(\infty, -\infty)$ is the time evolution operator from the far past to the far future.  This evolution operator is just
the $S$ matrix, so the expectation value is simply,
\begin{equation}
\begin{aligned}
\langle \outp1^\mu \rangle &= 
{}_\textrm{in}{\langle} \psi | S^\dagger \operator P_1^\mu S\, 
                             | \psi \rangle_\textrm{in}\,.
\end{aligned}
\end{equation}
We can insert a complete set of states and rewrite
the expectation value as,
\begin{equation}
\begin{aligned}
\langle \outp1^\mu \rangle 
&=\sum_X \int \df(\finalk_1)\, \df(\finalk_2) \; \finalk_1^\mu 
\; \bigl| \langle \finalk_1 \finalk_2 X |S| \psi\rangle\bigr|^2\,,
\end{aligned}
\label{p1Expectation}
\end{equation}
where we can think of the inserted states as the final state of a
scattering process.
In this equation, $X$ refers to any other particles which may be created.
The intermediate state containing $X$ also necessarily contains exactly one
particle each corresponding to fields~1 and~2.  Their momenta are denoted
by $\finalk_{1,2}$ respectively.
The sum over $X$ is a sum over all states, including $X$ empty, and includes
phase-space integrals for $X$ non-empty.
The phase-space integral over 
the momenta of particles 1 and 2 along with the sum over states $X$ is
what gives a complete sum over all states in the Hilbert space.
The expression~(\ref{p1Expectation}) hints at the possibility of evaluating the
momentum in terms of on-shell scattering amplitudes.

The physically interesting quantity is rather the change of momentum of the particle during the scattering, so we define,
\begin{equation}
\langle \Delta p_1^\mu \rangle =  \langle \psi |S^\dagger \, \operator P^\mu_1 \, S |\psi\rangle - \langle \psi | \, \operator P^\mu_1 \,  |\psi\rangle.
\end{equation}
This impulse is the difference between the expected outgoing and
the incoming momenta of 
particle 1. 
It is an on-shell observable, defined in both the quantum and the classical theories. 
Similarly, we can measure the impulse imparted to particle 2. In terms of the momentum operator, $\operator P_2^\mu$, 
of quantum field 2, this impulse is evidently,
\begin{equation}
\langle \Delta p_2^\mu \rangle =  \langle \psi |S^\dagger \, \operator P^\mu_2 \, S |\psi\rangle - \langle \psi | \, \operator P^\mu_2 \,  |\psi\rangle.
\end{equation}

Returning to the impulse on particle 1, we proceed by writing 
the scattering matrix in terms of the transition matrix $T$ via
$S = 1 + i T$, in order to make contact with the usual scattering amplitudes.
The no-scattering (unity) part of the $S$ matrix cancels in the impulse,
leaving behind only delta functions that identify the final-state momenta
with the initial-state ones in the wavefunction or its conjugate.
Using unitarity we obtain the result,
\begin{equation}
\langle \Delta p_1^\mu \rangle 
= \langle \psi | \, i [ \operator P_1^\mu, T ] \, | \psi \rangle + \langle \psi | \, T^\dagger [ \operator P_1^\mu, T] \, |\psi \rangle\,.
\label{eq:defl1}
\end{equation}

\subsection{Impulse in terms of amplitudes}

\def\ImpA{I_{(1)}^\mu}
\def\ImpB{I_{(2)}^\mu}
\def\ImpAcl{I_{(1),\class}^\mu}
\def\ImpBcl{I_{(2),\class}^\mu}
\def\ImpTcl{I_{\class}^\mu}
\newcommand{\ImpAsup}[1]{I_{(1)}^{\mu,#1}}
\newcommand{\ImpBsup}[1]{I_{(2)}^{\mu,#1}}
\newcommand{\ImpAclsup}[1]{I_{(1),\class}^{\mu,#1}}
\newcommand{\ImpBclsup}[1]{I_{(2),\class}^{\mu,#1}}

Having established a general expression for the impulse, we turn to expressing it in terms of scattering amplitudes.
It is convenient to work on the two terms in equation~\eqref{eq:defl1} separately. For ease of discussion, we define\begin{equation}
\begin{aligned}
\ImpA &\equiv \langle \psi | \, i [ \operator P_1^\mu, T ] \, | \psi \rangle\,, \\
\ImpB &\equiv \langle \psi | \, T^\dagger [ \operator P_1^\mu, T] \, |\psi \rangle\,,
\end{aligned}
\end{equation}
so that the impulse is $\langle \Delta p_1^\mu \rangle = \ImpA + \ImpB$. Expanding the wavefunction in the first term, $\ImpA$, we find
\begin{equation}
\begin{aligned}
\ImpA &= 
\int \! \df(\initialk_1)\df(\initialk_2)
                \df(\initialkc_1)\df(\initialkc_2)\;
   e^{i b \cdot (\initialk_1 - \initialkc_1)/\hbar} \, 
      \phi_1(\initialk_1) \phi_1^*(\initialkc_1) 
      \phi_2(\initialk_2) \phi_2^*(\initialkc_2) 
\\ &\hphantom{\df(\initialk_1)\df(\initialk_2)
                \df(\initialkc_1)\df(\initialkc_2)\;
   e^{i b \cdot (\initialk_1 - \initialkc_1)/\hbar} \,}\hspace*{-5mm}
\times i (\initialkc_1\!{}^\mu - \initialk_1^\mu) \, 
     \langle \initialkc_1 \initialkc_2 | \,T\, |
           \initialk_1 \initialk_2 \rangle
\\&= \int \! \df(\initialk_1)\df(\initialk_2)
                \df(\initialkc_1)\df(\initialkc_2)\;
   e^{i b \cdot (\initialk_1 - \initialkc_1)/\hbar} \, 
      \phi_1(\initialk_1) \phi_1^*(\initialkc_1) 
      \phi_2(\initialk_2) \phi_2^*(\initialkc_2) 
\\ &\hphantom{\df(\initialk_1)\df(\initialk_2)
                \df(\initialkc_1)\df(\initialkc_2)\;
   e^{i b \cdot (\initialk_1 - \initialkc_1)/\hbar} \,}\hspace*{-40mm}
\times i \int \df(\finalk_1)\df(\finalk_2)\;(\finalk_1^\mu-\initialk_1^\mu) \, 
   \langle \initialkc_1 \initialkc_2 | \finalk_1 \finalk_2 \rangle
         \langle \finalk_1 \finalk_2 | \,T\, |\initialk_1 \initialk_2 \rangle
\,,
\end{aligned}
\label{eq:defl2}
\end{equation}
where in the second form line we have re-inserted the final-state momenta $\finalk_i$ in
order to make manifest the phase independence of the result.
We label the states in the incoming wavefunction by $\initialk_{1,2}$, 
those in the conjugate ones by $\initialkc_{1,2}$. 
Let us now introduce the momentum shifts $q_i = \initialkc_i-\initialk_i$,
and then change variables in the integration from the $p_i'$ to the $q_i$. 
In these variables, the matrix element is,
\begin{equation}
\begin{aligned}
\langle \initialkc_1 \initialkc_2 | \,T\, | \initialk_1 \initialk_2 \rangle &= 
\Ampl(\initialk_1 \initialk_2 \rightarrow \initialkc_1\,, \initialkc_2)
 \del^{(4)}(\initialkc_1+\initialkc_2-\initialk_1-\initialk_2) 
\\&=
\Ampl(\initialk_1 \initialk_2 \rightarrow \initialk_1 + q_1\,, \initialk_2 + q_2)
 \del^{(4)}(q_1 + q_2)\,,
\end{aligned}
\end{equation}
yielding
\begin{equation}
\begin{aligned}
\ImpA &= \int \! \df(\initialk_1) \df(\initialk_2)
                 \df(q_1+\initialk_1)\df(q_2+\initialk_2)\;
\\&\qquad\times 
\phi_1(\initialk_1) \phi_1^*(\initialk_1 + q_1)
\phi_2(\initialk_2) \phi_2^*(\initialk_2+q_2) 
\, \del^{(4)}(q_1 + q_2)
\\&\qquad\times 
  \, e^{-i b \cdot q_1/\hbar} 
\,i q_1^\mu  \, \Ampl(\initialk_1 \initialk_2 \rightarrow 
                    \initialk_1 + q_1, \initialk_2 + q_2)\,
\,.
\end{aligned}
\label{eq:impulseGeneralTerm1a}
\end{equation}
We remind the reader of the shorthand notation 
introduced earlier for the phase-space measure,
\begin{equation}
\df(q_1+p_1) = \dd^4 q_1\; \del\bigl((p_1 + q_1)^2 - m_1^2\bigr)
   \Theta(p_1^0 + q_1^0)\,.
\end{equation} 
We can perform the integral over $q_2$ in \eqn{eq:impulseGeneralTerm1a}
using the four-fold delta function. Further relabeling $q_1 \rightarrow q$, we obtain
\begin{equation}
\begin{aligned}
\ImpA= \int \! \df(\initialk_1)\df(\initialk_2) &\dd^4 q  \; 
\del(2\initialk_1 \cdot q + q^2) \del(2 \initialk_2 \cdot q - q^2) 
   \Theta(\initialk_1^0+q^0)\Theta(\initialk_2^0-q^0)\\
&\times  e^{-i b \cdot q/\hbar} \phi_1(\initialk_1) \phi_1^*(\initialk_1 + q)
        \phi_2(\initialk_2) \phi_2^*(\initialk_2-q)
\\& \times  
\,  i q^\mu  \, \Ampl(\initialk_1 \initialk_2 \rightarrow 
                      \initialk_1 + q, \initialk_2 - q)\,.
\label{eq:impulseGeneralTerm1}
\end{aligned}
\end{equation}
Unusually for a physical observable, this contribution is linear in the amplitude.
We emphasize that the incoming and outgoing momenta of this amplitude
do \textit{not\/} correspond to the initial- and final-state momenta of the
scattering process, but rather both correspond to the initial-state momenta,
as they appear in the wavefunction and in its conjugate.  The momentum $q$
looks like a momentum transfer if we examine the amplitude alone, but for
the physical scattering process it represents a difference between the
momentum within the wavefunction and that in the conjugate.  We will
call it a `momentum mismatch'.  As indicated on the first line of
\eqn{eq:defl2}, we should think of this term as an interference of
a standard amplitude with an interactionless forward scattering.
Diagrammatically, we have learned that
\begin{equation}
\begin{aligned}
\hspace*{-10mm}
\ImpA = 
\mathlarger{{\int}} \df(\initialk_1)\df(\initialk_2) & \dd^4 q \, 
\del(2\initialk_1 \cdot q + q^2) \del(2 \initialk_2 \cdot q - q^2)
\Theta(\initialk_1^0+q^0)\Theta(\initialk_2^0-q^0) \, \\
& \times  e^{-i b \cdot q}  \,  iq^\mu \times\!\!\!\!
\begin{tikzpicture}[scale=1.0, 
                baseline={([yshift=-\the\dimexpr\fontdimen22\textfont2\relax]
                    current bounding box.center)},
        ] 
\begin{feynman}

	\vertex (b) ;
	\vertex [above left=1 and 0.66 of b] (i1) {$\phi_1(p_1)$};
	\vertex [above right=1 and 0.33 of b] (o1) {$\phi_1^*(p_1+q)$};
	\vertex [below left=1 and 0.66 of b] (i2) {$\phi_2(p_2)$};
	\vertex [below right=1 and 0.33 of b] (o2) {$\phi_2^*(p_2-q)$};
	
	\begin{scope}[decoration={
		markings,
		mark=at position 0.7 with {\arrow{Stealth}}}] 
		\draw[postaction={decorate}] (b) -- (o2);
		\draw[postaction={decorate}] (b) -- (o1);
	\end{scope}
	\begin{scope}[decoration={
		markings,
		mark=at position 0.4 with {\arrow{Stealth}}}] 
		\draw[postaction={decorate}] (i1) -- (b);
		\draw[postaction={decorate}] (i2) -- (b);
	\end{scope}
	
	\filldraw [color=white] (b) circle [radius=10pt];
	\filldraw [fill=allOrderBlue] (b) circle [radius=10pt];
\end{feynman}
\end{tikzpicture} 
\!\!\!\!.
\end{aligned}
\end{equation}

Turning to the second term, $\ImpB$, in the impulse, 
we again introduce a complete set of states labelled by 
$\finalk_1$, $\finalk_2$ and $X$ so that,
\begin{equation}
\begin{aligned}
\ImpB &= \langle \psi | \, T^\dagger [ \operator P_1^\mu, T] \, |\psi \rangle
\\&= \sum_X \int \! \df(\finalk_1) \df(\finalk_2)\; 
\langle \psi | \, T^\dagger | \finalk_1 \, \finalk_2 \, X \rangle 
\langle \finalk_1  \, \finalk_2 \, X | [\operator P_1^\mu, T] \, |\psi \rangle.
\end{aligned}
\end{equation}
As above, we can now expand the wavefunctions. 
We again label the states in the incoming wavefunction by $\initialk_{1,2}$, 
those in the conjugate ones by $\initialkc_{1,2}$, 
\begin{equation}
\begin{aligned}
\ImpB
&=
\sum_X \int \!\prod_{i = 1, 2}  \df(\finalk_i)  \df(\initialk_i) \df(\initialkc_i)
    \; \phi_i(\initialk_i) \phi^*_i(\initialkc_i) 
      e^{i b \cdot (\initialk_1 - \initialkc_1) / \hbar}
(\finalk_1^\mu - \initialk_1^\mu) \\
&\hspace*{5mm}\times \del^{(4)}(\initialk_1 + \initialk_2 
                                - \finalk_1 - \finalk_2 -\finalk_X) 
               \del^{(4)}(\initialkc_1+\initialkc_2 - \finalk_1 - \finalk_2 -\finalk_X)\\
&\hspace*{5mm}\times \Ampl(\initialk_1\,,\initialk_2 \rightarrow 
                      \finalk_1\,, \finalk_2 \,, \finalk_X)
   \Ampl^*(\initialkc_1\,, \initialkc_2 \rightarrow \finalk_1 \,, \finalk_2 \,, \finalk_X) 
               \,.
\end{aligned}
\label{eq:forcedef2}
\end{equation}
In this expression, $\finalk_X$ denotes the total momentum carried by 
particles in $X$.
The second term in the impulse can thus be interpreted as a weighted cut of an amplitude; the lowest order contribution is a weighted two-particle cut of a one-loop amplitude. 

In order to simplify $\ImpB$, let us
again define the momentum shifts
$q_i = \initialkc_i-\initialk_i$,
and change variables in the integration from the $\initialkc_i$ to the $q_i$,
so that,
\begin{equation}
\begin{aligned}
\ImpB
&=
\sum_X  \int \!\prod_{i = 1,2} \df(\finalk_i)  \df(\initialk_i) 
\df(q_i+\initialk_i)
    \; \phi_i(\initialk_i) \phi^*_i(\initialk_i+q_i) 
      e^{-i b \cdot q_1 / \hbar}
(\finalk_1^\mu - \initialk_1^\mu) \\
&\hspace*{5mm}\times \del^{(4)}(\initialk_1 + \initialk_2 
                                - \finalk_1 - \finalk_2 -\finalk_X) 
       \del^{(4)}(q_1+q_2)\\
&\hspace*{5mm}\times \Ampl(\initialk_1\,, \initialk_2 \rightarrow 
                      \finalk_1\,, \finalk_2 \,, \finalk_X)
\Ampl^*(\initialk_1+q_1\,, \initialk_2+q_2 \rightarrow \finalk_1 \,,\finalk_2 \,, \finalk_X)
               \,.
\end{aligned} \label{eq:impulseGeneralTerm2a}
\end{equation}
We can again perform the integral over $q_2$
using the four-fold delta function, and relabel $q_1 \rightarrow q$ to obtain,
\begin{equation}
\begin{aligned}
\ImpB
&=
\sum_X  \int \!\prod_{i = 1,2} \df(\finalk_i)  \df(\initialk_i) 
\dd^4 q\;
\del(2\initialk_1 \cdot q + q^2) \del(2 \initialk_2 \cdot q - q^2) 
   \Theta(\initialk_1^0+q^0)\Theta(\initialk_2^0-q^0)\\
&\hspace*{5mm}\times \phi_1(\initialk_1) \phi_2(\initialk_2)
    \; \phi^*_1(\initialk_1+q) \phi^*_2(\initialk_2-q) 
      e^{-i b \cdot q / \hbar}
(\finalk_1^\mu - \initialk_1^\mu) \\
&\hspace*{5mm}\times \del^{(4)}(\initialk_1 + \initialk_2 
                                - \finalk_1 - \finalk_2 -\finalk_X) 
\\ &\hspace*{5mm}\times \Ampl(\initialk_1\,, \initialk_2 \rightarrow 
                      \finalk_1\,, \finalk_2 \,, \finalk_X)
\Ampl^*(\initialk_1+q\,, \initialk_2-q \rightarrow \finalk_1 \,,\finalk_2 \,, \finalk_X)
               \,.
\end{aligned} 
\label{eq:impulseGeneralTerm2b}
\end{equation}
\def\xfer{w}
\def\xferb{\overline{w}}
\def\finalkb{\bar\finalk}
The momentum $q$ is again a momentum mismatch.
The momentum transfers $\xfer_i\equiv r_i-p_i$ will play an important role in
analyzing the classical limit, so its convenient to change variables 
to them from the final-state momenta $\finalk_i$,
\begin{equation}
\begin{aligned}
\ImpB
=
\sum_X  \int& \!\prod_{i = 1,2}  \df(\initialk_i) \dd^4 \xfer_i
\dd^4 q\;
\del(2p_i\cdot \xfer_i+\xfer_i^2)\Theta(p_i^0+\xfer_i^0)
\\&\hspace*{5mm}\times
\del(2\initialk_1 \cdot q + q^2) \del(2 \initialk_2 \cdot q - q^2) 
   \Theta(\initialk_1^0+q^0)\Theta(\initialk_2^0-q^0)\\
&\hspace*{5mm}\times \phi_1(\initialk_1) \phi_2(\initialk_2)
    \; \phi^*_1(\initialk_1+q) \phi^*_2(\initialk_2-q) 
\\&\hspace*{5mm}\times 
      e^{-i b \cdot q / \hbar}\,\xfer_1^\mu \;
\del^{(4)}(\xfer_1+\xfer_2+\finalk_X) 
\\ &\hspace*{5mm}\times 
 \Ampl(\initialk_1\,, \initialk_2 \rightarrow 
                      \initialk_1+\xfer_1\,, \initialk_2+\xfer_2 \,, \finalk_X)
\\ &\hspace*{5mm}\times 
 \Ampl^*(\initialk_1+q, \initialk_2-q \rightarrow 
   \initialk_1+\xfer_1 \,,\initialk_2+\xfer_2 \,, \finalk_X)
               \,.
\end{aligned} 
\label{eq:impulseGeneralTerm2}
\end{equation}
Diagrammatically, this second contribution to the impulse is
\begin{equation}
\begin{aligned}
\usetikzlibrary{decorations.markings}
\ImpB = 
\sum_X \mathlarger{\int} \!& \prod_{i = 1,2}  \df(\initialk_i) \dd^4 \xfer_i
\dd^4 q\; \del(2p_i\cdot \xfer_i+\xfer_i^2)\Theta(p_i^0+\xfer_i^0)
\\&\hspace*{5mm}\times
\del(2\initialk_1 \cdot q + q^2) \del(2 \initialk_2 \cdot q - q^2) 
   \Theta(\initialk_1^0+q^0)\Theta(\initialk_2^0-q^0)\\
& \hspace*{5mm}\times \; e^{-i b \cdot q / \hbar}\,\xfer_1^\mu \; \del^{(4)}(\xfer_1+\xfer_2+\finalk_X)  
\\ & \hspace*{5mm} \times \!\!\!\!\!
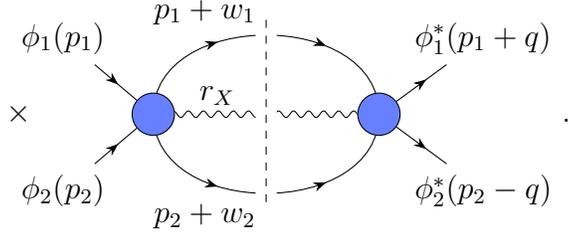
\begin{tikzpicture}[scale=1.0, 
                baseline={([yshift=-\the\dimexpr\fontdimen22\textfont2\relax]
                    current bounding box.center)},
        ] 
\begin{feynman}
\begin{scope}
	\vertex (ip1) ;
	\vertex [right=3 of ip1] (ip2);
	\node [] (X) at ($ (ip1)!.5!(ip2) $) {};
	\begin{scope}[even odd rule]
	\begin{pgfinterruptboundingbox} 
	\path[invclip] ($  (X) - (4pt, 35pt) $) rectangle ($ (X) + (4pt,35pt) $) ;
	\end{pgfinterruptboundingbox} 

	\vertex [above left=0.66 and 0.5 of ip1] (q1) {$ \phi_1(p_1)$};
	\vertex [above right=0.66 and 0.33 of ip2] (qp1) {$ \phi^*_1(p_1 + q)$};
	\vertex [below left=0.66 and 0.5 of ip1] (q2) {$ \phi_2(p_2)$};
	\vertex [below right=0.66 and 0.33 of ip2] (qp2) {$ \phi^*_2(p_2 - q)$};
	
	\diagram* {
		(ip1) -- [photon]  (ip2)
	};
	\begin{scope}[decoration={
		markings,
		mark=at position 0.4 with {\arrow{Stealth}}}] 
		\draw[postaction={decorate}] (q1) -- (ip1);
		\draw[postaction={decorate}] (q2) -- (ip1);
	\end{scope}
	\begin{scope}[decoration={
		markings,
		mark=at position 0.7 with {\arrow{Stealth}}}] 
		\draw[postaction={decorate}] (ip2) -- (qp1);
		\draw[postaction={decorate}] (ip2) -- (qp2);
	\end{scope}
	\begin{scope}[decoration={
		markings,
		mark=at position 0.34 with {\arrow{Stealth}},
		mark=at position 0.7 with {\arrow{Stealth}}}] 
		\draw[postaction={decorate}] (ip1) to [out=90, in=90,looseness=1.2] node[above left] {{$p_1 + w_1$}} (ip2);
		\draw[postaction={decorate}] (ip1) to [out=270, in=270,looseness=1.2]node[below left] {$p_2 + w_2$} (ip2);
	\end{scope}

	\node [] (Y) at ($(X) + (0,1.4)$) {};
	\node [] (Z) at ($(X) - (0,1.4)$) {};
	\node [] (k) at ($ (X) - (0.65,-0.25) $) {$\finalk_X$};

	\filldraw [color=white] ($ (ip1)$) circle [radius=8pt];
	\filldraw  [fill=allOrderBlue] ($ (ip1) $) circle [radius=8pt];
	
	\filldraw [color=white] ($ (ip2) $) circle [radius=8pt];
	\filldraw  [fill=allOrderBlue] ($ (ip2) $) circle [radius=8pt];
	
\end{scope}
\end{scope}
	  \draw [dashed] (Y) to (Z);
\end{feynman}
\end{tikzpicture} 
.
\end{aligned}
\end{equation}

\subsection{The momentum radiated during a collision}
\label{sec:radiatedmomentum}

A familiar classical observable is the energy radiated by an
accelerating particle, for example during a scattering process. More generally we can compute the four-momentum radiated. In quantum mechanics there is no precise 
prediction for the energy or the momentum radiated by localised particles;
we obtain a continuous spectrum if we measure a large number of events. 
However we can compute the expectation value of the four-momentum
radiated during a scattering process. This is a well-defined observable, and as we will see it is on-shell in the sense that it can be expressed in terms of on-shell amplitudes.

To define the observable, let us again surround the collision with detectors which 
measure outgoing radiation of some type. 
We may imagine two different contexts: scattering in electrodynamics with
radiation of photons, and gravitational scattering with radiation of gravitons.
In both cases, we will call the radiated particles `messengers'.
 Let $\operator K^\mu$ be the momentum operator 
for whatever field is radiated; then the expectation of the radiated momentum is
\begin{equation}
\begin{aligned}
\langle k^\mu \rangle &= {}_\textrm{out}{\langle} \psi | \operator K^\mu U(\infty, -\infty) \, | \psi \rangle_\textrm{in} \\
&= {}_\textrm{in}{\langle} \psi | \, U(\infty, -\infty)^\dagger \operator K^\mu U(\infty, -\infty) \, | \psi \rangle_\textrm{in},
\end{aligned}
\end{equation}
where $U(\infty, -\infty)$ is again the time evolution operator from the far past 
to the far future---that is, the $S$ matrix.  Once again we can anticipate that 
the radiation will be expressed in terms of amplitudes. 
Again rewriting $S = 1 + i T$, the expectation becomes,
\begin{align}
\Rad^\mu \equiv \langle k^\mu \rangle &= {}_\textrm{in}{\langle} \psi | \, S^\dagger \operator K^\mu S \, | \psi \rangle_\textrm{in}
= {}_\textrm{in}\langle \psi | \, T^\dagger \operator K^\mu T \, | \psi \rangle_\textrm{in},
\end{align}
because $\operator K^\mu |\psi \rangle_\textrm{in} = 0$ (there are no quanta of radiation in the incoming state). 

\def\dfg{\df{\gamma}}
We can insert a complete set of states $|X k \finalk_1\finalk_2\rangle$ 
containing at least one radiated messenger
of momentum $k$, and
write the expectation value of the radiated momentum as follows,
\begin{equation}
\begin{aligned}
\Rad^\mu &= \sum_X \int \df(k) \df(\finalk_1) \df(\finalk_2)\;
\langle \psi | \, T^\dagger \, | k \, \finalk_1 \, \finalk_2 \, X \rangle 
  k_X^\mu \langle k\, \finalk_1 \, \finalk_2 \, X \,| \, T \, | \psi \rangle\\
&= \sum_X \int \df(k) \df(\finalk_1) \df(\finalk_2)\;
 k_X^\mu \bigl|\langle k\, \finalk_1 \, \finalk_2 \, X \,| \, T \,
              | \psi \rangle\bigr|^2\,,
\end{aligned}
\label{eq:radiationTform}
\end{equation}
In this expression, $X$ can again be empty, and $k_X^\mu$ is the sum of the explicit
messenger momentum $k^\mu$ and the momenta of any messengers in the state $X$.
Notice that we are including explicit phase-space integrals for particles 1 and 2, consistent with our assumption that the number of these particles is conserved during the process. The state $| k \rangle$ describes a
radiated messenger;
the phase space integral over $k$ implicitly includes a sum over its helicity.

\def\squeeze{\hspace*{-1.5pt}}
Expanding the initial state, we find that the expectation value of the radiated momentum is given by,
\begin{equation}
\begin{aligned}
\Rad^\mu = \sum_X &\int \df(k) \df(\finalk_1) \df(\finalk_2)\,
k_X^\mu \bigg| \int \df(\initialk_1)\df(\initialk_2)\, 
   e^{i b \cdot \initialk_1/\hbar} \phi_1(\initialk_1) \phi_2(\initialk_2) 
\\ & \times \Ampl(\initialk_1\,, \initialk_2 \rightarrow 
                  \finalk_1\,, \finalk_2\,, k\,, \finalk_X) 
            \del^{(4)}(\initialk_1 + \initialk_2 - \finalk_1 - \finalk_2 - k - \finalk_X) \bigg|^2
\\= \sum_X &\int \df(k) \prod_{i=1,2}\df(\finalk_i) 
          \df(\initialk_i)\df(\initialkc_i)\;
\phi_i(\initialk_i) \phi_i^*(\initialkc_i)\, k_X^\mu \,
e^{i b \cdot (\initialk_1-\initialkc_1)/\hbar} 
\\&\times \Ampl(\initialk_1\,, \initialk_2 \rightarrow 
                \finalk_1\,, \finalk_2\,, k\,, \finalk_X)
 \del^{(4)}(\initialk_1 + \initialk_2 - \finalk_1 - \finalk_2 - k - \finalk_X) 
\\&\qquad\times \Ampl^*(\initialkc_1\,, \initialkc_2 \rightarrow 
                          \finalk_1\,, \finalk_2\,, k\,, \finalk_X) 
\del^{(4)}(\initialkc_1 + \initialkc_2 - \finalk_1 - \finalk_2 - k - \finalk_X)\,.
\end{aligned}
\label{eq:ExpectedMomentum}
\end{equation}
We can again introduce momentum transfers, $q_i=\initialkc_i-\initialk_i$, 
and trade
the integrals over $\initialkc_i$ for integrals over the $q_i$.  One of the
four-fold $\delta$ functions will again become $\del^{(4)}(q_1+q_2)$, and we can
use it to perform the $q_2$ integrations.  
We again relabel $q_1\rightarrow q$.
The integration leaves behind
a pair of on-shell $\delta$ functions and positive-energy $\Theta$
functions just as in \eqns{eq:impulseGeneralTerm1}{eq:impulseGeneralTerm2},
\begin{equation}
\begin{aligned}
\Rad^\mu = \sum_X \int &\df(k) \prod_{i=1,2}\df(\finalk_i) \df(\initialk_i) \dd^4 q\;
\phi_1(\initialk_1) \phi_2(\initialk_2) 
   \phi_1^*(\initialk_1+q) \phi_2^*(\initialk_2-q) \,
\\&\times 
\del(2\initialk_1 \cdot q + q^2) \del(2 \initialk_2 \cdot q - q^2) 
   \Theta(\initialk_1{}^0+q^0)\Theta(\initialk_2{}^0-q^0)
\\&\times 
 k_X^\mu \, e^{-i b \cdot q/\hbar} 
 \,\del^{(4)}(\initialk_1 + \initialk_2 
                   - \finalk_1 - \finalk_2 - k - \finalk_X)%
\\&\times 
  \Ampl(\initialk_1\,, \initialk_2 \rightarrow \finalk_1\,, \finalk_2\,, k\,, \finalk_X)
  \Ampl^*(\initialk_1+q\,, \initialk_2-q \rightarrow 
          \finalk_1\,, \finalk_2\,, k\,, \finalk_X)
\,.
\end{aligned}
\label{eq:ExpectedMomentum2b}
\end{equation}
We emphasise that this is an all-orders expression: the amplitude 
$\Ampl(\initialk_1,\squeeze \initialk_2 \squeeze\rightarrow
\squeeze \finalk_1,\squeeze \finalk_2,\squeeze k,\squeeze \finalk_X)$ 
includes all loop corrections, though of course it can be expanded in 
perturbation theory. 
The corresponding real-emission contributions are present in the sum
over states $X$.   If we truncate the amplitude at a fixed order in
perturbation theory, we should similarly truncate the sum over states.
Given that the expectation value is expressed in terms of an on-shell amplitude, it is also appropriate to regard this observable as a fully on-shell quantity.
As in \eqns{eq:impulseGeneralTerm1}{eq:impulseGeneralTerm2}, $q$ represents a 
momentum mismatch
rather than a momentum transfer.  Here too, the scattering
momentum transfers $\xfer_i = r_i-p_i$
will play an important role in our later discussion of the classical limit,
and it is convenient to change variables from the $r_i$ to make use of them,
\begin{equation}
\begin{aligned}
\Rad^\mu = \sum_X \int \df(k)& \prod_{i=1,2} \df(\initialk_i) \dd^4\xfer_i\dd^4 q\;
\del(2p_i\cdot \xfer_i+\xfer_i^2)\Theta(p_i^0+\xfer_i^0)
\\&\times 
\del(2\initialk_1 \cdot q + q^2) \del(2 \initialk_2 \cdot q - q^2) 
   \Theta(\initialk_1{}^0+q^0)\Theta(\initialk_2{}^0-q^0)
\\&\times 
\phi_1(\initialk_1) \phi_2(\initialk_2) 
   \phi_1^*(\initialk_1+q) \phi_2^*(\initialk_2-q) \,
\\& \times  k_X^\mu \, e^{-i b \cdot q/\hbar}  \del^{(4)}(\xfer_1+\xfer_2+ k+ \finalk_X)%
\\&\times 
  \Ampl(\initialk_1\,, \initialk_2 \rightarrow 
        \initialk_1+\xfer_1\,, \initialk_2+\xfer_2\,, k\,, \finalk_X)
\\&\times 
  \Ampl^*(\initialk_1+q\,, \initialk_2-q \rightarrow 
          \initialk_1+\xfer_1\,, \initialk_2+\xfer_2\,, k\,, \finalk_X)
\,.
\end{aligned}
\label{eq:ExpectedMomentum2}
\end{equation}

It can be useful to represent the observables diagrammatically. Two equivalent expressions for the radiated momentum are helpful:
\begin{equation}
\usetikzlibrary{decorations.markings}
\usetikzlibrary{positioning}
\begin{aligned}
\Rad^\mu \!=\!
& \sum_X  \mathlarger{\int} \df(k)\df(\finalk_1)\df(\finalk_2)\;
  k_X^\mu 
\\& \times\left| \mathlarger{\int} \df(\initialk_1) \df(\initialk_2)\; 
      e^{i b \cdot \initialk_1/\hbar} \, 
   \del^{(4)}(\initialk_1 + \initialk_2 - \finalk_1 - \finalk_2 - k - \finalk_X) \hspace*{-13mm}
\begin{tikzpicture}[scale=1.0, 
                baseline={([xshift=-5cm,yshift=-\the\dimexpr\fontdimen22\textfont2\relax]
                    current bounding box.center)},
        ] 
\begin{feynman}
	\vertex (b) ;
	\vertex [above left=of b] (i1) {$\phi_1(\initialk_1)$};
	\vertex [above right=of b] (o1) {$\finalk_1$};
	\vertex [above right =0.2 and 1.4 of b] (k) {$k$};
    \vertex [below right =0.2 and 1.4 of b] (X) {$\finalk_X$};
	\vertex [below left=1 and 1 of b] (i2) {${\phi_2(\initialk_2)}$};
	\vertex [below right=1 and 1 of b] (o2) {$\finalk_2$};
	\diagram* {
        (b) -- [photon, photonRed] (k)
        (b) -- [boson] (X)
	};
	\begin{scope}[decoration={
		markings,
		mark=at position 0.7 with {\arrow{Stealth}}}] 
		\draw[postaction={decorate}] (b) -- node [right=4pt] {}(o2);
		\draw[postaction={decorate}] (b) -- node [left=4pt] {} (o1);
	\end{scope}
	\begin{scope}[decoration={
		markings,
		mark=at position 0.42 with {\arrow{Stealth}}}] 
		\draw[postaction={decorate}] (i1) -- node [left=4pt] {} (b);
		\draw[postaction={decorate}] (i2) --node [right=4pt] {} (b);
	\end{scope}	
	\filldraw [color=white] (b) circle [radius=10pt];
	\filldraw [fill=allOrderBlue] (b) circle [radius=10pt];	
\end{feynman}
\end{tikzpicture}
 \right|^2,
\end{aligned}
\label{RadiationPerfectSquare}
\end{equation}
which is a direct pictorial interpretation of equation~\eqref{eq:ExpectedMomentum}, and
\begin{equation}
\usetikzlibrary{decorations.markings}
\usetikzlibrary{positioning}
\begin{aligned}
\Rad^\mu =
\sum_X \mathlarger{\int} \df(k) \prod_{i = 1, 2} \df(\finalk_i)
& \df(\initialk_i) \df(\initialkc_i)\; k_X^\mu  \, e^{i b \cdot (\initialk_1 - \initialkc_1)/\hbar}   \\
& \times \del^{(4)}(\initialk_1 + \initialk_2 - \finalk_1 - \finalk_2 - k - \finalk_X) \\
& \times \del^{(4)}(\initialkc_1 + \initialkc_2 - \finalk_1 - \finalk_2 - k - \finalk_X) \\
&\times
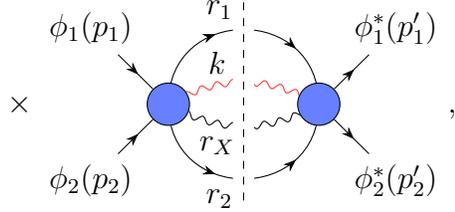
\begin{tikzpicture}[scale=1.0, 
                baseline={([yshift=-\the\dimexpr\fontdimen22\textfont2\relax]
                    current bounding box.center)},
        ] 
\begin{feynman}
\begin{scope}
	\vertex (ip1) ;
	\vertex [right=2 of ip1] (ip2);
    \node [] (X) at ($ (ip1)!.5!(ip2) $) {};
	\begin{scope}[even odd rule]
	\begin{pgfinterruptboundingbox} 
	\path[invclip] ($  (X) - (4pt, 30pt) $) rectangle ($ (X) + (4pt,30pt) $) ;
	\end{pgfinterruptboundingbox} 

	\vertex [above left=0.66 and 0.33 of ip1] (q1) {$ \phi_1(\initialk_1)$};
	\vertex [above right=0.66 and 0.33 of ip2] (qp1) {$ \phi^*_1(\initialkc_1)$};
	\vertex [below left=0.66 and 0.33 of ip1] (q2) {$ \phi_2(\initialk_2)$};
	\vertex [below right=0.66 and 0.33 of ip2] (qp2) {$ \phi^*_2(\initialkc_2)$};
	
	\diagram* {
		(ip1) -- [photon, out=30, in=150, photonRed]  (ip2)
		(ip1) -- [photon, out=330, in=210]  (ip2)
	};

	\begin{scope}[decoration={
		markings,
		mark=at position 0.4 with {\arrow{Stealth}}}] 
		\draw[postaction={decorate}] (q1) -- (ip1);
		\draw[postaction={decorate}] (q2) -- (ip1);
	\end{scope}
	\begin{scope}[decoration={
		markings,
		mark=at position 0.7 with {\arrow{Stealth}}}] 
		\draw[postaction={decorate}] (ip2) -- (qp1);
		\draw[postaction={decorate}] (ip2) -- (qp2);
	\end{scope}
	\begin{scope}[decoration={
		markings,
		mark=at position 0.38 with {\arrow{Stealth}},
		mark=at position 0.74 with {\arrow{Stealth}}}] 
		\draw[postaction={decorate}] (ip1) to [out=90, in=90,looseness=1.7] node[above left] {{$ \finalk_1$}} (ip2);
		\draw[postaction={decorate}] (ip1) to [out=270, in=270,looseness=1.7]node[below left] {${\finalk_2}$} (ip2);
	\end{scope}

	\node [] (Y) at ($(X) + (0,1.5)$) {};
	\node [] (Z) at ($(X) - (0,1.5)$) {};
	\node [] (k) at ($ (X) - (0.35,-0.55) $) {$k$};
	\node [] (x) at ($ (X) - (0.35,0.55) $) {$\finalk_X$};

	\filldraw [color=white] ($ (ip1)$) circle [radius=8pt];
	\filldraw  [fill=allOrderBlue] ($ (ip1) $) circle [radius=8pt];
		
	\filldraw [color=white] ($ (ip2) $) circle [radius=8pt];
	\filldraw  [fill=allOrderBlue] ($ (ip2) $) circle [radius=8pt];
	
\end{scope} 
\end{scope}
	  \draw [dashed] (Y) to (Z);
\end{feynman}
\end{tikzpicture},
\end{aligned}
\end{equation}
which demonstrates that we can think of the expectation value as the weighted cut of a loop amplitude.
As $X$ can be empty, the lowest-order contribution arises from the 
weighted cut of a two-loop 
amplitude.

\subsection{Conservation of momentum}
\label{sect:allOrderConservation}

The expectation of the radiated momentum is not completely independent of the impulse. 
In fact the relation between these quantities is physically rich. In the classical electrodynamics of point particles, for example, the impulse is due to a 
combination of the usual Lorentz force 
and the ALD radiation reaction (more precisely,
 the total time integrals of these forces). The Lorentz force exchanges momentum between particles 1 and 2, while the radiation reaction accounts for the irreversible loss of momentum due to radiation. Of course, the ALD force is a notably subtle issue in the classical theory.
 
In the quantum theory, there can be no question of violating conservation of momentum, so the quantum observables we have defined must already include all the effects which would classically be attributed to both the Lorentz and ALD forces. In particular it must be the case that our definitions respect conservation of momentum. It is easy to demonstrate this formally to all orders using our definitions. Later, in section~\ref{sec:ALD}, we will indicate how the radiation reaction is included in the impulse more explicitly.

Our scattering processes involve two incoming particles. Consider, then,
\begin{equation}
\begin{aligned}
\langle \Delta p_1^\mu \rangle + \langle \Delta p_2^\mu \rangle &= 
\langle \psi | i [ \operator P_1^\mu + \operator P_2^\mu, T ] | \psi \rangle 
 + \langle \psi | T^\dagger [ \operator P_1^\mu + \operator P_2^\mu, T ] | \psi \rangle \\
&= \bigl\langle \psi \big| i \bigl[ \textstyle{\sum_i} \operator P_i^\mu, T \bigr] 
\big| \psi \bigr\rangle 
+ \langle \psi | T^\dagger [ \operator P_1^\mu + \operator P_2^\mu, T ] | \psi \rangle,
\end{aligned}
\end{equation}
where the sum $\sum \operator P_i^\mu$ is over all momentum operators in the theory. The second equality above holds because $\operator P_i^\mu | \psi \rangle = 0$ for $i \neq 1,2$; only quanta of fields 1 and 2 are present in the incoming state. Next, we use the fact that the total momentum is time independent, in other
words
\begin{equation}
\Bigl[ \sum \operator P_i^\mu, T \Bigr] = 0,
\end{equation}
where the sum extends over all fields. Consequently,
\begin{equation}
\langle \psi | i [ \operator P_1^\mu + \operator P_2^\mu, T ] | \psi \rangle =  
\bigl\langle \psi \big| i \bigl[ \textstyle{\sum_i} \operator P_i^\mu, T \bigr] \big| 
\psi \bigr\rangle = 0.
\label{eq:commutatorVanishes}
\end{equation}
Thus the first term $\langle \psi | i [ \operator P_1^\mu, T ] | \psi \rangle$ in 
the impulse~(\ref{eq:defl1}) describes only the exchange of momentum between 
particles~1 and~2; in this sense it is associated with the classical Lorentz force (which 
shares this property) rather than with the classical ALD force (which does not). The second 
term in the impulse, on the other hand, includes radiation. To make the situation as clear 
as possible, let us restrict attention to the case where the only other momentum operator is $\operator K^\mu$, the momentum operator for the messenger field. 
Then we know that $[ \operator P_1^\mu + \operator P_2^\mu + \operator K^\mu, T] = 0$, and conservation of 
momentum at the level of expectation values is easy to demonstrate:
\begin{equation}
\langle \Delta p_1^\mu \rangle + \langle \Delta p_2^\mu \rangle = 
- \langle \psi | T^\dagger [ \operator K^\mu, T ] | \psi \rangle = 
- \langle \psi | T^\dagger  \operator K^\mu T  | \psi \rangle = 
- \langle k^\mu \rangle = - \Rad^\mu\,,
\end{equation}
once again using the fact that there are no messengers in the incoming state.

In the classical theory, radiation reaction is a subleading effect, entering for two-body scattering at order $e^6$ in perturbation theory in electrodynamics. This is also the case in the quantum theory.
To see why, we again expand the operator product in the second term of \eqn{eq:defl1} using a complete set of states:
\begin{equation}
 \langle \psi | \, T^\dagger [ \operator P_1^\mu, T] \, |\psi \rangle = \sum_X \int \! \df(\finalk_1)\df(\finalk_2)\;
 \langle \psi | \, T^\dagger | \finalk_1 \, \finalk_2 \, X\rangle 
 \langle \finalk_1 \, \finalk_2 \, X | [\operator P_1^\mu, T] \, |\psi \rangle\,.
\end{equation}
The sum over $X$ is over all states, including an implicit integral over their momenta and a sum over any other quantum numbers.
The inserted-state momenta of particles 1 and~2 (necessarily present) are labeled by $\finalk_i$, and the corresponding integrations
over these momenta by $\df(\finalk_i)$.  These will ultimately become integrations over the final-state momenta in
the scattering.
To make the loss of momentum due to radiation explicit at this level, we note that
\begin{equation}
 \langle \psi | \, T^\dagger [ \operator P_1^\mu +  \operator P_2^\mu, T] \, |\psi \rangle 
  =
 -\sum_X \int \! \df(\finalk_1)\df(\finalk_2)\;
 \langle \psi | \, T^\dagger | \finalk_1 \, \finalk_2 \, X\rangle 
 \langle \finalk_1 \, \finalk_2 \, X | \, \operator P_X^\mu T \,  |\psi \rangle\,,
\end{equation}
where $\operator P_X$ is the sum over momentum operators of all quantum fields other than the scalars~1 and 2. 
The sum over all states $X$ will contain, for example, terms where the state $X$ 
includes messengers of momentum $k^\mu$ along with other massless particles.
We can further restrict attention to the contributions of the messenger's momentum 
to $\operator P_X^\mu$. 
This contribution produces a net change of momentum of particle 1 given by
\begin{equation}
-\sum_X \int \!  \df(k) \df(\finalk_1)\df(\finalk_2)\; k^\mu \, 
   \langle \psi | \, T^\dagger | k \, \finalk_1 \, \finalk_2 \, X\rangle
    \langle k\, \finalk_1 \, \finalk_2 \, X | \, T \,  |\psi \rangle 
= - \langle k^\mu \rangle\,,
\end{equation} 
with the help of equation~\eqref{eq:radiationTform}. Thus we explicitly see the net loss of 
momentum due to radiating messengers. This quantity
is suppressed by factors of $g$ because of the 
additional state. The lowest order case corresponds to $X = \emptyset$; as there are two 
quanta in $|\psi \rangle$, we must compute the modulus squared of a five-point tree 
amplitude. The term is proportional to $g^6$, where $g$ is the coupling in the
elementary three-point amplitude; 
as far as the impulse is concerned, it is a next-to-next-to-leading
order (NNLO) effect. 
Other particles in the state $X$, and other contributions to its momentum, describe higher-order effects.

\section{Classical Point Particles}
\label{sec:PointParticleLimit}

The two observables we have discussed  --- the impulse and
the expectation value of the radiated momentum --- are designed to be well-defined in both the quantum and the classical theories. 
As we approach the classical limit, these expectation values should reduce to the classical impulse and the classical radiated momentum. 
This should ensure that we are able to explore the $\hbar \rightarrow 0$ limit.  

We have already discussed in \sect{RestoringHBar} how
to make explicit the factors of $\hbar$ in the observables.  We must still discuss the issue of suitable wavefunctions. We must first ensure that the chosen wavefunctions have the
desired classical limit.  At that point, we could in principle perform the full quantum
calculation, using the specific wavefunction we choose, and expand 
in the $\hbar\rightarrow 0$
limit at the end.  However, we also want to choose wavefunctions that allow us to approach
the limit as early as possible in the calculation, without relying on 
their detailed properties.  This will lead us to impose stronger constraints on the
choice than the mere existence of a suitable classical limit.

\subsection{Wavefunctions}
\label{subsec:Wavefunctions}

\def\pcl{\breve{p}}
\def\ucl{{u}}
\def\spread{\sigma^2}
Heuristically, the wavefunctions for the scattered particles must satisfy
two separate conditions.  We will take these to be wavepackets,
characterized by a smearing or spread in 
momenta\footnote{Evaluating positions and uncertainties therein in 
relativistic field theory is a bit delicate, and we will not consider
the question in this article.}.  That spread should not be too large,
so that the interaction with the other particle cannot peer into the
details of the wavepacket.  At the same time, the details of the
wavepacket should not be sensitive to quantum effects.   

Let us ground our intuition about scales by first examining nonrelativistic
wavefunctions.  An example of a minimum-uncertainty wavefunction 
in momentum space (ignoring normalization) for a particle of mass $m$
growing sharper in the $\hbar\rightarrow 0$
limit has the form,
\begin{equation}
\exp\left( -\frac{\v{p}\mskip1mu{}^2}{2 \hbar m \lcomp/ \lpack^2}\right)
= \exp\left( -\frac{\v{p}\mskip1mu{}^2}{2m^2 \lcomp^2/\lpack^2}\right)\,,
\label{NonrelativisticMomentumSpaceWavefunction}
\end{equation}
where $\lcomp$ is the particle's Compton wavelength, and where $\lpack$
is an additional parameter with dimensions of length.
We can obtain
the conjugate in position space by Fourier-transforming,
\begin{equation}
\exp\left( -\frac{(\v{x}-\v{x}_0)^2}{2 \lpack^2}\right)\,.
\end{equation}
The precision with which we know the particle's location is given
by $\lpack$, which we could take as an intrinsic measure of the wavefunction's
spread.

\def\uapprox{\breve{u}}
This suggests that in considering relativistic wavefunctions, we should also take
the dimensionless parameter controlling the approach to the classical
limit in momentum space to be the square of the ratio of the Compton wavelength $\lcomp$
to the intrinsic spread~$\lpack$,
\begin{equation}
\xi \equiv \biggl(\frac{\lcomp}{\lpack}\biggr){\vphantom{\frac{\lcomp}{\lpack}}}^2\,.
\end{equation}
We obtain the classical result by studying the behavior
of observables as $\xi\rightarrow 0$.
Towards
the limit the wavefunctions must be sharply peaked around the classical
value for the momenta, $\pcl_i = m_i \ucl_i$ with the classical four-velocities
$\ucl_i$ normalized to $\ucl_i^2 = 1$.  We can express this requirement through the
conditions,
\begin{equation}
\begin{aligned}
\langle p_i^\mu\rangle &= \int \df(p_i)\; p_i^\mu\, |\phi(p_i)|^2 = 
  m_i \uapprox_i^\mu f_{p,i}(\xi)\,,
\\ f_{p,i}(\xi) &= 1+\Ord(\xi^{\beta'})\,,
\\ \uapprox_i\cdot \ucl_i &= 1+\Ord(\xi^{\beta''})\,,
\\ \spread(p_i)/m_i^2 &=
\langle \bigl(p_i-\langle p_i\rangle\bigr){}^2\rangle/m_i^2
\\&= \bigl(\langle p_i^2\rangle-\langle p_i\rangle{}^2\bigr)/m_i^2
= c_\Delta \xi^\beta\,,
\end{aligned}
\label{eq:expectations}
\end{equation}
where $c_\Delta$ is a constant of order unity,
and the $\beta$s are simple rational exponents. (The integration measure for $p_i$ enforces 
$\langle p_i^2\rangle = m_i^2$.)
For the simplest wavefunctions, $\beta=1$.  
This spread around the classical value is 
not necessarily positive, as the difference $p_i-\langle p_i^\mu\rangle$ 
may be spacelike, and the expectation of its Lorentz square possibly negative.
For that reason, we should resist the usual temptation of taking its square root
to obtain a variance.

Because of
the phase-space integrals over the initial-state momenta, which enforce
the on-shell conditions $p_i^2=m_i^2$, the only Lorentz invariant built out
of each $p_i$ is constant, and so the wavefunction cannot usefully depend on 
it.  This means the wavefunction must depend on at least one four-vector
parameter.  The simplest wavefunctions will depend on exactly one four-vector,
which we can think of as the (classical)
four-velocity $\ucl$ of the corresponding particle.  It can depend only
on the dimensionless combination $p\cdot \ucl/m$ in addition to the
parameter $\xi$.  The simplest form will be a function of 
these two in the combination $p\cdot\ucl/(m\xi)$, so that
large deviations from $m \,\ucl$ will be suppressed in a classical quantity.
The wavefunction will have additional dependence
on $\xi$ in its normalization.

To see the meaning of the constraints more quantitatively, let us examine
$\ImpA$~(\ref{eq:impulseGeneralTerm1}) more closely. It has the form
of an amplitude integrated over the on-shell phase space for both of the
incoming momenta, subject to additional $\delta$ function constraints
--- and then weighted by a phase $e^{-ib\cdot q/\hbar}$ dependent on the
momentum mismatch $q$, and finally integrated over
all $q$.  
As one nears the classical limit, the wavefunction and its conjugate should
both represent the particle, that is they should be sharply peaked,
and in addition their overlap should be $\Ord(1)$, up to corrections of
$\Ord(\xi)$.  
The amplitude will vary slowly on the scale of the wavefunction when one is close
to the limit.
We will integrate the momentum mismatch $q$ over all possible values,
so it is somewhat a matter of taste how we normalize it.  Nonetheless,
if we take $q_0$ to be a `characteristic' value of $q$, 
requiring the overlap to be
$\Ord(1)$ is equivalent to requiring that $\phi^*(p+q)$ does not differ
much from $\phi^*(p)$, which in turn requires that the derivative at $p$ is
small or that,
\begin{equation}
\frac{q_0\cdot\ucl_i}{m\xi} \ll 1\,.
\label{qConstraint}
\end{equation}
If we scale $q$ by $1/\hbar$, replacing
the momentum by a wavenumber, this constraint takes the following form,
\begin{equation}
\qb_0\cdot\ucl_i\,\lpack \ll \sqrt{\xi}.
\label{qbConstraint}
\end{equation}
We next examine another rapidly varying factor that appears in all our integrands,
the delta functions in $q$ arising from the on-shell constraints on the conjugate
momenta $\initialkc_i$. 
These delta functions, appearing
in eqns.~(\ref{eq:impulseGeneralTerm1},\ref{eq:impulseGeneralTerm2},\ref{eq:ExpectedMomentum2}), take the form,
\begin{equation}
\del(2p_i\cdot q+q^2) = \frac1{\hbar m_i}\del(2\qb\cdot u_i+\lcomp \qb^2)\,.
\label{universalDeltaFunction}
\end{equation}
The integration over the initial momenta $\initialk_i$ and the initial
wavefunctions will smear out these
delta functions to sharply peaked functions whose scale is the same order
as the original wavefunctions.  As $\xi$ gets smaller, this function
will turn back into a delta function imposed on the $\qb$ integration.
In addition to its dependence on $\xi$, it
will depend on two additional dimensionless ratios,
\begin{equation}
\lcomp \sqrt{-\qb^2}
\qquad
\textrm{and}\qquad
\frac{\qb\cdot u_i}{\sqrt{-\qb^2}}\,.
\end{equation}
(The characteristic momentum mismatch $q$ is necessarily spacelike.)
Let us call $1/\sqrt{-\qb^2}$ a `scattering length' $\lscatt$.  In terms
of this length, our two dimensionless ratios are,
\begin{equation}
\frac{\lcomp}{\lscatt}
\qquad
\textrm{and}\qquad
{\qb\cdot u_i}\,\lscatt\,.
\end{equation}
The argument of the delta function is polynomial in the two ratios, so we expect
them to be constrained to be of order an appropriate power of the spread $\xi$.
The direction-averaging implicit in the integration over the $\initialk_i$
will lead to a constraint on two positive quantities built out of the ratios,
so in general we expect both to be constrained independently.  As
suggested again by the nonrelativistic limit, the smallest
reasonable power of $\xi$ we can imagine emerging as a constraint from the later
$\qb$ integration is one-half,
\begin{equation}
\begin{aligned}
\frac{\lcomp}{\lscatt} &\lesssim \sqrt{\xi}\,,
\\{\qb\cdot u_i}\,\lscatt &\lesssim \sqrt{\xi}\,.
\end{aligned}
\label{deltaConstraints}
\end{equation}
If we take a higher power of $\xi$, the constraints would grow stronger.
If we had not already scaled out a factor of $\hbar$ from $q$, these constraints would
make it natural to do so.  Combining the second constraint with \eqn{qbConstraint},
we obtain the constraint $\lpack\ll\lscatt$.  The first constraint is weaker,
$\lpack\lesssim\lscatt$, but on physical grounds as well we should expect the
stronger one: only if the wavefunction spread is much smaller than the scattering
length can we expect the interaction's probing of the internals of the particle
to be negligible.

Combining the stronger constraint with $\xi\ll 1$,
we obtain our first version of the `Goldilocks' inequalities,
\begin{equation}
\lcomp \ll \lpack \ll \lscatt\,.
\label{Goldilocks1}
\end{equation}
As we shall see later in the explicit evaluation of $\ImpA$, 
$\lscatt\sim \sqrt{-b^2}$; this follows on dimensional grounds along with
the observation that the integrals in $\ImpA$ lead to no large 
parameter-free dimensionless
numbers.  This gives us the second version
 of the `Goldilocks' requirement,
\begin{equation}
\lcomp\ll \lpack\ll \sqrt{-b^2}\,. 
\label{Goldilocks2}
\end{equation}

The combination of \eqn{qbConstraint} and the second constraint
in \eqn{deltaConstraints} yields a stronger restriction than the first
one in \eqn{deltaConstraints}.
We should not expect a similar strengthening of the second restriction; the sharp
peaking alone will not force the left-hand side to be much smaller than the right-hand
side.  This means that we should expect $\qb\cdot u_i$ to be smaller than, but still of
order, $\sqrt{\xi}/\lscatt$.  If we compare the two terms in the argument to the
delta function~(\ref{universalDeltaFunction}), we see that the second term,
\begin{equation}
\lcomp \qb^2 \sim \frac{\lcomp}{\lscatt} \frac1{\lscatt} \ll \frac{\sqrt{\xi}}{\lscatt}\,,
\end{equation}
so that $\lcomp \qb^2 \ll \qb\cdot u_i$, and the second term should be negligible.
There is one caveat to the implied simplification, which we will mention below.

\def\lclass{{\rho}_{\textit{cl}}}

In computing the classical observable, we cannot simply set $\xi=0$.
Indeed, we don't even want to fully take the $\xi\rightarrow 0$ limit.
Rather, we want to take the leading term in that limit.  This term may
in fact be proportional to a power of $\xi$.  To understand this, we should take
note of one additional length scale in the problem, namely the classical
radius of the point particle.  In electrodynamics, this is $\lclass=e^2/(4\pi m)$.
However,
\begin{equation}
\lclass = \frac{\hbar e^2}{4\pi\hbar m} = \alpha\lcomp\,,
\end{equation}
where $\alpha$ is the usual, dimensionless, electromagnetic coupling.
Dimensionless ratios of $\lclass$ to other length scales will be the
expansion parameters in classical observables; but as this relation shows,
they too will vanish in the $\xi\rightarrow 0$ limit.
There are really three dimensionless parameters we must consider: $\xi$;
$\lpack/\lscatt$; and $\lclass/\lscatt$.  We want to retain the full dependence
on the latter, while considering only effects independent of the first two.

\def\expchange{\langle\Delta p_i\rangle}
Under the influence of a perturbatively weak interaction (such as
electrodynamics or gravity) below the particle-creation threshold, 
we expect a wavepacket's shape to be distorted slightly, but not radically
changed by the scattering.  We would expect the outgoing particles to
be characterized by wavepackets similar to those of the incoming particles.
However, using a wavepacket basis of states for the state sums in
\sect{QFTsetup} would be cumbersome, inconvenient, and computationally
less efficient than the plane-wave states we used.  We expect the
narrow peaking of the wavefunction to impose constraints on the momentum
transfers as they appear in higher-order corrections to the
impulse $\ImpB$, and in the leading contribution to the radiation
reaction $\Rad^\mu$~(\ref{eq:ExpectedMomentum2}); 
but we will need to see this narrowness indirectly,
via assessments of the spread as in \eqn{eq:expectations}
rather than directly through
the presence of a wavefunction (or wavefunction mismatch) factors in
our observables.  We can estimate the spread $\spread(\finalk_i)$
in a final-state
momentum $\finalk_i$ as follows,
\begin{equation}
\begin{aligned}
\spread(\finalk_i)/m_i^2 &= 
   \langle\bigl(\finalk_i-\langle\finalk_i\rangle\bigr)^2\rangle/m_i^2
\\&= \bigl(\langle \finalk_i^2\rangle-\langle\finalk_i\rangle{}^2\bigr)/m_i^2
\\&= 1-\bigl(\langle\initialk_i\rangle+\expchange\bigr){}^2/m_i^2
\\&= \spread(p_i)/m_i^2 -\langle\Delta p_i\rangle\cdot 
\bigl(2\langle\initialk_i\rangle+\expchange\bigr)/m_i^2\,.
\end{aligned}
\end{equation}
So long as $\expchange/m_i\lesssim\spread(\initialk_i)/m_i^2$,
 the second term will not greatly increase
the result, and the spread
in the final-state momentum will be of the same order as that in the initial-state
momentum.   Whether this condition holds depends on the details of the wavefunction.
Even if it is violated, so long as $\expchange/m_i \lesssim c'_\Delta \xi^{\beta'''}$
with $c'_\Delta$ a constant of $\Ord(1)$, then
the final-state momentum will have a narrow spread towards the limit.  (It would be broader
than the initial-state momentum spread, but that does not affect the applicability of
our results.)

The magnitude of $\expchange$ can be determined perturbatively.  The 
leading-order value comes from $\ImpA$, with $\ImpB$ and radiative corrections
contributing yet-smaller corrections.  As we shall see, these computations
reveal $\expchange/m_i$ to scale like $\sqrt{\xi}$ and be numerically much smaller.
This in turn implies that for perturbative consistency, the `characteristic'
values of momentum transfers
$w_i$ inside the definition of $R$ must also be very small compared to $m_i\sqrt{\xi}$.
(This constraint is in fact much weaker than implied by the leading-order
value of $\expchange$.)
Just as for $q_0$ in \eqn{qConstraint}, we should scale these momentum transfers
by $1/\hbar$, replacing them by wavenumbers $\xferb_i$.  The corresponding scattering
lengths $\lscatt' = \sqrt{-w_{1,2}^2}$ must again satisfy $\lscatt'\gg \lpack$.
If we now examine the energy-momentum-conserving delta function 
in \eqn{eq:ExpectedMomentum2}, rewritten in terms of momentum transfers,
\begin{equation}
\del^{(4)}(w_1+w_2 + k + \finalk_X)\,,
\end{equation}
we see that all radiated momenta $k$ and $\finalk_X$ must likewise be small
compared to $m_i\sqrt{\xi}$: all their energy components must be positive and
hence no cancellations are possible inside the delta function.  The typical
values of these momenta should again by scaled by $1/\hbar$ and replaced
by wavenumbers.  

What about loop integrations?  As we integrate the loop
momentum $\ell$ over all values, it is again
a matter of taste how we scale it.  If it is the momentum of a (virtual) massless
line, however, unitarity considerations suggest that as the natural
scaling is to remove a
factor of $\hbar$ in real-emission contributions, we should likewise do so for
virtual lines.  More generally, we should scale those differences of the loop momentum
with external legs that correspond to massless particles, and replace them
by wavenumbers.  Moreover, unitarity considerations also suggest that we should
choose the loop momentum to be that of a massless line in the loop, if there is
one.
\def\ellb{\bar\ell}

In general, we
may not be able to approach the $\hbar\rightarrow0$ limit of each contribution
to an observable separately, because they may contain terms which are
singular, having too many inverse powers of $\hbar$.  We find
that such singular terms meet one of two fates: they are multiplied by
functions which vanish in the regime of validity of the limit; or they
cancel in the sum over all contributions.  We cannot yet offer a general argument
that such troublesome terms necessarily disappear in one of these two
manners.  We can treat independently
contributions whose singular terms ultimately cancel
in the sum, so long as we expand each contribution in a Laurent series in $\hbar$.

Integrand factors that appear uniformly in all contributions --- that is, factors which 
appear directly in a final expression after cancellation of terms singular in
the $\hbar\rightarrow 0$ limit --- can benefit from applying two simplifications
to the integrand: setting $p_i$ to $m_i\ucl_i$, and truncating at the lowest
order in $\hbar$ or $\xi$.  For other factors, we must be careful to expand in
a Laurent series.  As mentioned above, inside the on-shell delta functions 
$\del(2p_i\cdot \qb\pm \hbar \qb^2)$ we can
neglect the $\hbar \qb^2$ term; this is true so long as
the factors multiplying these delta functions are not singular in $\hbar$.
If they are indeed nonsingular (after summing over terms),
we can safely neglect the second term inside such delta functions, and
replace them by $\del(2p_1\cdot \qb)$.  A similar argument allows us to neglect the
$\hbar \qb^0$ term inside the positive-energy theta functions; the $\qb$ integration
then becomes independent of them.  Similar arguments, and caveats, apply to the
squared momentum-transfer terms $\hbar \xferb_i^2$ appearing inside on-shell delta 
functions in higher-order contributions, along with the energy 
components $\xferb_i^0$ appearing
inside positive-energy theta functions.  They can be neglected so long as the accompanying
factors are not singular in $\hbar$.  If accompanying factors \textit{are} singular
as $\hbar\rightarrow 0$, then we may need to retain such formally suppressed
$\hbar \qb^2$ or $\hbar \xferb_i^2$ terms inside delta functions.
We will see an example of this in the calculation of the NLO contributions to
the impulse in \sect{sec:examples}.

It will be convenient to introduce a notation to allow us to manipulate 
integrands under the eventual approach to the $\hbar\rightarrow0$ limit; 
we will use large angle
brackets for the purpose,
\begin{equation}
\Lexp f(\initialk_1,\initialk_2,\ldots) \Rexp \equiv 
\int \df(\initialk_1)\df(\initialk_2)\;|\phi_1(\initialk_1)|^2\,|\phi_2(\initialk_2)|^2\,
  f(\initialk_1,\initialk_2,\ldots) \,,
\label{eq:angleBrackets}
\end{equation}
where the integration over both $\initialk_1$ and $\initialk_2$ is implicit.  Within
the angle brackets, we have approximated $\phi(p+q)\simeq \phi(p)$, and 
when evaluating the integrals (implicit in the large angle brackets), 
we will also set $p_i\simeq
m_i \ucl_i$, along with the other simplifications discussed 
above.

\subsection{An example wavefunction}
\label{sec:exampleWavefunction}
\def\Norm{\mathcal{N}}
It will be helpful to look at the scales that arise in calculations in
the context of a specific example for a wavefunction.  For this purpose we
take a linear exponential,
\begin{equation}
\phi(p_1) = \Norm m_1^{-1}\exp\biggl[-\frac{p_1\cdot u}{\hbar\lcomp/\lpack^2}\biggr]
= \Norm m_1^{-1}\exp\biggl[-\frac{p_1\cdot u}{m_1\xi}\biggr]\,.
\label{LinearExponential}
\end{equation}
In this function, $u$ is a dimensionless four-vector; we will ultimately
identify it as the four-velocity of particle~1, and normalize it to $u^2=1$.
(For the moment, we keep its normalization general.)  
As discussed in the previous section, $\lpack$ characterizes the `intrinsic'
spread of the wavepacket.  
In spite of the linearity of the exponent in $p_1$, this function
gives rise to the Gaussian of \eqn{NonrelativisticMomentumSpaceWavefunction} 
in the nonrelativistic limit in the rest frame of $u$.
(The wavefunction shares some features with relativistic wavefunctions discussed
in ref.~\cite{RelativisticWavefunctions}.)

The normalization 
condition~(\ref{WavefunctionNormalization}) requires,
\begin{equation}
\Norm = \frac{2\sqrt2\pi}{\xi^{1/2} K_1^{1/2}(2/\xi)}
\,,
\end{equation}
where $K_1$ is a modified Bessel function of the second kind.
For details of this computation and following ones, see appendix~\ref{app:wavefunctions}.  
We can compute the spread of the wavepacket straightforwardly, obtaining
\begin{equation}
\frac{\langle (\Delta p_1)^2\rangle}{\langle p_1^2\rangle} = 
1-\frac{K_2^2(2/\xi)}{K_1^2(2/\xi)}\,.
\end{equation}
As we approach the classical region, where $\xi\rightarrow 0$, the wavefunction
indeed becomes sharply peaked, as
\begin{equation}
\frac{\langle (\Delta p_1)^2\rangle}{\langle p_1^2\rangle} \rightarrow -\frac32 \xi + 
\Ord(\xi^2)\,.
\end{equation}

\def\IntOne{T_1}
Next, let us consider the implications of the on-shell delta function.
Examine a
wavefunction integral similar to $\ImpA$, but with a simpler integrand,
\begin{equation}
\IntOne = \int \df(p_1)\;\phi(p_1)\phi^*(p_1+q)\,\del(2 p_1\cdot q+q^2)\,.
\end{equation}
With $\phi$ chosen to be the linear exponential~(\ref{LinearExponential}),
this integral simplifies,
\begin{equation}
\IntOne = \frac{1}{\hbar m_1} \eta_1(\qb;p_1)\,\int \df(p_1)\;
\del(2 p_1\cdot\qb/m_1+\hbar\qb^2/m_1)\,|\phi(p_1)|^2\,,
\end{equation}
where we have also replaced $q\rightarrow\hbar \qb$, and where
\begin{equation}
\eta_1(\qb;p_1) = \exp\biggl[-\frac{\hbar\qb\cdot u}{m_1\xi}\biggr]\,.
\end{equation}

The remaining integrations in $\IntOne$ are evaluated in 
appendix~\ref{app:wavefunctions}, yielding
\begin{equation}
\IntOne = 
\frac1{4 \hbar m_1\sqrt{(\qb\cdot u)^2-\qb^2}\,K_1(2/\xi)} 
 \exp\biggl[-\frac2{\xi}\frac{\sqrt{(\qb\cdot u)^2-\qb^2}}{\sqrt{-\qb^2}}
                               \sqrt{1-\hbar^2\qb^2/(4m_1^2)}\biggr]
\,.
\end{equation}
(The wavenumber transfer is necessarily spacelike, so that $-\qb^2 > 0$.)

As we approach the $\hbar,\xi\rightarrow 0$ limit, we may expect this function to be concentrated
in a small region in $\qb$.  Towards the limit, the dependence on the magnitude is just
given by the prefactor.  To understand the behavior in the boost and angular
degrees of freedom, we may note that 
\begin{equation}
\frac1{K_1(2/\xi)} \sim \frac2{\sqrt{\pi}\sqrt{\xi}} \exp\biggl[\frac2{\xi}\biggr]\,,
\end{equation}
and that $\hbar\sqrt{\xi}$ is of order $\xi$, so that
overall $\IntOne$ has the form,
\begin{equation}
\frac1{\xi}\exp\biggl[-\frac{f(\qb)}{\xi}\biggr]\,,
\label{LimitForm}
\end{equation}
which becomes a delta function as $\xi\rightarrow 0$ limit.  The more
detailed discussion in the appendix 
shows that it has the form,
\begin{equation}
\delta(\qb\cdot u)\,,
\end{equation}
as anticipated in the previous section.

\subsection{Classical impulse}
\label{sec:classicalImpulse}

We have written the impulse in terms of two terms, $\langle \Delta p_1^\mu \rangle = \ImpA + \ImpB$, and expanded these in terms of wavefunctions in equations~\eqref{eq:impulseGeneralTerm1} and ~\eqref{eq:impulseGeneralTerm2}. We will now discuss the classical limit of these terms in detail, applying the rules discussed in
\sect{subsec:Wavefunctions}.

\def\class{\textrm{cl}}
We begin with the first and simplest term in the impulse, $\ImpA$, given in 
\eqn{eq:impulseGeneralTerm1}, here recast in the notation of
\eqn{eq:angleBrackets} in preparation,
\begin{equation}
\begin{aligned}
\ImpAcl = i \Lexp \int \!\dd^4 q  \; 
&\del(2\initialk_1 \cdot q + q^2) \del(2 \initialk_2 \cdot q - q^2) 
   \Theta(\initialk_1^0+q^0)\Theta(\initialk_2^0-q^0)\\
& \times  e^{-i b \cdot q/\hbar} 
\, q^\mu  \, \Ampl(\initialk_1 \initialk_2 \rightarrow 
                      \initialk_1 + q, \initialk_2 - q)\,\Rexp\,.
\end{aligned}
\label{eq:impulseGeneralTerm1recast}
\end{equation}
Rescale $q \rightarrow \hbar\qb$; drop
the $q^2$ inside the on-shell delta functions;
 and also remove the overall
factor of $g^2$ and accompanying $\hbar$s from the amplitude, to obtain 
the leading-order (LO) contribution to the classical impulse,
\begin{equation}
\begin{aligned}
\DeltaPlo \equiv \ImpAclsup{(0)} = i \frac{g^2}{4} \Lexp \hbar^2 \int \!\dd^4 \qb  \; 
&\del(\qb\cdot p_1) \del(\qb\cdot p_2) 
\\& \hspace*{-5mm}\times  
e^{-i b \cdot \qb} 
\, \qb^\mu  \, \AmplB^{(0)}(p_1,\,p_2 \rightarrow 
                      p_1 + \hbar\qb, p_2 - \hbar\qb)\,\Rexp\,.
\end{aligned}
\label{eq:impulseGeneralTerm1classicalLO}
\end{equation}
We denote by $\AmplB^{(L)}$ the reduced $L$-loop amplitude,
that is the $L$-loop amplitude with a factor of $g/\sqrt{\hbar}$ 
removed
for every interaction: in the electromagnetic case,
this removes a factor of $e/\sqrt{\hbar}$, while
in the gravitational case, we would remove a
factor of $\kappa/\sqrt{\hbar}$.  In general, this rescaled 
fixed-order amplitude depends only on
$\hbar$-free ratios of couplings; in pure electrodynamics or gravitational theory,
it is independent of couplings.  In pure electrodynamics, it depends on
the charges of the scattering particles.
While it is free of the powers of $\hbar$ 
discussed in \sect{RestoringHBar}, it will in general still scale with an overall
power of $\hbar$ thanks to dependence on momentum mismatches or transfers.
As we shall see in the next section, additional inverse powers of $\hbar$ emerging
from $\AmplB$ will cancel the $\hbar^2$ prefactor and yield a nonvanishing result.

As a reminder, while this contribution to a physical observable is linear in an
amplitude, it arises from an expression involving wavefunctions multiplied by their
conjugates.  This is reflected in the fact that both the `incoming' and `outgoing'
momenta in the amplitude here are in fact initial-state momenta.
Any phase which could be introduced by hand in the initial state 
would thus cancel out of the observable.

\def\Acl{\Ampl_\textrm{cl}}
The LO classical impulse is special in that only the 
first term~(\ref{eq:impulseGeneralTerm1}) contributes.
In general, it is only the sum of the two terms in \eqn{eq:defl1}
that has a well-defined classical limit.  
We may write this sum as
\begin{equation}
\ImpTcl = 
i 
\Lexp \hbar^{-2}
\int \!\dd^4 q  \; \del(2\initialk_1 \cdot q + q^2) \del(2 \initialk_2 \cdot q - q^2) 
   \Theta(\initialk_1^0+q^0)\Theta(\initialk_2^0-q^0) \; e^{-i b \cdot q/\hbar} \; \impKer \Rexp \,,
\label{eq:partialClassicalLimitNLO}
\end{equation}
where the \textit{impulse kernel\/} $\impKer$ is defined as,
\begin{equation}
\begin{aligned}
\impKer &\equiv  \hbar^2 q^\mu \, \Ampl(\initialk_1 \initialk_2 \rightarrow 
                      \initialk_1 + q, \initialk_2 - q)
\\&\hphantom{\equiv}\hspace*{3mm} 
-i \hbar^2 \sum_X \int \!\prod_{i = 1,2}  \dd^4 \xfer_i
\del(2p_i\cdot \xfer_i+\xfer_i^2)\Theta(p_i^0+\xfer_i^0)
\\&\hphantom{\equiv}\hspace*{25mm}\times 
      \xfer_1^\mu\,
\del^{(4)}(\xfer_1+\xfer_2+\finalk_X) 
\\ &\hphantom{\equiv}\hspace*{25mm}\times 
 \Ampl(\initialk_1 \initialk_2 \rightarrow 
                      \initialk_1+\xfer_1\,, \initialk_2+\xfer_2 \,, \finalk_X)
\\ &\hphantom{\equiv}\hspace*{25mm}\times 
 \Ampl^*(\initialk_1+q, \initialk_2-q \rightarrow 
   \initialk_1+\xfer_1 \,,\initialk_2+\xfer_2 \,, \finalk_X)
\,.
\end{aligned}
\label{FullImpulse}
\end{equation}
The prefactor in \eqn{eq:partialClassicalLimitNLO} and the normalization
of $\impKer$ are chosen so that the latter is $\Ord(\hbar^0)$ in the
classical limit.
At leading order, 
the only contribution comes from the tree-level four-point amplitude in the
first term, and after passing to the classical limit, we 
recover \eqn{eq:impulseGeneralTerm1classicalLO} as expected.
At next-to-leading order (NLO), both terms contribute.  The contribution from
the first term is from the one-loop amplitude, while that from the second
term has $X=\emptyset$, so that both the amplitude and conjugate inside the
integral are tree-level four-point amplitudes.

Focus on the NLO contributions, and
pass to the classical limit. As discussed in \sect{subsec:Wavefunctions} 
we may neglect the $q^2$ terms
in the delta functions present in \eqn{eq:partialClassicalLimitNLO} so long as any singular terms in the impulse
kernel cancel. 
We then rescale $q \rightarrow \hbar\qb$;  
and remove an overall
factor of $g^4$ and accompanying $\hbar$s from the amplitudes.
In addition, we may rescale $\xfer\rightarrow \hbar\xferb$. 
However, since singular terms may be present in the individual summands of the 
impulse kernel --- in general, they will cancel against singular terms emerging
from the loop integration in the first term in
\eqn{FullImpulse} ---  
we are not entitled to
drop the $w^2$ inside the on-shell delta functions. We obtain,
\begin{equation}
\DeltaPnlo= i\frac{g^4}{4}\Lexp \int \!\dd^4 \qb  \; \del(\initialk_1 \cdot \qb ) 
     \del(\initialk_2 \cdot \qb) 
   \; e^{-i b \cdot \qb} \;  \impKerCl \Rexp ,
\label{eq:classicalLimitNLO}
\end{equation}
where,
\begin{equation}
\begin{aligned}
\impKerCl &= \hbar \qb^\mu \, \AmplB^{(1)}(\initialk_1 \initialk_2 \rightarrow 
                      \initialk_1 + \hbar\qb  , \initialk_2 -  \hbar\qb)
\\&\hphantom{=} 
  -i \hbar^3 \int \! \dd^4 \xferb \; 
 \del(2p_1\cdot \xferb+ \hbar\xferb^2)\del(2p_2\cdot \xferb- \hbar\xferb^2)  \; \xferb^\mu \;
\\&\hspace*{15mm}\times 
 \AmplB^{(0)}(\initialk_1\,, \initialk_2 \rightarrow 
              \initialk_1+ \hbar \xferb\,, \initialk_2- \hbar\xferb)
\\&\hspace*{15mm}\times 
 \AmplB^{(0)*}(\initialk_1+ \hbar\qb\,, \initialk_2-  \hbar\qb \rightarrow 
               \initialk_1+\hbar\xferb  \,,\initialk_2- \hbar\xferb) .
\label{eq:impKerClDef}
\end{aligned}
\end{equation}
Once again, we will see in the next section that additional inverse powers of $\hbar$ will arise from the amplitudes, and will yield a finite and nonvanishing answer
in the classical limit.

\subsection{Classical radiation}
\label{sec:classicalradiation}
\def\Radcl{K^\mu}

Our starting point for obtaining a prediction for the classical limit of
the momentum emitted in radiation is \eqn{eq:ExpectedMomentum2}, which we recast in 
the notation of \eqn{eq:angleBrackets},
\begin{equation}
\begin{aligned}
\Rad^\mu_\class = \sum_X \Lexp \int \df(k)& \prod_{i=1,2} \dd^4\xfer_i\dd^4 q\;
\del(2p_i\cdot \xfer_i+\xfer_i^2)\Theta(p_i^0+\xfer_i^0)
\\&\times 
\del(2\initialk_1 \cdot q + q^2) \del(2 \initialk_2 \cdot q - q^2) 
   \Theta(\initialk_1{}^0+q^0)\Theta(\initialk_2{}^0-q^0)
\\& \times  k_X^\mu \, e^{-i b \cdot q/\hbar}  \del^{(4)}(\xfer_1+\xfer_2+ k+ \finalk_X)%
\\&\times 
  \Ampl(\initialk_1 \initialk_2 \rightarrow 
        \initialk_1+\xfer_1\,, \initialk_2+\xfer_2\,, k\,, \finalk_X)
\\&\times 
  \Ampl^*(\initialk_1+q, \initialk_2-q \rightarrow 
          \initialk_1+\xfer_1\,, \initialk_2+\xfer_2\,, k\,, \finalk_X)\,\Rexp
\,.
\end{aligned}
\label{eq:ExpectedMomentum2recast}
\end{equation}
Recall that $k_X^\mu$ is the sum of the messenger momentum $k^\mu$
and the momenta of other messengers in $X$.
We will determine the classical limit using precisely the same logic as in the previous subsection.
Let us again focus on the leading contribution, with $X=\emptyset$.  Once again,
rescale $q \rightarrow \hbar\qb$, and drop
the $q^2$ inside the on-shell delta functions.
Here, remove an overall
factor of $g^6$ and accompanying $\hbar$s from the amplitude and its conjugate.
In addition, rescale the momentum transfers
$\xfer\rightarrow \hbar\xferb$ and the radiation
momenta, $k\rightarrow\hbar k$. At leading order there is no sum, so 
there will be no hidden cancellations, and we may drop the $\xfer_i^2$ inside
the on-shell delta functions to obtain,
\def\kb{\bar k}
\begin{equation}
\begin{aligned}
\Rad^{\mu,(0)}_\class = 
g^6
\Lexp \hbar^4 \int &\df(\kb) \prod_{i=1,2} \dd^4\xferb_i\dd^4 \qb\;
\del(2\xferb_i\cdot p_i)
\del(2\qb\cdot p_1) \del(2\qb\cdot p_2) 
\\& \times  \kb^\mu \, e^{-i b \cdot \qb}  \del^{(4)}(\xferb_1+\xferb_2+ \kb)
\\&\times 
  \AmplB^{(0)}(\initialk_1 \initialk_2 \rightarrow 
        \initialk_1+\hbar\xferb_1\,, \initialk_2+\hbar\xferb_2\,, \hbar\kb)
\\&\times 
  \AmplB^{(0)*}(\initialk_1+\hbar \qb, \initialk_2-\hbar \qb \rightarrow 
          \initialk_1+\hbar\xferb_1\,, \initialk_2+\hbar\xferb_2\,, \hbar\kb)\,\Rexp
\,.
\end{aligned}
\label{eq:ExpectedMomentum2classicalLO}
\end{equation}
We will make use of this expression below to verify that momentum is conserved
as expected.

One disadvantage of this expression for the leading order radiated momentum is that it is no longer in a form of an 
integral over a perfect square, such as
shown in \eqn{RadiationPerfectSquare}.
Nevertheless we can recast \eqn{eq:ExpectedMomentum2recast} in such a form.
To do so, perform a change of variable, including in the wavefunctions. To begin, it is helpful to write \eqn{eq:ExpectedMomentum2recast} as
\begin{equation}
\begin{aligned}
\Rad^\mu_\class = \sum_X \prod_{i=1,2}  \int \! \df(\initialk_i) & |\phi_i(\initialk_i)|^2 \int \! \df(k) \df( \xfer_i+\initialk_i) \df(q_i+\initialk_i) \; 
\\& \times \del^{(4)}(\xfer_1+\xfer_2+ k+ \finalk_X) \del^{(4)}(q_1 + q_2) \, e^{-i b \cdot q_1/\hbar} \, k_X^\mu \,   %
\\&\times 
  \Ampl(\initialk_1\,, \initialk_2 \rightarrow 
        \initialk_1+\xfer_1\,, \initialk_2+\xfer_2\,, k\,, \finalk_X)
\\&\times 
  \Ampl^*(\initialk_1+q_1\,, \initialk_2 + q_2 \rightarrow 
          \initialk_1+\xfer_1\,, \initialk_2+\xfer_2\,, k\,, \finalk_X)\,
\,.
\end{aligned}
\end{equation}
\def\tinitialk{\tilde\initialk}
\def\txfer{\tilde\xfer}
\def\tq{\tilde q}
We will now re-order the integration and perform a change of variables. Let us define 
$\tinitialk_i=\initialk_i - \txfer_i$, $\tq_i = q_i + \txfer_i$, and
$\txfer_i = - \xfer_i$, changing
variables from $\initialk_i$ to $\tinitialk_i$, from $q_i$ to $\tq_i$,
and from $\xfer_i$ to $\txfer_i$,
\begin{equation}
\begin{aligned}
\Rad^\mu_\class = \sum_X \prod_{i=1,2}  \int \! &\df(\tinitialk_i)  \df(k) 
\df(\txfer_i+\tinitialk_i) \df (\tq_i+\tinitialk_i) 
 |\phi_i(\tinitialk_i+\txfer_i)|^2 \; 
\\& \times \del^{(4)}(\tilde \xfer_1+ \tilde \xfer_2- k- \finalk_X) 
\del^{(4)}(\tq_1 + \tq_2 - k - \finalk_X) 
\\&\times e^{-i b \cdot (\tq_1 - \txfer_1)/\hbar} \, k_X^\mu \,
\\&\times 
  \Ampl(\tinitialk_1 + \txfer_1\,, \tinitialk_2 + \txfer_2\rightarrow 
        \tinitialk_1\,, \tinitialk_2\,, k\,, \finalk_X)
\\&\times 
  \Ampl^*(\tinitialk_1+ \tq_1\,, \tinitialk_2 + \tq_2 \rightarrow 
          \tinitialk_1\,, \tinitialk_2\,, k\,, \finalk_X)\,
\,.
\end{aligned}
\end{equation}
As the $\txfer_i$ implicitly carry a factor of $\hbar$,
just as argued in \sect{subsec:Wavefunctions}
for the momentum mismatch $q$, we may
neglect the shift in the wave functions.  Dropping the tildes, 
and associating the $\xfer_i$ integrals with $\Ampl$
and the $q_i$ integrals with $\Ampl^*$, our expression is 
revealed as an integral over a perfect square,
\begin{equation}
\begin{aligned}
\Rad^\mu_\class &= \sum_X \prod_{i=1,2}  \int \! \df(\initialk_i)  |\phi_i( \initialk_i)|^2 \int \! \df(k) \, k_X^\mu
\\& \hspace*{20mm}\times
\biggl | \int \! \df( \xfer_i+\initialk_i)  \; 
 \del^{(4)}( \xfer_1+  \xfer_2- k- \finalk_X)
\\& \hspace*{25mm}\times   e^{i b \cdot  \xfer_1/\hbar}  \,   
  \Ampl( \initialk_1 +  \xfer_1,  \initialk_2 +  \xfer_2\rightarrow 
         \initialk_1\,,  \initialk_2\,, k\,, \finalk_X) \biggr|^2
         \\
& = \sum_X \prod_{i=1,2} \Lexp \int \! \df(k) \, k_X^\mu
\biggl | \int \! \df(\xfer_i + \initialk_i)  \; 
 \del^{(4)}( \xfer_1+  \xfer_2- k- \finalk_X)\\
& \hspace*{25mm} \times   e^{i b \cdot  \xfer_1/\hbar}  \,   
  \Ampl( \initialk_1 +  \xfer_1,  \initialk_2 +  \xfer_2\rightarrow 
         \initialk_1\,,  \initialk_2\,, k\,, \finalk_X) \biggr|^2 \Rexp
         \,.
\label{eq:radiatedMomentumClassicalAllOrder}
\end{aligned}
\end{equation}
The perfect-square structure allows us to define a \textit{radiation kernel\/},
\begin{equation}
\begin{aligned}
\RadKer(k, \finalk_X)
&\equiv \hbar^{3/2} \prod_{i = 1, 2}  \int \! \df( \initialk_i +  \xfer_i)  \; 
 \del^{(4)}( \xfer_1+  \xfer_2- k- \finalk_X) \\
& \qquad \qquad \times  e^{i b \cdot  \xfer_1/\hbar}  \,   
  \Ampl( \initialk_1 +  \xfer_1,  \initialk_2 +  \xfer_2\rightarrow 
         \initialk_1\,,  \initialk_2\,, k\,, \finalk_X), \\
&= \hbar^{3/2}\prod_{i = 1, 2} \int \! \dd^4 \xfer_i \; 
     \del(2 p_i \cdot \xfer_i + \xfer_i^2) \Theta(\initialk_i^0+\xfer_i^0)
     \del^{(4)}( \xfer_1+  \xfer_2- k- \finalk_X) \\
& \qquad \qquad \times  e^{i b \cdot  \xfer_1/\hbar}  \,   
  \Ampl( \initialk_1 +  \xfer_1,  \initialk_2 +  \xfer_2\rightarrow 
         \initialk_1\,,  \initialk_2\,, k\,, \finalk_X)
\label{eq:defOfR}
\end{aligned}
\end{equation}
so that
\begin{equation}
\begin{aligned}
\Rad^\mu_\class &=  \sum_X \hbar^{-3}\Lexp \int \! \df(k) \, k_X^\mu
\left |\RadKer(k, \finalk_X) \right|^2 \Rexp
         \,.
\label{eq:radiatedMomentumClassical}
\end{aligned}
\end{equation}
The prefactor along with the normalization of $\RadKer$ are again chosen
so that the classical limit of the radiation kernel will be of $\Ord(\hbar^0)$.
As we will see in \sect{sec:genClassicalRadiation}, this expression is related to a classical all-order formula for the radiated momentum.

Let us now focus once more on the leading contribution, with $X=\emptyset$.  As usual,
rescale $\xfer \rightarrow \hbar\xferb$, and as above we may drop
the $\xfer^2$ inside all of the on-shell delta functions.
Again, remove an overall
factor of $g^6$ and accompanying $\hbar$s from the amplitude and its conjugate.
We choose to express the leading-order classical radiated momentum in terms 
of a leading-order radiation kernel,
\begin{equation}
\begin{aligned}
\RadKerCl(\wn k) 
& \equiv \hbar^2 \prod_{i = 1, 2} \int \! \dd^4 \xferb_i \, \del(2p_i \cdot \xferb_i) 
\del^{(4)}( \xferb_1+  \xferb_2- \wn k)
e^{i b \cdot  \xferb_1}  
\\& \hspace{70pt} \times  \AmplB^{(0)}( \initialk_1 +  \hbar \xferb_1,  \initialk_2 + \hbar \xferb_2\rightarrow 
         \initialk_1\,,  \initialk_2\,, \hbar \wn k)\,,
\label{eq:defOfRLO}
\end{aligned}
\end{equation}
so that the leading-order momentum radiated is simply,
\begin{equation}
\begin{aligned}
\Rad^{\mu, (0)}_\class &= g^6 \Lexp \int \! \df(\wn k) \, \wn k^\mu
\left | \RadKerCl(\wn k) \right|^2 \Rexp
         \,.
\label{eq:radiatedMomentumClassicalLO}
\end{aligned}
\end{equation}

\subsection{Conservation of momentum}
\label{sect:classicalConservation}

Conservation of momentum certainly holds to all orders, as we saw in \sect{sect:allOrderConservation}.  In the classical theory,
momentum conservation was however historically a controversial issue.
It is thus worth making sure that we have not spoiled this critical physical property 
in our classical impulse discussion in \sect{sec:classicalImpulse}, or in our classical
radiation discussion, \sect{sec:classicalradiation}. One might worry, for example, that
there is a subtlety with the order of limits.

There is no issue at LO and NLO for the impulse, because
\begin{equation}
\DeltaPlo + \DeltaPloTwo = 0 ,\quad \DeltaPnlo + \DeltaPnloTwo = 0.
\end{equation}
These follow straightforwardly from the definitions, \eqn{eq:impulseGeneralTerm1classicalLO} and \eqn{eq:classicalLimitNLO}. The essential point is that the amplitudes entering
into these orders in the  impulse conserve momentum for four particles. 
At LO, for example, using \eqn{eq:impulseGeneralTerm1classicalLO} 
the impulse on particle 2 can be written as
\begin{equation}
\begin{aligned}
\DeltaPloTwo= i \frac{g^2}{4} \Lexp \hbar^2 \int \!\dd^4 \qb_1  \dd^4 \qb_2 \; 
&\del(\qb_1\cdot p_1) \del(\qb_1\cdot p_2)  \del^{(4)}(\qb_1 + \qb_2)
\\& \hspace*{-5mm}\times  
e^{-i b \cdot \qb_1} 
\, \qb_2^\mu  \, \AmplB^{(0)}(p_1,\,p_2 \rightarrow 
                      p_1 + \hbar\qb_1, p_2 + \hbar\qb_2)\,\Rexp.
\end{aligned}
\end{equation}
In this equation, conservation of momentum at the level of the four point amplitude $\AmplB^{(0)}(p_1,\,p_2 \rightarrow p_1 + \hbar\qb_1, p_2 + \hbar\qb_2)$ is expressed by the presence of the four-fold delta function $\del^{(4)}(\qb_1 + \qb_2)$. Using this delta function, we may replace $\qb_2^\mu$ with $- \qb_1^\mu$ and then integrate over $\qb_2$, once again using the delta function. The result is manifestly $-\DeltaPlo$,  \eqn{eq:impulseGeneralTerm1classicalLO}. A similar calculation goes through at 
NLO.

In this sense, the scattering is conservative at LO and at NLO. 
At NNLO, however, we must take
radiative effects into account.  This backreaction is entirely described by $\ImpB$.
As indicated in \eqn{eq:commutatorVanishes}, $\ImpA$ is always conservative.
From our perspective here, this is because it involves only four-point amplitudes. 
Thus to understand conservation of momentum we need to investigate $\ImpB$. The lowest order case in which a five point amplitude can enter $\ImpB$ is at NNLO. Let us restrict attention to this lowest order case, taking the additional state $X$ to be a messenger.

Using precisely the methods of previous subsections, the lowest order term in $\ImpB$ involving one messenger is, in the classical regime,
\begin{equation}
\hspace*{-3mm}\begin{aligned}
\ImpBclsup{(\textrm{rad})} = g^6
\Lexp \hbar^{4}\!\int \! d\Phi(\wn k) & \prod_{i = 1,2} \dd^4\xferb_i\, 
\dd^4 \qb_1 \dd^4 \qb_2 \;\del(2 \xferb_i\cdot p_i)
\del(2 \qb_1\cdot p_1) \del(2 \qb_2\cdot p_2)
\\&\backdentA\times  
      e^{-i b \cdot \qb_1}\,\xferb_1^\mu\,
\del^{(4)}(\xferb_1+\xferb_2 + \bar k) \del^{(4)}(\qb_1+\qb_2)
\\&\backdentA\times \AmplB^{(0)}(\initialk_1\,, \initialk_2 \rightarrow 
                      \initialk_1+\hbar\xferb_1\,, \initialk_2+\hbar\xferb_2, \hbar\kb)
\\&\backdentA\times 
    \AmplB^{(0)*}(\initialk_1+\hbar \qb_1\,, \initialk_2 + \hbar \qb_2 \rightarrow 
        \initialk_1+\hbar\xferb_1\,, \initialk_2+\hbar\xferb_2, \hbar\kb)
               \,\Rexp\,.
\end{aligned} 
\label{eq:nnloImpulse}
\end{equation}
To see that this balances the radiated momentum, we use \eqn{eq:ExpectedMomentum2classicalLO}. The structure of the expressions are almost identical; conservation of momentum holds because the factor $\kb^\mu$ in \eqn{eq:ExpectedMomentum2classicalLO} is balanced by $\xferb_1^\mu$ in \eqn{eq:nnloImpulse} and $\xferb_2^\mu$ in the equivalent expression for particle 2.

Thus conservation of momentum continues to hold in our expressions once we have passed to the classical limit, at least through NNLO. At this order there is non-zero momentum
radiated, so momentum conservation is non-trivial from the classical point of view. We will 
see by explicit calculation in later sections that our classical impulse correctly 
incorporates the impulse from the ALD force in addition to the Lorentz force.

\section{Examples}
\label{sec:examples}

To build confidence in the formalism developed in prior
sections, let us use it to calculate some classical point-particle observables
explicitly. We will work in the context of scalar electrodynamics, with Lagrangian
\begin{equation}
\mathcal{L} = -\frac14 F^{\mu\nu} F_{\mu\nu} + \sum_{i = 1, 2} \left[ (D^\mu \Phi_i)^\dagger D_\mu \Phi_i - m_i^2 \Phi_i^\dagger \Phi_i \right].
\end{equation}
We take the charges of the fields $\Phi_i$ to be $Q_i$, with the electromagnetic coupling
the usual $e$.

We will begin by studying the impulse at leading and next-to-leading order. Later,
in section~\ref{sec:EDimpulse}, we will study the same quantity by iterating the classical equations of motion. The
fact that this iterative classical approach to the impulse is reminiscent of Feynman diagrams was recently highlighted
by Damour~\cite{Damour:2017zjx}. Following our discussion of the impulse, we will
discuss the momentum radiated before turning to momentum conservation.

\subsection{Leading-order electromagnetic impulse}

We begin by computing the impulse, $\DeltaPlo$, on particle 1 at leading order.
At this order, only $\ImpA$ contributes, as expressed in 
\eqn{eq:impulseGeneralTerm1classicalLO}.
To evaluate the impulse, we must first compute the $2\rightarrow 2$ tree-level scattering amplitude.
The reduced amplitude $\AmplB^{(0)}$ is,
\begin{equation}
i\AmplB^{(0)}(p_1 p_2 \rightarrow p_1+\hbar\qb\,, p_2-\hbar\qb) =
\begin{tikzpicture}[scale=1.0, baseline={([yshift=-\the\dimexpr\fontdimen22\textfont2\relax] current bounding box.center)}, decoration={markings,mark=at position 0.6 with {\arrow{Stealth}}}]
\begin{feynman}
	\vertex (v1);
	\vertex [below = 1 of v1] (v2);
	\vertex [above left=0.6 and 0.66 of v1] (i1) {$p_1$};
	\vertex [above right=0.6 and 0.33 of v1] (o1) {$p_1+\hbar \qb$};
	\vertex [below left=0.6 and 0.66 of v2] (i2) {$p_2$};
	\vertex [below right=0.6 and 0.33 of v2] (o2) {$p_2-\hbar \qb$};
	\draw [postaction={decorate}] (i1) -- (v1);
	\draw [postaction={decorate}] (v1) -- (o1);
	\draw [postaction={decorate}] (i2) -- (v2);
	\draw [postaction={decorate}] (v2) -- (o2);
	\diagram*{(v1) -- [photon] (v2)};
\end{feynman}	
\end{tikzpicture}
\!\!\!\!\!\!\!\!= i Q_1 Q_2 \frac{4 p_1 \cdot p_2 + \hbar^2 \qb^2}
                                                 {\hbar^2 \qb^2}.
\label{ReducedAmplitude1}
\end{equation}
We can neglect the second term in the numerator,
 which is subleading in the classical limit.
 
Substituting this expression into \eqn{eq:impulseGeneralTerm1classicalLO}, we obtain,
\begin{equation}
\begin{aligned}
\DeltaPlo = i e^2 Q_1 Q_2 \Lexp \int \!\dd^4 \qb  \; 
&\del(\qb\cdot p_1) \del(\qb\cdot p_2) 
e^{-i b \cdot \qb} \frac{p_1 \cdot p_2}{\qb^2}
\, \qb^\mu\,\Rexp\,.
\end{aligned}
\label{eq:impulseClassicalLOa}
\end{equation}
As promised, the leading-order expression is independent of $\hbar$.
Evaluating the $p_{1,2}$ integrals,
in the process applying the simplifications explained in \sect{subsec:Wavefunctions},
namely replacing $p_i\rightarrow m_i\ucl_i$, 
we find that,
\begin{equation}
\begin{aligned}
\DeltaPlo = i e^2 Q_1 Q_2 \int \!\dd^4 \qb  \; 
&\del(\qb\cdot \ucl_1) \del(\qb\cdot \ucl_2) 
e^{-i b \cdot \qb} \frac{\ucl_1 \cdot \ucl_2}{\qb^2}
\, \qb^\mu\,.
\end{aligned}
\label{eq:impulseClassicalLO}
\end{equation}
This expression has intriguing similarities to quantities that arise in the high-energy
limit of two-body scattering~\cite{Eikonal1,Amati:1990xe,%
Eikonal2,DAppollonio:2010krb,%
Eikonal3,Luna:2016idw,Collado:2018isu}.
The eikonal approximation used there is known to
exponentiate, and it would be interesting to explore this connection further.

It is straightforward to perform the integral over $\qb$ 
in \eqn{eq:impulseClassicalLO} to obtain an explicit 
expression for the leading order impulse. To do so, we work in the rest frame of
particle 1, so that $\ucl_1 = (1, 0, 0, 0)$. Without loss of generality we can orientate
the spatial coordinates in this frame so that particle 2 is moving along the $z$ axis, 
with proper velocity $\ucl_2 = (\gamma, 0, 0, \gamma \beta)$. We have introduced
the standard Lorentz gamma factor $\gamma = \ucl_1 \cdot \ucl_2$ and the
velocity parameter $\beta$ satisfying $\gamma^2 ( 1- \beta^2 ) = 1$. In terms
of these variables, the impulse is
\begin{equation}
\begin{aligned}
\DeltaPlo &= i e^2 Q_1 Q_2 \int \!\dd^4 \qb  \;
\del(\qb^0) \del(\gamma \qb^0 - \gamma \beta \qb^3) \;
e^{-i b \cdot \qb} \frac{\gamma}{\qb^2}
\, \qb^\mu \\
&= -i \frac{e^2 Q_1 Q_2 }{4\pi^2 |\beta|}\int \! d^2 \qb  \;
e^{i \v{b} \cdot \v{\qb}_\perp} \frac{1}{\v \qb_\perp^2}
\, \qb^\mu \, ,
\end{aligned}
\end{equation}
where $\qb^0 = \qb^3 = 0$ and the non-vanishing components of $\qb^\mu$ in the $xy$ plane of our
corrdinate system are $\v \qb_\perp$.
It remains to perform the two dimensional integral over $\v \qb_\perp$, which is easily done using polar coordinates. Let the magnitude
of $\v \qb_\perp$ be $\chi$ and orient the $x$ and $y$ axes so that $\v b \cdot \v \qb_\perp = | \v b| \chi \cos \theta$. Then
the non-vanishing components of $\qb^\mu$ are $\qb^\mu = (0, \chi \cos \theta, \chi \sin \theta, 0)$ and the impulse is
\begin{equation}
\begin{aligned}
\DeltaPlo &= -i \frac{e^2 Q_1 Q_2 }{4\pi^2 |\beta|}\int_0^\infty d \chi \; \chi \int_{-\pi}^\pi d \theta   \;
e^{i | \v b| \chi \cos \theta} \frac{1}{\chi^2}
\, (0, \chi \cos \theta, \chi \sin \theta, 0) \\
&= -i \frac{e^2 Q_1 Q_2 }{4\pi^2 |\beta|}\int_0^\infty d \chi \; \int_{-\pi}^\pi d \theta   \;
e^{i | \v b| \chi \cos \theta} 
\, (0, \cos \theta, \sin \theta, 0) \\
&= \frac{e^2 Q_1 Q_2 }{2\pi |\beta|}\int_0^\infty d \chi \; 
J_1 ( |\v b| \chi) \; \hat{\v b} \\ 
&= \frac{e^2 Q_1 Q_2 }{2\pi |\beta|} \; \frac{\hat{\v b}}{| \v b|} \, ,
\end{aligned}
\end{equation}
where $\hat {\v b}$ is the spatial unit vector in the direction of the impact parameter. To restore manifest Lorentz invariance, note that
\begin{equation}
\frac{1}{| \beta|} = \frac{\gamma}{\sqrt{\gamma^2 - 1}}\,, 
\quad \frac{\hat{\v b}}{|\v b|} = - \frac{b^\mu}{b^2}\,.
\end{equation}
(Recall that $b^\mu$ is spacelike, so $-b^2>0$.)
With this input, we may write the impulse as,
\begin{equation}
\DeltaPlo  
= -\frac{e^2 Q_1 Q_2}{2\pi} \frac{\gamma}{\sqrt{\gamma^2 - 1}} \frac{b^\mu}{b^2}\,.
\end{equation}

In the non-relativistic limit this should match a familiar formula: the expansion of the Rutherford scattering angle $\theta(b)$ as a function of the impact
parameter. To keep things simple, we consider Rutherford scattering of a light particle 
(for example, an electron) off a heavy particle (a nucleus). Taking particle 1
to be the moving light particle, particle 2 is very heavy and we work in its rest frame. 
Expanding the textbook Rutherford result to order $e^2$, we find
\begin{equation}
\theta(b) = 2 \tan^{-1} \frac{e^2 Q_1 Q_2}{4 \pi m v^2 b} \simeq \frac{e^2 Q_1 Q_2}{2 \pi m v^2 b},
\end{equation}
where $v$ is the non-relativistic velocity of the particle.
To recover this simple result from equation~\eqref{eq:impulseClassicalLO}, recall that in the non-relativistic limit $\gamma \simeq 1 + v^2/2$. The scattering angle, at this order, is simply $\Delta v/v$.  We will make use of this frame in later sections as well.

We note in passing that the second term in the numerator of
\eqn{ReducedAmplitude1} is a quantum correction.  It will ultimately be suppressed
by $\lcomp^2/b^2$, and in addition would contribute only a contact interaction,
as it leads to a $\delta^{(2)}(b)$ term in the impulse.

\subsection{Next-to-leading order impulse}
\label{sec:nloQimpulse}

At the next order in perturbation theory, order $e^4$, a well-defined classical impulse is only obtained by combining all terms in the impulse $\langle \Delta p_1^\mu \rangle$ of order $e^4$. As we discussed in \sect{sec:classicalImpulse}, both $\ImpA$ and $\ImpB$ contribute. We found in \eqn{eq:classicalLimitNLO} that the impulse is a simple integral over an impulse kernel $\impKerCl$, defined in \eqn{eq:impKerClDef}, which has a well-defined classical limit. 

The determination of the impulse kernel at this order
requires us to compute the four-point one-loop amplitude along with
a cut amplitude, that is an integral over a term quadratic in the tree amplitude. 
As the one-loop amplitude in electrodynamics is simple, we compute it explicitly
using on-shell renormalised perturbation theory in Feynman gauge.

The contributions to the impulse in the quantum theory can be
divided into three classes, according to the prefactor in the charges they
carry: $C_1$, those proportional to $Q_1^3 Q_2$; $C_2$, 
those to $Q_1^2 Q_2^2$; and $C_3$, those to $Q_1 Q_2^3$.
The first class can be further subdivided into $C_{1a}$, terms which would be proportional
to $Q_1 (Q_1^2+n_s Q_3^2) Q_2$ were we to add $n_s$ species of a third scalar with
charge $Q_3$, and into
$C_{1b}$, terms which would retain the simple $Q_1^3 Q_2$ prefactor.  Likewise,
the last class can be further subdivided into $C_{3a}$,
terms which would be proportional
to $Q_1 (Q_2^2+n_s Q_3^2) Q_2$, and into $C_{3b}$,
those whose prefactor would remain simply $Q_1 Q_2^3$.

Classes $C_{1a}$ and $C_{3a}$ consist of photon self-energy corrections along with
renormalization counterterms.  They appear only in the one-loop
corrections to the four-point amplitude, in the first
term in the impulse kernel $\impKerCl$.
We will discuss them in detail in \sect{PurelyQuantum}
below.  As one may suspect \textit{ab ipso initio\/}, they give no contribution
in the classical limit.  Likewise, classes $C_{1b}$ and $C_{3b}$ consist of vertex
corrections, wavefunction renormalization, and their counterterms.  We will not
discuss them in detail, but they too give no contribution in the classical limit.

This leaves us with contributions of class $C_2$; these appear in both terms
in the impulse kernel.  These contributions to the
one-loop amplitude in the first term take the form,
\begin{equation}
\begin{aligned}
i \AmplB^{(1)}(p_1p_2 \rightarrow p_1' p_2') &= \begin{tikzpicture}[scale=1.0, baseline={([yshift=-\the\dimexpr\fontdimen22\textfont2\relax] current bounding box.center)}] 
\begin{feynman}
	\vertex (b) ;
	\vertex [above left=1 and 0.66 of b] (i1) {$p_1$};
	\vertex [above right=1 and 0.33 of b] (o1) {$p_1+q$};
	\vertex [below left=1 and 0.66 of b] (i2) {$p_2$};
	\vertex [below right=1 and 0.33 of b] (o2) {$p_2-q$};
	\begin{scope}[decoration={
		markings,
		mark=at position 0.7 with {\arrow{Stealth}}}] 
		\draw[postaction={decorate}] (b) -- (o2);
		\draw[postaction={decorate}] (b) -- (o1);
	\end{scope}
	\begin{scope}[decoration={
		markings,
		mark=at position 0.4 with {\arrow{Stealth}}}] 
		\draw[postaction={decorate}] (i1) -- (b);
		\draw[postaction={decorate}] (i2) -- (b);
	\end{scope}	
	\filldraw [color=white] (b) circle [radius=10pt];
	\draw [pattern=north west lines, pattern color=patternBlue] (b) circle [radius=10pt];
	\filldraw [fill=white] (b) circle [radius=6pt];
\end{feynman}
\end{tikzpicture} 
=\\
&\hspace*{-20mm}
\begin{tikzpicture}[scale=1.0, baseline={([yshift=-\the\dimexpr\fontdimen22\textfont2\relax] current bounding box.center)}] 
\begin{feynman}
	\vertex (ip1) ;
	\vertex [right=0.7 of ip1] (ip2);
	\vertex [above=0.7 of ip1] (ip3);
	\vertex [above=0.7 of ip2] (ip4);
	\vertex [above left=0.66 and 0.33 of ip3] (i1) {};
	\vertex [above right=0.66 and 0.33 of ip4] (o1) {};
	\vertex [below left=0.66 and 0.33 of ip1] (i2) {};
	\vertex [below right=0.66 and 0.33 of ip2] (o2) {};
	\draw (i1) -- (ip3) -- (ip4) -- (o1);
	\draw (i2) -- (ip1) -- (ip2) -- (o2);
	\diagram*{
		(ip1) -- [photon] (ip3);
		(ip2) -- [photon] (ip4);
	};
\end{feynman}
\end{tikzpicture}
+
\begin{tikzpicture}[scale=1.0, baseline={([yshift=-\the\dimexpr\fontdimen22\textfont2\relax] current bounding box.center)}] 
\begin{feynman}
	\vertex (ip1) ;
	\vertex [right=0.7 of ip1] (ip2);
	\vertex [above=0.7 of ip1] (ip3);
	\vertex [above=0.7 of ip2] (ip4);
	\vertex [above left=0.66 and 0.33 of ip3] (i1) {};
	\vertex [above right=0.66 and 0.33 of ip4] (o1) {};
	\vertex [below left=0.66 and 0.33 of ip1] (i2) {};
	\vertex [below right=0.66 and 0.33 of ip2] (o2) {};
	\draw (i1) -- (ip3) -- (ip4) -- (o1);
	\draw (i2) -- (ip1) -- (ip2) -- (o2);
	\diagram*{(ip1) -- [photon] (ip4);};
	\filldraw [color=white] ($ (ip1) !.5! (ip4) $) circle [radius = 1.2pt];
	\diagram*{(ip2) -- [photon] (ip3);};
\end{feynman}
\end{tikzpicture}
+
\begin{tikzpicture}[scale=1.0, baseline={([yshift=-\the\dimexpr\fontdimen22\textfont2\relax] current bounding box.center)}] 
\begin{feynman}
	\vertex (i1) ;
	\vertex [right=2 of i1] (i2);
	\vertex [below=2 of i1] (o1);
	\vertex [below=2 of i2] (o2);
	\vertex [below right=0.7 of i1] (v1);
	\vertex [below left=0.7 of i2] (v2);
	\vertex [above right=0.7 and 1 of o1] (v3);
	\draw (i1) -- (v1) -- (v2) -- (i2);
	\draw (o1) --(v3) -- (o2);
	\diagram*{
		(v1) -- [photon] (v3);
		(v2) -- [photon] (v3);
	};
\end{feynman}
\end{tikzpicture}
+
\begin{tikzpicture}[scale=1.0, baseline={([yshift=-\the\dimexpr\fontdimen22\textfont2\relax] current bounding box.center)}] 
\begin{feynman}
	\vertex (o1) ;
	\vertex [right=2 of o1] (o2);
	\vertex [below=2 of o1] (i1);
	\vertex [below=2 of o2] (i2);
	\vertex [above right=0.7 of i1] (v1);
	\vertex [above left=0.7 of i2] (v2);
	\vertex [below right=0.7 and 1 of o1] (v3);
	\draw (i1) -- (v1) -- (v2) -- (i2);
	\draw (o1) --(v3) -- (o2);
	\diagram*{
		(v1) -- [photon] (v3);
		(v2) -- [photon] (v3);
	};
\end{feynman}
\end{tikzpicture}
+
\begin{tikzpicture}[scale=1.0, baseline={([yshift=-\the\dimexpr\fontdimen22\textfont2\relax] current bounding box.center)}] 
\begin{feynman}
	\vertex (o1) ;
	\vertex [right=2 of o1] (o2);
	\vertex [below=2 of o1] (i1);
	\vertex [below=2 of o2] (i2);
	\vertex [above right=0.7 and 1 of i1] (v1);
	\vertex [below right=0.7 and 1 of o1] (v2);
	\draw (i1) -- (v1) -- (i2);
	\draw (o1) --(v2) -- (o2);
	\diagram*{
		(v1) -- [photon, half left] (v2);
		(v1) -- [photon, half right] (v2);
	};
\end{feynman}
\end{tikzpicture}
.
\end{aligned}
\label{OneLoopImpulse}
\end{equation}
In each contribution, we count powers of $\hbar$ following the rules in
\sect{subsec:Wavefunctions}, replacing $\ell\rightarrow\hbar\ellb$
and $q\rightarrow\hbar \qb$.  In the double-seagull contribution, we will get
four powers from the loop measure, and four inverse powers from the two
photon propagators.  Overall, we will not get enough inverse powers to 
compensate the power in front of the
integral in \eqn{eq:impKerClDef}, and so the seagull will die in
the classical limit.  The remaining diagrams will contribute in the
limit, and we discuss them in \sects{Triangles}{Boxes}.  We discuss the
contributions from the second term in the impulse kernel in 
\sect{CutBoxes}, and combine terms in \sect{CombiningTerms}.

\subsubsection{Purely Quantum Contributions}
\label{PurelyQuantum}

Let us begin with the first and last classes of contributions described in
the beginning of this section.  These correspond to vertex and photon self-energy
terms, including counterterms.
Consider, for example, the photon self-energy terms, focussing on internal scalars of 
mass $m$ and charge $Q_i$. Define the self-energy via,
\begin{equation}
 Q_i^2 \Pi(q^2) \left( q^2 \eta^{\mu\nu} - q^\mu q^\nu \right) \equiv \feynmandiagram [inline = (a.base), horizontal=a to b, horizontal=c to d] { a -- [photon, momentum'=\(q\)] b -- [fermion, half left] c -- [fermion, half left] b -- [draw = none] c -- [photon] d};
\; + \!\!\! \feynmandiagram[inline = (a.base), horizontal=a to b]{a -- [photon, momentum'=\(q\)] c -- [out=45, in=135, loop, min distance=2cm]c -- [photon] b};
\!\!\! + \;\; \feynmandiagram[inline = (a.base), layered layout, horizontal=a to b] { a -- [photon, momentum'=\(q\)] b [crossed dot] -- [photon] c}; \,,
\label{SelfEnergyContributions}
\end{equation}
where we have made the projector required by gauge invariance manifest, but have 
not included factors of the electromagnetic coupling $e$.
We have extracted the charges $Q_i$ for later convenience.
The contribution of the photon self-energy to the 
reduced four-point amplitude is
\begin{equation}
\AmplB_\Pi = {Q_1 Q_2 Q_i^2} \frac{(2p_1 + \hbar \qb) \cdot (2p_2 -\hbar \qb)}
                                  {\hbar^2 \qb^2} \Pi(\hbar^2 \qb^2)\,.
\end{equation}
The counterterm is adjusted to impose the renormalisation condition that $\Pi(0) = 0$, 
required in order to match the identification of the electromagnetic
coupling $e$ with its classical counterpart.
As a power series in the dimensionless ratio $q^2 / m^2 = \hbar^2 \qb^2 / m^2$, 
which is of order $\lcomp^2 / b^2$,
\begin{equation}
\Pi(q^2) = \hbar^2 \Pi'(0) \frac{\qb^2}{m^2} 
+ \mathcal{O}\biggl(\frac{\lcomp^4}{b^4} \biggr)\,.
\end{equation}
The renormalization condition is essential in eliminating possible contributions
of $\Ord(\hbar^0)$.
One way to see that $\AmplB_\Pi$ is a purely quantum correction is to follow
the powers of $\hbar$.   
As $\Pi(q^2)$ is of order $\hbar^2$, $\AmplB_\Pi$ is of order $\hbar^0$. 
This gives a contribution of $\Ord(\hbar)$ to the impulse kernel~(\ref{eq:impKerClDef}),
which in turn gives a contribution of $\Ord(\hbar)$ to the impulse, as can
be seen in \eqn{eq:classicalLimitNLO}.

Alternatively, one can consider the contribution of these graphs to $\Delta p / p$. Counting each factor of $\qb$ as of order $b$, and using $\Pi(q^2) \sim \lcomp^2 / b^2$, it is easy to see that these self-energy graphs yield a contribution to $\Delta p / p$ of order $\alpha^2 \hbar^3 / (mb)^3 \sim (\lclass^2 / b^2) \,( \lcomp / b)$.

The renormalisation of the vertex is similarly a purely quantum effect.

\subsubsection{Triangles}
\label{Triangles}

We turn next to contributions which do survive
in the classical limit.  Let us first examine the two triangle diagrams in
\eqn{OneLoopImpulse}.  They are related by swapping particles~1 and~2.
 The first diagram is,
\begin{equation}
i T_{12} = 
\begin{tikzpicture}[scale=1.0, baseline={([yshift=-\the\dimexpr\fontdimen22\textfont2\relax] current bounding box.center)}, decoration={markings,mark=at position 0.6 with {\arrow{Stealth}}}] 
\begin{feynman}
	\vertex (i1) {$p_1$};
	\vertex [right=2.5 of i1] (i2) {$p_1 + q$};
	\vertex [below=2.5 of i1] (o1) {$p_2$};
	\vertex [below=2.5 of i2] (o2) {$p_2 - q$};

	\vertex [below right=1.1 of i1] (v1);
	\vertex [below left=1.1 of i2] (v2);
	\vertex [above right=1.1 and 1.25 of o1] (v3);

	\draw [postaction={decorate}] (i1) -- (v1);
	\draw [postaction={decorate}] (v1) -- (v2);
	\draw [postaction={decorate}] (v2) -- (i2);
	\draw [postaction={decorate}] (o1) -- (v3);
	\draw [postaction={decorate}] (v3) -- (o2);

	\diagram*{
		(v3) -- [photon, momentum=\(\ell\)] (v1);
		(v2) -- [photon] (v3);
	};
\end{feynman}
\end{tikzpicture}
= -2 Q_1^2 Q_2^2 \int \!\dd^D \ell \frac{(2p_1 + \ell) \cdot (2 p_1 + q + \ell)}
{\ell^2 (\ell - q)^2 (2p_1 \cdot \ell + \ell^2 + i \epsilon)}.
\end{equation}
In this integral, we use a dimensional regulator in a standard way ($D=4-2\epsilon$)
in order to regulate potential divergences.
We have retained an explicit $i \epsilon$ in the massive scalar propagator,
because it will play an important role below.

To extract the classical contribution of this integral to the amplitude, we recall from \sect{subsec:Wavefunctions} that we should set $q =  \hbar\qb$ and $\ell =  \hbar\ellb$, 
and therefore that the components of $q$ and $\ell$ are all small compared 
to $m$. 
Consequently, the triangle simplifies to,
\begin{equation}
T_{12} = \frac{4 i Q_1^2 Q_2^2 m_1^2}{\hbar} \int \dd^4 \bar \ell \frac{1}{\bar \ell^2 (\bar \ell - \bar q)^2 (p_1 \cdot \bar \ell + i \epsilon)}.
\label{eq:triangleIntermediate1}
\end{equation}
Here, we have taken the limit $D\rightarrow 4$, as the integral
is now free of divergences.
Notice that we have exposed one additional inverse power of $\hbar$. Comparing to the definition of $\impKerCl$ in \eqn{eq:impKerClDef}, we see that this inverse power of $\hbar$ will cancel against the explicit factor of $\hbar$ in $\ImpAclsup{(1)}$, signaling a classical contribution to the impulse.

At this point we employ a simple trick which simplifies the loop integral appearing in \eqn{eq:triangleIntermediate1}, and which will be of great help in simplifying the more complicated box topologies below. The on-shell condition for the outgoing particle 1 requires that $p_1 \cdot \qb = - \hbar \qb^2/2$, so replace 
$\ellb \rightarrow \ellb' = \qb - \ellb$ in $T_{12}$:
\begin{equation}
\begin{aligned}
T_{12} &= -\frac{4 i Q_1^2 Q_2^2 m_1^2}{\hbar} \int \dd^4 \bar \ell' \frac{1}{\bar \ell'^2 (\bar \ell' - \bar q)^2 (p_1 \cdot \bar \ell' + \hbar \qb^2 - i \epsilon)} \\
&= -\frac{4 i Q_1^2 Q_2^2 m_1^2}{\hbar} \int \dd^4 \bar \ell' \frac{1}{\bar \ell'^2 (\bar \ell' - \bar q)^2 (p_1 \cdot \bar \ell' - i \epsilon)} + \mathcal{O}(\hbar^0),
\end{aligned}
\end{equation}
Because of the linear power of $\hbar$ appearing in \eqn{eq:impKerClDef},
the second term
is in fact a quantum correction. We therefore neglect it, and write,
\begin{equation}
T_{12}= -\frac{4 i Q_1^2 Q_2^2 m_1^2}{\hbar} \int \dd^4 \bar \ell \frac{1}{\bar \ell^2 (\bar \ell - \bar q)^2 (p_1 \cdot \bar \ell - i \epsilon)} \, ,
\end{equation}
where we have dropped the prime on the loop momentum: $\ell' \rightarrow \ell$. Comparing with our previous expression, \eqn{eq:triangleIntermediate1}, for the triangle, the net result of these replacements has simply been to introduce an overall sign while, crucially, also switching the sign of the $i \epsilon$ term. 
Symmetrising over the two expressions for $T_{12}$, we learn that
\begin{equation}
T_{12}= \frac{2 Q_1^2 Q_2^2 m_1^2}{\hbar} \int \dd^4 \bar \ell \frac{\del(p_1 \cdot \bar \ell)}{\bar \ell^2 (\bar \ell - \bar q)^2} ,
\end{equation}
using the identity
\begin{equation}
\frac{1}{x-i \epsilon} - \frac{1}{x+i \epsilon} = i \del(x).
\label{eq:deltaPoles}
\end{equation}

The second triangle contributing to the amplitude, $T_{21}$, can be obtained from $T_{12}$ simply by interchanging the labels 1 and 2:
\begin{equation}
T_{21} = \frac{2 Q_1^2 Q_2^2 m_2^2}{\hbar} \int \! \dd^4 \bar \ell \,
\frac{\del(p_2 \cdot \bar \ell)}{\bar \ell^2 (\bar \ell - \bar q)^2}.
\end{equation}
These triangles contribute to the impulse kernel via
\begin{equation}
\begin{aligned}
\impKerCl \big|_\mathrm{triangle} &= \hbar\qb^\mu (T_{12} + T_{21}) 
\\&= {2 Q_1^2 Q_2^2}
 \; \bar q^\mu \int \! \frac{\dd^4 \bar \ell}{\bar \ell^2 (\bar \ell - \bar q)^2} \left(m_1^2\del(p_1 \cdot \bar \ell) + m_2^2\del(p_2 \cdot \bar \ell) \right).
\end{aligned}
\end{equation}
Recall that we must integrate over the wavefunctions
in order to obtain the classical impulse from the impulse kernel. 
As we have discussed in \sect{subsec:Wavefunctions}, 
because the inverse power of $\hbar$ here is canceled by the linear
power present explicitly in \eqn{eq:classicalLimitNLO}, 
we may evaluate the wavefunction integrals by replacing the $p_i$ by their classical
values $m_i \ucl_i$. The result for the contribution to the kernel is,
\begin{equation}
\begin{aligned}
\impKerTerm1 &\equiv
{2 Q_1^2 Q_2^2} \qb^\mu \!\int \! \dd^4 \ellb\;
\frac{1}{\ellb^2 (\ellb - \qb)^2} 
\biggl(m_1{\del(\ucl_1 \cdot \ellb)} 
+ m_2{\del(\ucl_2 \cdot \ellb)} \biggr)\,.
\end{aligned}
\label{TriangleContribution}
\end{equation}
One must still integrate this expression over $\qb$ as in \eqn{eq:classicalLimitNLO} to
obtain the contribution to the impulse.

\subsubsection{Boxes}
\label{Boxes}

\def\tp{\!+\!}
\def\tm{\!-\!}
\def\td{\!\cdot\!}
The one-loop amplitude also includes boxes, crossed boxes, and 
the NLO contribution to the impulse includes as well a term quadratic in the tree 
amplitude which we can think of as the cut of a one-loop box. Because of the
power of $\hbar$ in front of the first term in \eqn{eq:impKerClDef},
we need to extract the contributions of all of these quantities at order $1/\hbar$. 
However, as we will see, each individual diagram also contains singular terms 
of order $1/\hbar^2$.  We might fear that these terms pose an obstruction to the 
very existence of a classical limit of the observable in which we are interested. 
As we will see, this fear is misplaced, as these singular terms cancel completely,
leaving a well-defined classical result. It is straightforward to evaluate the individual
contributions, but making the cancellation explicit requires some care. 
We begin with the box:
\begin{equation}
\begin{aligned}
\hspace*{-7mm}i B &= \hspace*{-2mm}
\begin{tikzpicture}[scale=1.0, baseline={([yshift=-\the\dimexpr\fontdimen22\textfont2\relax] current bounding box.center)}, decoration={markings,mark=at position 0.6 with {\arrow{Stealth}}}] 
\begin{feynman}
	\vertex (i1) {$p_1$};
	\vertex [right=2.5 of i1] (o1) {$p_1 + q$};
	\vertex [below=2.5 of i1] (i2) {$p_2$};
	\vertex [below=2.5 of o1] (o2) {$p_2 - q$};
	\vertex [below right=1.1 of i1] (v1);
	\vertex [below left=1.1 of o1] (v2);
	\vertex [above right=1.1 of i2] (v3);
	\vertex [above left=1.1 of o2] (v4);
	\draw [postaction={decorate}] (i1) -- (v1);
	\draw [postaction={decorate}] (v1) -- (v2);
	\draw [postaction={decorate}] (v2) -- (o1);
	\draw [postaction={decorate}] (i2) -- (v3);
	\draw [postaction={decorate}] (v3) -- (v4);
	\draw [postaction={decorate}] (v4) -- (o2);
	\diagram*{
		(v3) -- [photon, momentum=\(\ell\)] (v1);
		(v2) -- [photon] (v4);
	};
\end{feynman}
\end{tikzpicture} 
\hspace*{-9mm}
= Q_1^2 Q_2^2 \int \! \dd^D \ell \;
\frac{(2 p_1 \tp \ell) \td (2p_2 \tm \ell)\,(2 p_1 \tp q\tp \ell) \td (2 p_2\tm q\tm \ell)}
   {\ell^2 (\ell \tm q)^2 (2 p_1 \cdot \ell \tp \ell^2 \tp i \epsilon)
    (-2p_2 \cdot \ell \tp \ell^2 \tp i \epsilon)}
 =\hspace*{-8mm}
 \\[-2mm]
\\&\hspace*{-5mm} \frac{Q_1^2 Q_2^2}{\hbar^{2+2\epsilon}} \!\!\int \! \dd^D \ellb\;
 \frac{\bigl[4 p_1\td p_2\tm 2\hbar(p_1\tm p_2)\td\ellb\tm \hbar^2\ellb^2\bigr]
       \bigl[4 p_1\td p_2\tm 2\hbar(p_1\tm p_2)\td(\ellb\tp \qb)\tm \hbar^2(\ellb\tp \qb)^2\bigr]}
{\ellb^2 (\ellb - \qb)^2 (2 p_1 \cdot \ellb + \hbar\ellb^2 + i \epsilon)
 (-2p_2 \cdot \ellb + \hbar\ellb^2 + i \epsilon)}\,,\hspace*{-8mm}
\end{aligned}
\end{equation}
where as usual, we have set $q = \hbar \qb$, $\ell = \hbar \ellb$. 
We get four powers of $\hbar$ from changing variables in the measure, but six inverse 
powers from the propagators\footnote{We omit fractional powers of $\hbar$ in this counting as they will disappear when we take $D \rightarrow 4$.}. We thus encounter an apparently singular $1/\hbar^2$ 
leading behaviour.  We must extract both this singular, $\Ord(1/\hbar^2)$, term 
as well as the terms contributing in the classical limit, which here are $\Ord(1/\hbar)$.
Consequently, we must take care to remember that the on-shell delta functions
enforce $\qb \cdot p_1 = - \hbar \qb^2 / 2$ and $\qb \cdot p_2 = \hbar \qb^2 / 2$. 

Performing a Laurent expansion in $\hbar$, truncating after order $1/\hbar$,
and separating different orders in $\hbar$, we find 
that the box is given by,
\begin{equation}
\begin{aligned}
B &= B_{-1}+B_0\,,
\\ B_{-1} &= \frac{4 i Q_1^2 Q_2^2}{\hbar^{2+2\epsilon}} (p_1 \cdot p_2)^2
\int \frac{\dd^D \ellb}{\ellb^2 (\ellb - \qb)^2
               (p_1 \cdot \ellb + i \epsilon)(p_2 \cdot \ellb - i \epsilon)} \,,
\\ B_{0} &= -\frac{2i Q_1^2 Q_2^2}{\hbar^{1+2\epsilon}} p_1 \cdot p_2 
\int \frac{\dd^D \ellb}{\ellb^2 (\ellb - \qb)^2
                        (p_1 \cdot \ellb + i \epsilon)(p_2 \cdot \ellb - i \epsilon)}
\\& \hspace*{40mm}\times
 \biggl[ 2{(p_1 - p_2)\cdot \ellb}
+ \frac{ (p_1 \cdot p_2) \ellb^2}{(p_1 \cdot \ellb + i \epsilon)} 
- \frac{(p_1 \cdot p_2) \ellb^2}{(p_2 \cdot \ellb - i \epsilon)}\biggl]\,.
\end{aligned}
\label{BoxExpansion}
\end{equation}
Note that pulling out a sign from one of the denominators has given the appearance of
flipping the sign of one of the denominator $i\epsilon$ terms.  We must also
bear in mind that the integral in $B_{-1}$ is itself \textit{not\/} $\hbar$-independent,
so that we will later need to expand it as well.

Similarly, the crossed box is
\begin{equation}
\begin{aligned}
i C &= 
\begin{tikzpicture}[scale=1.0, baseline={([yshift=-\the\dimexpr\fontdimen22\textfont2\relax] current bounding box.center)}, decoration={markings,mark=at position 0.6 with {\arrow{Stealth}}}] 
\begin{feynman}
	\vertex (i1) {$p_1$};
	\vertex [right=2.5 of i1] (o1) {$p_1 + q$};
	\vertex [below=2.5 of i1] (i2) {$p_2$};
	\vertex [below=2.5 of o1] (o2) {$p_2 - q$};
	\vertex [below right=1.1 of i1] (v1);
	\vertex [below left=1.1 of o1] (v2);
	\vertex [above right=1.1 of i2] (v3);
	\vertex [above left=1.1 of o2] (v4);
	\draw [postaction={decorate}] (i1) -- (v1);
	\draw [postaction={decorate}] (v1) -- node [above] {$\scriptstyle{p_1 + \ell}$} (v2);
	\draw [postaction={decorate}] (v2) -- (o1);
	\draw [postaction={decorate}] (i2) -- (v3);
	\draw [postaction={decorate}] (v3) -- (v4);
	\draw [postaction={decorate}] (v4) -- (o2);
	\diagram*{(v3) -- [photon] (v2);};
	\filldraw [color=white] ($ (v3) !.5! (v2) $) circle [radius = 2pt];
	\diagram*{	(v1) -- [photon] (v4);};
\end{feynman}
\end{tikzpicture} 
\\
&= Q_1^2 Q_2^2 \int \! \dd^D \ell\, \frac{(2 p_1 + \ell) \cdot (2p_2 - 2q + \ell)(2 p_1 +q+ \ell) \cdot (2 p_2 -q + \ell)}{\ell^2 (\ell - q)^2 (2 p_1 \cdot \ell + \ell^2 + i \epsilon)(2p_2 \cdot (\ell-q) + (\ell-q)^2 + i \epsilon)}
\\
&= \frac{Q_1^2 Q_2^2}{\hbar^{2+2\epsilon}} \!\! \int \! \dd^D \ellb\, 
\frac{(2 p_1 + \hbar\ellb) \cdot (2p_2 - 2\hbar\qb + \hbar\ellb)\,
      (2 p_1 +\hbar\qb+ \hbar\ellb) \cdot (2 p_2 -\hbar\qb + \hbar\ellb)}
       {\ellb^2 (\ellb - \qb)^2 (2 p_1 \cdot \ellb + \hbar\ellb^2 + i \epsilon)
         (2p_2 \cdot (\ellb-\qb) + \hbar(\ellb-\qb)^2 + i \epsilon)}.
\end{aligned}
\end{equation}
Using the on-shell conditions to simplify $p_i\cdot \qb$ terms in the denominator
and numerator, and once
again expanding in powers of $\hbar$, truncating after order $1/\hbar$, 
and separating different orders in $\hbar$,  we find,
\begin{equation}
\begin{aligned}
C &= C_{-1}+C_0\,,
\\ C_{-1} &= -\frac{4i Q_1^2 Q_2^2}{\hbar^{2+2\epsilon}} (p_1 \cdot p_2)^2
 \!\int \!\frac{\dd^D \ellb}{\ellb^2(\ellb - \qb)^2} 
 \frac{1}
      {(p_1 \cdot \ellb + i \epsilon)(p_2 \cdot \ellb + i \epsilon)} 
\\ C_{0} &= -\frac{2i Q_1^2 Q_2^2}{\hbar^{1+2\epsilon}} p_1 \cdot p_2
 \!\int \!\frac{\dd^D \ellb}{\ellb^2(\ellb - \qb)^2
                    (p_1 \cdot \ellb + i \epsilon)(p_2 \cdot \ellb + i \epsilon)} 
\\&\hspace*{37mm}\times
    \biggl[2  (p_1 + p_2) \cdot \ellb
 - \frac{(p_1 \cdot p_2) \ellb^2}{(p_1 \cdot \bar \ell + i \epsilon)} 
 - \frac{(p_1 \cdot p_2) [(\ellb - \qb)^2 - \qb^2]}
            {(p_2 \cdot \ellb + i \epsilon)}\biggr]\,.\hspace*{-20mm}
\end{aligned}
\label{CrossedBoxExpansion}
\end{equation}
Comparing the expressions for the $\Ord(1/\hbar^2)$ terms in the box and the crossed box, 
$B_{-1}$ and $C_{-1}$ respectively, we see that there is only a partial cancellation of 
the singular, $\mathcal{O}(1/\hbar^2)$, term in the reduced amplitude $\AmplB^{(1)}$. 
The impulse kernel, \eqn{eq:impKerClDef}, does contain another term, which is quadratic 
in the tree-level reduced amplitude  $\AmplB^{(0)}$.  We will see below that taking this 
additional contribution into account leads to a complete cancellation of the singular term; 
but the classical limit does not exist for each of these terms separately.

\subsubsection{Cut Box}
\label{CutBoxes}

\def\cutbox{\cut{B}}
In order to see the cancellation of the singular term we must incorporate
the term in the impulse kernel which is quadratic in tree amplitudes.  This
contribution can be viewed as proportional to the cut of the one-loop box,
weighted by the loop momentum $\hbar \xferb^\mu$:
\begin{equation}
\cutbox^\mu = -i\hbar^2\int \! \dd^4 \xferb \, \xferb^\mu \, \del(2 p_1 \cdot \xferb + \hbar \xferb^2) \del(2p_2 \cdot \xferb - \hbar \xferb^2) \times
\begin{tikzpicture}[scale=1.0, baseline={([yshift=-\the\dimexpr\fontdimen22\textfont2\relax] current bounding box.center)}, decoration={markings,mark=at position 0.6 with {\arrow{Stealth}}}] 
\begin{feynman}
	\vertex (i1) {$p_1$};
	\vertex [right=2.5 of i1] (o1) {$p_1 + \hbar \qb$};
	\vertex [below=2.5 of i1] (i2) {$p_2$};
	\vertex [below=2.5 of o1] (o2) {$p_2 - \hbar\qb$};
	\node [] (cutTop) at ($ (i1)!.5!(o1) $) {};
	\node [] (cutBottom) at ($ (i2)!.5!(o2) $) {};
	\vertex [below right=1.1 of i1] (v1);
	\vertex [below left=1.1 of o1] (v2);
	\vertex [above right=1.1 of i2] (v3);
	\vertex [above left=1.1 of o2] (v4);
	\draw [postaction={decorate}] (i1) -- (v1);
	\draw (v1) -- (v2);
	\draw [postaction={decorate}] (v2) -- (o1);
	\draw [postaction={decorate}] (i2) -- (v3);
	\draw (v3) -- (v4);
	\draw [postaction={decorate}] (v4) -- (o2);
	\filldraw [color=white] ($  (cutTop) - (3pt, 0) $) rectangle ($ (cutBottom) + (3pt,0) $) ;
	\draw [dashed] (cutTop) -- (cutBottom);
	\diagram*{
		(v3) -- [photon, momentum=\(\hbar\xferb\)] (v1);
		(v2) -- [photon] (v4);
	};
\end{feynman}
\end{tikzpicture}\!\!\!\!\!\!\!\!\!\!\!\!,
\end{equation}
where an additional factor of $\hbar$ in the second term of \eqn{eq:impKerClDef} will
be multiplied into \eqn{CombiningBoxes} below, as it parallels the factor in the first
term of \eqn{eq:impKerClDef}.
Evaluating the Feynman diagrams, we obtain,
\begin{equation}
\begin{aligned}
\cutbox^\mu = -i\frac{Q_1^2 Q_2^2}{\hbar^2} \int \! \dd^4 \xferb \, 
\del(2 p_1 \cdot \xferb + \hbar \xferb^2) \del(2p_2 \cdot \xferb - \hbar \xferb^2) \,
  \frac{\xferb^\mu}{\xferb^2 (\xferb - \qb)^2} \\
 \times
 (2 p_1 + \hbar\xferb ) \cdot (2p_2 - \xferb \hbar)\,
 (2 p_1 + \hbar\qb + \hbar\xferb ) \cdot (2 p_2 - \hbar\qb  -  \hbar\xferb)\, .
\end{aligned}
\end{equation}
As in the previous subsection, expand in $\hbar$, and truncate after order $1/\hbar$,
so that,
\begin{equation}
\begin{aligned}
\cutbox^\mu &= \cutbox_{-1}^\mu + \cutbox_{0}^\mu\,,
\\ \cutbox_{-1}^\mu &=  -\frac{4i Q_1^2 Q_2^2}{\hbar^2} (p_1 \cdot p_2)^2  
 \!\int \frac{\dd^4 \ellb \; \ellb^\mu}{\ellb^2 (\ellb - \qb)^2}
   \del(p_1 \cdot \ellb) \del(p_2 \cdot \ellb) \,, 
\\ \cutbox_{0}^\mu &=  -\frac{2i Q_1^2 Q_2^2}{\hbar} (p_1 \cdot p_2)^2  
 \!\int \frac{\dd^4 \ellb \; \ellb^\mu}{\ellb^2 (\ellb - \qb)^2}\,
 {\ellb^2} \bigl(\del'(p_1 \cdot \ellb) \del(p_2 \cdot \ellb) 
                        - \del(p_1 \cdot \ellb) \del'(p_2 \cdot \ellb) \bigr)\,.
\end{aligned}
\label{CutBoxExpansion}
\end{equation}
We have relabeled $\xferb\rightarrow\ellb$ in order to line up terms
more transparently with corresponding ones in the box and crossed box contributions.

\subsubsection{Combining Contributions}
\label{CombiningTerms}

We are now in a position to assemble the elements computed in the three
previous subsections in order to obtain the NLO contributions
to the impulse kernel $\impKerCl$, and thence the NLO contributions
to the impulse using \eqn{eq:classicalLimitNLO}.  
Let us begin by examining the singular terms.
We must combine the terms from the box, crossed box, and cut box.

We can simplify the cut-box contribution $\cutbox_{-1}^\mu$ 
by exploiting the linear change of variable $\ellb' = \qb - \ellb$,
\begin{equation}
\begin{aligned}
\cutbox_{-1}^\mu 
&= -\frac{4i Q_1^2 Q_2^2}{\hbar^2} (p_1 \cdot p_2)^2  
\!\int \dd^4 \ellb' \;\frac{ (\qb^\mu - \ellb'^\mu)}{\ellb'^2 (\ellb' - \qb)^2}
 \del(p_1 \cdot \ellb'-p_1\cdot\qb) \del(p_2 \cdot \ellb'-p_2\cdot \qb)
\\&= -\frac{4i Q_1^2 Q_2^2}{\hbar^2} (p_1 \cdot p_2)^2  
\!\int \dd^4 \ellb' \;\frac{ (\qb^\mu - \ellb'^\mu)}{\ellb'^2 (\ellb' - \qb)^2}
 \del(p_1 \cdot \ellb'+\hbar\qb^2/2) \del(p_2 \cdot \ellb'-\hbar\qb^2/2)
\\&= -\frac{2i Q_1^2 Q_2^2}{\hbar^2} (p_1 \cdot p_2)^2 \qb^\mu
\! \int \dd^4 \ellb \;\frac{\del(p_1 \cdot \ellb) \del(p_2 \cdot \ellb)}
           {\ellb^2 (\ellb - \qb)^2}
  +\Ord(1/\hbar)\,,
 \end{aligned}
\label{CutSingular}
\end{equation}
where we have used the on-shell conditions to replace 
$p_1\cdot\qb\rightarrow -\hbar\qb^2/2$
and $p_2\cdot\qb\rightarrow \hbar\qb^2/2$, and
where the last line arises from averaging over the two equivalent expressions for 
$\cutbox_{-1}^\mu$.

We may similarly simplify the singular terms from the box and cross box. Indeed, using the identity~\eqref{eq:deltaPoles} followed by the linear change of variable we have,
\begin{equation}
\begin{aligned}
B_{-1} + C_{-1} &= 
-\frac{4 Q_1^2 Q_2^2}{\hbar^{2+2\epsilon}} (p_1 \cdot p_2)^2
\!\int \frac{\dd^D \ellb}{\ellb^2 (\bar \ell - \bar q)^2} 
 \frac{1}{(p_1 \cdot \ellb + i \epsilon)} \del(p_2 \cdot \ellb)
\\&= \frac{4 Q_1^2 Q_2^2}{\hbar^{2+2\epsilon}} (p_1 \cdot p_2)^2
\!\int \frac{\dd^D \ellb'}{\ellb'^2 (\ellb' - \qb)^2}
 \frac{1}{(p_1 \cdot \ellb'+\hbar\qb^2/2 - i \epsilon)} 
  \del(p_2 \cdot \ellb'-\hbar\qb^2/2) 
\\&= \frac{2i Q_1^2 Q_2^2}{\hbar^2} (p_1 \cdot p_2)^2
\!\int \frac{\dd^4 \ellb\;\del(p_1 \cdot \ellb)\del(p_2 \cdot \ellb) }
       {\ellb^2 (\ellb - \qb)^2} + \Ord(1/\hbar)
\end{aligned}
\label{BoxSingular}
\end{equation}
where we have averaged over equivalent forms,
and then used \eqn{eq:deltaPoles} a second time in obtaining the last line.
At the very end, we took $D\rightarrow 4$.

Combining \eqns{CutSingular}{BoxSingular}, we find that the potentially
singular contributions to the impulse kernel in the classical limit are,
\begin{equation}
\begin{aligned}
\impKerCl &\big|_\textrm{singular} =
\hbar\qb^\mu (B_{-1}+C_{-1}) +\hbar\cutbox^\mu_{-1} 
\\&\hspace*{-6mm} = \frac{2i Q_1^2 Q_2^2}{\hbar} (p_1 \cdot p_2)^2\qb^\mu \Biggl[
\int \frac{\dd^4 \ellb\;\del(p_1 \cdot \ellb)\del(p_2 \cdot \ellb) }
       {\ellb^2 (\ellb - \qb)^2}
- \!\int \dd^4 \ellb \;\frac{\del(p_1 \cdot \ellb) \del(p_2 \cdot \ellb)}
           {\ellb^2 (\ellb - \qb)^2}\Biggr]
  +\Ord(\hbar^0)
\\&\hspace*{-6mm}=\Ord(\hbar^0)\,.
\end{aligned}
\label{CombiningBoxes}
\end{equation}
The dangerous terms cancel, leaving only well-defined contributions.

\def\Cancelling{Z}
It remains to extract the $\Ord(1/\hbar)$ terms from the box, crossed box,
and cut box contributions, and to combine them with 
the triangles~(\ref{TriangleContribution}), which are
of this order.  In addition to $B_0$ from \eqn{BoxExpansion}, $C_0$
from \eqn{CrossedBoxExpansion}, and $\cutbox^\mu_0$ from \eqn{CutBoxExpansion}, 
we must also include 
the $\Ord(1/\hbar)$ terms
left implicit in \eqns{CutSingular}{BoxSingular}.  In the former contributions,
we can now set $p_{1,2}\cdot \qb = 0$, as the $\hbar$ terms in the on-shell
delta functions would give rise to contributions of $\Ord(\hbar^0)$ to the
impulse kernel, which in turn will give contributions of $\Ord(\hbar)$ to
the impulse.  In combining all these terms, we make use of summing over
an expression and the expression after the linear change of variables;
the identity~(\ref{eq:deltaPoles}); and the identity,
\begin{equation}
\del'(x) = \frac{i}{(x-i\epsilon)^2} - \frac{i}{(x+i\epsilon)^2}\,.
\end{equation}
We find that,
\begin{equation}
\begin{aligned}
\hbar\qb^\mu &(B_0 + C_0) +\bigl[\hbar\qb^\mu (B_{-1} + C_{-1})\bigr]\big|_{\Ord(\hbar^0)}
 = 
\\& {2 Q_1^2 Q_2^2} \, (p_1 \cdot p_2)^2 \qb^\mu
\\&\hspace*{10mm}\times \!\int \!\frac{\dd^4 \ellb}{\ellb^2 (\ellb \tm \qb)^2}  
\biggl(\del(p_2 \td \ellb) 
          \frac{\ellb \td (\ellb \tm\qb) }{(p_1 \td \ellb \tp i \epsilon )^2} 
     + \del(p_1 \td \ellb) 
          \frac{\ellb \td (\ellb \tm\qb) }{(p_2 \td \ellb \tm i \epsilon )^2}\biggr) 
          + \Cancelling^\mu
\,,
\\ &\hspace*{-5mm}
\hbar\cutbox^\mu_0 +\bigl[\hbar\cutbox^\mu_{-1}\bigr]\big|_{\Ord(\hbar^0)}
 = 
\\& -\!2 iQ_1^2 Q_2^2 \, (p_1 \cdot p_2)^2 
\\&\hspace*{10mm}\times\! \int\! \frac{\dd^4 \ellb}{\ellb^2 (\ellb \tm \qb)^2}  
  \ellb^\mu \, \ellb \td (\ellb \tm \qb)\,
   \bigl( \del^\prime(p_1 \td \ellb) \del(p_2 \td \ellb) - \del^\prime(p_2 \td \ellb)
          \del(p_1 \td \ellb)\bigr) - \Cancelling^\mu \,,
\end{aligned}
\end{equation}
where we have now taken $D\rightarrow4$, and where the quantity $\Cancelling^\mu$ is,
\begin{equation}
\begin{aligned}
\Cancelling^\mu = i Q_1^2& Q_2^2 (p_1 \cdot p_2)^2 \qb^\mu 
\\&\times \!\int \! \frac{\dd^4 \ellb}{\ellb^2 (\ellb \tm \qb)^2} 
\;(2 \ellb \cdot \qb - \ellb^2 )
 \bigl( \del^\prime(p_1 \td \ellb) \del(p_2 \td \ellb) - \del^\prime(p_2 \td \ellb)
          \del(p_1 \td \ellb)\bigr) \, .
\end{aligned}
\end{equation}

Finally, we integrate over the external wavefunctions. 
The possible singularity in $\hbar$ has canceled, so 
as discussed in \sect{subsec:Wavefunctions}, 
we perform the integrals by replacing the momenta $p_i$ with their 
classical values $m_i \ucl_i$, so that the box-derived contribution is,
\begin{equation}
\begin{aligned}
\impKerTerm2 &\equiv
{2 Q_1^2 Q_2^2} \,\gamma^2 \qb^\mu
\\&\hspace*{10mm}\times\! \int \!\frac{\dd^4 \ellb}{\ellb^2 (\ellb \tm \qb)^2}  
\biggl(m_2\del(\ucl_2 \td \ellb) 
          \frac{\ellb \td (\ellb \tm\qb) }{(\ucl_1 \td \ellb \tp i \epsilon )^2} 
     + m_1\del(\ucl_1 \td \ellb) 
          \frac{\ellb \td (\ellb \tm\qb) }{(\ucl_2 \td \ellb \tm i \epsilon )^2}\biggr) 
\,,
\end{aligned}
\end{equation}
while that from the cut box is,
\begin{equation}
\impKerTerm3 \equiv
{-2 i Q_1^2 Q_2^2} \, \gamma^2 
\! \int\! \frac{\dd^4 \ellb}{\ellb^2 (\ellb \tm \qb)^2}  
  \ellb^\mu \, \ellb \td (\ellb \tm \qb)\,
   \bigl( m_2\del^\prime(\ucl_1 \td \ellb) \del(\ucl_2 \td \ellb) 
         - m_1\del^\prime(\ucl_2 \td \ellb) \del(\ucl_1 \td \ellb)\bigr)\,.
\end{equation}
In both contributions, we have dropped the $\Cancelling^\mu$ term which cancels
between the two.
The full impulse kernel is given by the sum $\impKerTerm1+\impKerTerm2+\impKerTerm3$,
and the impulse by,
\begin{equation}
\begin{aligned}
\DeltaPnlo &=
  \frac{i e^4}{4}\hbar \int \! \dd^4 \qb \, \del(\qb \cdot \ucl_1) \del(\qb \cdot \ucl_2) 
                   e^{-i \qb\cdot b} \impKerCl 
\\&= \frac{i e^4}{4}\hbar\int \! \dd^4 \qb \, \del(\qb \cdot \ucl_1) \del(\qb \cdot \ucl_2) 
 e^{-i \qb\cdot b} 
  \left( \impKerTerm1 + \impKerTerm2 + \impKerTerm3 \right)
\\&= \frac{i Q_1^2 Q_2^2 e^4}{2} 
\int \! \frac{\dd^4 \ellb}{\ellb^2 (\ellb - \qb)^2}\dd^4 \qb 
 \del(\qb \cdot \ucl_1) \del(\qb \cdot \ucl_2) 
                   e^{-i \qb\cdot b}
\\&\hspace*{15mm}\times\biggl[
\qb^\mu \bigl( \frac{\del(\ucl_1 \cdot \ellb)}{m_2}
                + \frac{\del(\ucl_2 \cdot \ellb)}{m_1} \bigr)
\\&\hspace*{15mm} \hphantom{\times\biggl[}
 +\gamma^2\qb^\mu \biggl(\frac{\del(\ucl_2 \td \ellb)}{m_1}
          \frac{\ellb \td (\ellb -\qb) }{(\ucl_1 \td \ellb + i \epsilon )^2} 
     + \frac{\del(\ucl_1 \td \ellb)}{m_2}
          \frac{\ellb \td (\ellb -\qb) }{(\ucl_2 \td \ellb - i \epsilon )^2}\biggr) 
\\&\hspace*{15mm} \hphantom{\times\biggl[} -
  i \gamma^2\ellb^\mu \, \ellb \td (\ellb - \qb)\,
   \biggl( \frac{\del^\prime(\ucl_1 \td \ellb) \del(\ucl_2 \td \ellb)}{m_1}
        - \frac{\del^\prime(\ucl_2 \td \ellb) \del(\ucl_1 \td \ellb)}{m_2}\biggr)\biggr]\,.
\end{aligned}
\label{NLOImpulse}
\end{equation}

\subsubsection{On-Shell Cross Check}

As we have seen, careful inclusion of boxes, crossed boxes as well as cut boxes are necessary to determine the impulse in the classical regime.
This may seem to be at odds with other work on the classical limit of amplitudes, which often emphasises the particular importance of triangle diagrams to the classical potential at next to leading order. However, in the context of the potential, the partial cancellation between boxes and crossed boxes is well-understood~\cite{Donoghue:1996mt}, and it is because of this fact that triangle diagrams are particularly important. The residual phase is known to exponentiate so that it does not effect classical physics. Meanwhile, the relevance of the subtraction of iterated (cut) diagrams has been discussed in~\cite{Sucher:1994qe,BjerrumBohr:2002ks,Neill:2013wsa}.

Nevertheless in the case of the impulse it may seem that the various boxes play a more significant role, as they certainly contribute to the classical result for the impulse. In fact, it is easy to see that these terms must be included to recover a physically sensible result. The key observation is that the final momentum, $\finalk_1^\mu$, of the outgoing particle after a classical scattering process must be on shell, $\finalk_1^2 = m_1^2$.

We may express the final momentum in terms of the initial momentum and the impulse, so that
\begin{equation}
\finalk_1^\mu = p_1^\mu + \Delta p_1^\mu\,.
\end{equation}
The on-shell condition is then
\begin{equation}
(\Delta p_1)^2 + 2 p_1 \cdot \Delta p_1 = 0\,.
\end{equation}
At order $e^2$, this requirement is satisfied trivially. At this order 
$(\Delta p_1)^2$ is negligible, while
\begin{equation}
p_1 \cdot \Delta p_1 = i m_1 e^2 Q_1 Q_2 \int \! \dd^4 \qb \,
 \del(\qb \cdot \ucl_1) \del(\qb \cdot \ucl_2) 
  \, e^{-i \qb \cdot b} \, \qb \cdot \ucl_1 \frac{\ucl_1 \cdot \ucl_2 }{\qb^2} = 0\,,
\end{equation}
using our result for the LO impulse in \eqn{eq:impulseClassicalLO}.

The situation is less trivial at order $e^4$, as neither $p_1 \cdot \Delta p_1$ 
nor $(\Delta p_1)^2$ vanish. In fact, at this order we may 
use \eqn{eq:impulseClassicalLO} once again to find that
\begin{equation}
\begin{aligned}
(\Delta p_1)^2 = - e^4 & Q_1^2 Q_2^2 \, (\ucl_1 \cdot \ucl_2)^2 \\
& \times \int \! \dd^4 \qb \dd^4 \qb' \, \del(\qb \cdot \ucl_1) 
\del(\qb \cdot \ucl_2) \del(\qb' \cdot \ucl_1) \del(\qb' \cdot \ucl_2) 
 \, e^{-i (\qb + \qb') \cdot b} \,  \frac{\qb \cdot \qb'}{\qb^2 \, \qb'^2}.
\end{aligned}
\label{DeltaPsquared}
\end{equation}
Meanwhile, to evaluate $p_1 \cdot \Delta p_1$ we must turn to our NLO result for the impulse, \eqn{NLOImpulse}. Thanks to the delta functions present in the impulse, we 
find a simple expression:
\begin{equation}
\begin{aligned}
2 p_1 \cdot & \Delta p_1 = e^4 Q_1^2 Q_2^2 \, (\ucl_1 \cdot \ucl_2)^2 \\
&\quad \times \int \! \dd^4 \qb \, \del(\qb \cdot \ucl_1) \del(\qb \cdot \ucl_2) 
\, e^{-i \qb\cdot b} \int \! \dd^4 \ellb \; 
\ellb \cdot \ucl_1 \, \del'(\ellb \cdot \ucl_1) \del(\ellb \cdot \ucl_2) \,
 \frac{\ellb \cdot (\ellb - \qb)}{\ellb^2 (\ellb - \qb)^2}\,.
\end{aligned}
\label{eq:pDotDeltaP}
\end{equation}
To simplify this expression, it may be helpful to imagine working in the restframe of the timelike vector $u_1$. Then, the $\ellb$ integral involves the distribution $\ellb_0 \, \del'(\ellb_0)$, while $\qb_0 = 0$. Thus the $\ellb_0$ integral has the form
\begin{equation}
\int \! \dd \ellb_0 \, \ellb_0 \, \del'(\ellb_0) \, f(\ellb_0{}^2) = -\int \! \dd \ellb_0 \, \del(\ellb_0) \, f(\ellb_0{}^2)\,.
\end{equation}
Using this observation, we may simplify equation~\eqref{eq:pDotDeltaP} to find
\begin{equation}
\begin{aligned}
2 p_1 \cdot \Delta p_1 &= -e^4 Q_1^2 Q_2^2 \, (\ucl_1 \cdot \ucl_2)^2 \\
& \hspace{40pt} \times 
\int \! \dd^4 \qb \dd^4 \ellb \, \del(\qb \cdot \ucl_1) \del(\qb \cdot \ucl_2) 
\del(\ellb \cdot \ucl_1) \del(\ellb \cdot \ucl_2)  
e^{-i \qb\cdot b}  \frac{\ellb \cdot (\ellb - \qb)}{\ellb^2 (\ellb - \qb)^2}  \\
&=  e^4 Q_1^2 Q_2^2 \, (\ucl_1 \cdot \ucl_2)^2 \\
& \hspace{40pt} \times
\int \! \dd^4 \ellb \, \dd^4 \qb' \;
 \del(\ellb \cdot \ucl_1) \del(\ellb \cdot \ucl_2) \del(\qb' \cdot \ucl_1) 
 \del(\qb' \cdot \ucl_2) \, e^{-i (\ellb+\qb')\cdot b} 
 \frac{\ellb \cdot \qb'}{\ellb^2 \qb'^2}\,,
\end{aligned}
\end{equation}
where in the last line we set $\qb' = \qb - \ellb$. 
This expression is equal but opposite to \eqn{DeltaPsquared}, and so
the final momentum is on shell as it must be.

It is worth remarking that the part of the NLO impulse that is relevant in this cancellation arises solely from the cut boxes. One can therefore view this phenomenon as an analogue of the removal of iterations of the tree in the potential.

\subsection{Radiation}
\label{sec:LOradiation}

The LO and NLO impulse are conservative in the sense that momentum is simply exchanged from particle 1 to particle 2 at these orders; it is only at NNLO that momentum radiated away back-reacts on the impulse. We will study this back-reaction in the next section, but first we turn to a direct computation of the radiated momentum.

The relevance of a classical limit of a scattering amplitude to what we are calling the 
radiation kernel was previously discussed by one of the authors and his 
collaborators~\cite{Luna:2017dtq}. The main advantage of the 
present discussion of radiation is that we have constructed a first-principles definition of 
the radiated momentum  \eqn{eq:radiatedMomentumClassical} in terms of an
on-shell scattering amplitude, \eqn{eq:defOfR}.
Our goal in this section is to compute the LO radiation kernel explicitly in 
electrodynamics.  Later, in section~\ref{sec:classicalLOradiation}, we will see that the 
radiation kernel has the classical interpretation of a current.

\begin{figure}[t]
	\centering
		\begin{tikzpicture}[decoration={markings,mark=at position 0.6 with {\arrow{Stealth}}}]
		\begin{feynman}
		\vertex (v1);
		\vertex [above left=1 and 0.66 of v1] (i1) {$\initialk_1+\xfer_1$};
		\vertex [above right=1 and 0.8 of v1] (o1) {$\initialk_1$};
		\vertex [right=1.2 of v1] (k) {$k$};
		\vertex [below left=1 and 0.66 of v1] (i2) {$\initialk_2+\xfer_2$};
		\vertex [below right=1 and 0.8 of v1] (o2) {$\initialk_2$};
		\draw [postaction={decorate}] (i1) -- (v1);
		\draw [postaction={decorate}] (v1) -- (o1);
		\draw [postaction={decorate}] (i2) -- (v1);
		\draw [postaction={decorate}] (v1) -- (o2);
		\diagram*{(v1) -- [photon] (k)};
		\filldraw [color=white] (v1) circle [radius=10pt];
		\draw [pattern=north west lines, pattern color=patternBlue] (v1) circle [radius=10pt];
		\end{feynman}	
		\end{tikzpicture} 
	\caption{The amplitude $\Ampl^{(0)}(\initialk_1+\xfer_1\,,\initialk_2+\xfer_2\rightarrow
	  \initialk_1\,,\initialk_2\,,k)$ appearing in the radiation kernel at leading order.}
	\label{fig:5points}
\end{figure}

\def\pol{\varepsilon}
The amplitude appearing in the radiation kernel is a five-point, tree amplitude (figure~
\ref{fig:5points}) which is easily computed.
We find that the radiation kernel is,
\begin{equation}
\begin{aligned}
\RadKerCl(\kb) &= {4} \int \! \dd^4 \xferb_1 \dd^4 \xferb_2 \;
\del(2p_1\cdot\xferb_1) \del(2p_2\cdot\xferb_2) \del^{(4)}(\kb - \xferb_1 - \xferb_2) \,
 e^{i\xferb_1 \cdot b} 
\\&\hphantom{=}\hspace*{15mm}
\times \biggl\{ \frac{Q_1^2Q_2^{\vphantom{2}}}{\xferb_2^2} 
\biggl[-p_2\cdot\pol + \frac{(p_1\cdot p_2)(\xferb_2\cdot\pol)}{p_1\cdot\kb} 
+ \frac{(p_2\cdot\bar{k})(p_1\cdot\pol)}{p_1\cdot\kb} 
\\&\hspace*{40mm} 
 - \frac{(\kb\cdot\xferb_2)(p_1\cdot p_2)(p_1\cdot\pol)}{(p_1\cdot\kb)^2}\biggr] 
\vphantom{\frac{(p_1\cdot p_2)(\xferb_2\cdot\pol_h)}{p_1\cdot\bar{k}}}
+ (1 \leftrightarrow 2)\biggr\} \,,
\label{eq:Rcalculation}
\end{aligned}
\end{equation}
where $\pol$ is the polarization vector for the emitted photon.
As the quantities $p_i \cdot \kb$ do not vanish on the support of the integrals in the radiation kernel, we have ignored the $i \epsilon$ factors in the massive propagators. 

This lowest order 
radiation kernel is of $\Ord(\hbar^0)$, so we may now perform the integrals over the 
particle wavefunctions, which effectively replaces the momenta $p_i$ in 
the radiation kernel by their classical values $m_i \ucl_i$. 
\begin{equation}
\begin{aligned}
\RadKerCl(\kb) &\rightarrow \frac{1}{m_1} \int \! \dd^4 \xferb_1 \dd^4 \xferb_2 \;
\del(\ucl_1\cdot\xferb_1) \del(\ucl_2\cdot\xferb_2) 
\del^{(4)}(\kb - \xferb_1 - \xferb_2) \, e^{i\xferb_1 \cdot b} 
\\& \hphantom{\rightarrow}\hspace*{15mm}\times\biggl\{
\frac{Q_1^2Q_2^{\vphantom{2}}}{\xferb_2^2} 
\biggl[-\ucl_2\cdot\pol + \frac{(\ucl_1\cdot \ucl_2)(\xferb_2\cdot\pol)}{\ucl_1\cdot\kb} 
+ \frac{(\ucl_2\cdot\kb)(\ucl_1\cdot\pol)}{\ucl_1\cdot\kb} 
\\&\hspace*{40mm} 
- \frac{(\kb\cdot\xferb_2)(\ucl_1\cdot \ucl_2)(\ucl_1\cdot\pol)}{(\ucl_1\cdot\kb)^2}\bigg] 
+ (1 \leftrightarrow 2)\biggr\} \,.
\label{eq:RcalculationAfterWFs}
\end{aligned}
\end{equation}
We will see this expression once again in section~\ref{sec:classicalLOradiation}, appearing as a classical current.

\subsection{Momentum Conservation and Radiation Reaction}
\label{sec:ALD}

We have already seen that conservation of momentum holds exactly (in \sect{sect:allOrderConservation}) and in our classical expressions (in \sect{sect:classicalConservation}). Let us now make sure that there is no subtlety in these discussions by explicit calculation.

To do so, we calculate the part of the NNLO impulse $\ImpBclsup{(\textrm{rad})}$ which encodes radiation reaction, defined in \eqn{eq:nnloImpulse}.
The two amplitudes appearing in equation~\eqref{eq:nnloImpulse} are in common with the amplitudes relevant for the radiated momentum, equation~\eqref{eq:defOfRLO}, though they are evaluated at slightly different kinematics. 
It will be convenient to change the sign of $\xferb_i$ here; with
that change, the amplitudes are:
\begin{equation}
\begin{aligned}
\AmplB^{(0)}&(\initialk_1 \initialk_2 \rightarrow 
              \initialk_1-\hbar\xferb_1\,, \initialk_2-\hbar\xferb_2 \, , \kb) =  
\\&\hspace*{-3mm}\frac{4Q_1^2 Q_2^{\vphantom{2}}}{\hbar^{2}\, \xferb_2^2} 
\bigg[-p_2\td\pol + \frac{(p_1\td p_2)(\xferb_2\td\pol)}{p_1\cdot\bar{k}} 
 + \frac{(p_2\cdot\kb)(p_1\td\pol)}{p_1\td\kb} 
 - \frac{(\kb\td\xferb_2)(p_1\td p_2)(p_1\td\pol)}{(p_1\cdot\kb)^2}\bigg] 
\\&\hspace*{-3mm}
 + ( 1 \leftrightarrow 2),
\end{aligned}
\end{equation}
and
\begin{equation}
\begin{aligned}
\AmplB^{(0)*}&(\initialk_1+\hbar \qb_1\,, \initialk_2 + \hbar \qb_2 \rightarrow 
               \initialk_1-\hbar\xferb_1\,, \initialk_2-\hbar\xferb_2 \,, \kb)
= 
\\&\hspace*{-3mm}\frac{4Q_1^2 Q_2^{\vphantom{2}}}{\hbar^{2}\, \xferb_2'^2} 
\bigg[\tm p_2\td\pol^*
 \tp \frac{(p_1\td p_2)(\xferb'_2\td\pol^*)}{p_1\cdot\kb} 
\tp \frac{(p_2\td\kb)(p_1\td\pol^*)}{p_1\cdot\kb} 
\tm \frac{(\kb\td\xferb'_2)(p_1\td p_2)(p_1\td\pol^*)}{(p_1\cdot\kb)^2}\bigg] 
\\&\hspace*{-3mm} + ( 1 \leftrightarrow 2),
\end{aligned}
\end{equation}
where we find it convenient to define $\xferb_i' = \qb_i + \xferb_i$ (after the
change of sign). 

We can now write the impulse 
contribution as,
\begin{equation}
\begin{aligned}
	\ImpBclsup{(\textrm{rad})} = -e^6 \Lexp \int \! d\Phi(\kb) 
	  \prod_{i = 1,2} \int \dd^4\xferb_i\, \dd^4 \xferb'_i \; \xferb_1^\mu \; 
	\mathcal{X}(\xferb_1, \xferb_2, \kb) \mathcal{X}^*(\xferb'_1, \xferb'_2, \kb)
	\Rexp\,  ,
\label{eq:impulseNNLO}
\end{aligned} 
\end{equation}
where 
\begin{equation}
\begin{aligned}
\mathcal{X}(\xferb_1, \xferb_2, \kb) &= {4}  \, \del(2 \xferb_1\cdot p_1)
 \del(2 \xferb_2\cdot p_2) \del^{(4)}(\bar k - \xferb_1 - \xferb_2) 
 \, e^{i b \cdot  \xferb_1}
  \\ 
& \hphantom{=}\hspace*{10mm} \times 
\biggl\{Q_1^2 Q_2^{\vphantom{2}} \frac{\pol_{\mu} }{\xferb_2^2}
 \bigg[-p_2^\mu + \frac{p_1\cdot p_2 \, \xferb_2^\mu}{p_1\cdot\kb} 
 + \frac{p_2\cdot\kb \, p_1^\mu}{p_1\cdot\kb} 
\\&\hphantom{=}\hspace*{35mm}
 - \frac{(\kb\cdot\xferb_2)(p_1\cdot p_2) \, p_1^\mu}{(p_1\cdot\kb)^2}\bigg] 
  + (1 \leftrightarrow 2)\biggr\}\,.
\end{aligned}
\label{eq:X1}  
\end{equation}
This expression is directly comparable to those for radiated momentum: \eqn{eq:impulseNNLO}, and the equivalent impulse contribution to particle 2, balance the radiated momentum \eqn{eq:radiatedMomentumClassicalLO} using $\xferb_1^\mu + \xferb_2^\mu = \bar k^\mu$, provided that the radiation kernel, \eqn{eq:Rcalculation}, is related to integrals over $\mathcal{X}$. Indeed this relationship holds: the integrations present in the radiation kernel are supplied by the $\xferb_i$ and $\xferb'_i$ integrals in \eqn{eq:impulseNNLO}; these integrations disentangle in the sum of impulses on particles 1 and 2 when we impose $\xferb_1^\mu + \xferb_2^\mu = \bar k^\mu$, and 
then form the square of the radiation kernel.

It is interesting to compare this radiated momentum with the situation in traditional
formulations of classical physics, where one must include the ALD radiation reaction force 
by hand in order to enforce momentum conservation. Because the situation is simplest when 
only one particle is dynamical, let us take the mass $m_2$ to be very large compared to 
$m_1$ in the remainder of 
this section, and work in particle 2's rest frame.  In this frame,
it does not radiate, and the only radiation reaction is on particle 1 --- the radiated
momentum is precisely balanced by 
the impulse on particle 1 due to the ALD force. We can therefore continue our discussion 
with reference to our expression for radiated momentum, \eqn{eq:radiatedMomentumClassicalLO} 
and the radiation kernel, \eqn{eq:Rcalculation}. In this situation we may also simplify the 
kernels by dropping the $(1 \leftrightarrow 2)$ instruction: notice that the explicit terms 
in the kernel of equation~\eqref{eq:Rcalculation} are linear in $m_2$. Terms obtained by 
symmetrising in particle labels are therefore linear in $m_1$, and so are subdominant when $m_2 \gg m_1$. 

\newcommand{\phInt}{Y^\mu_\textrm{r}} 
We will compute the impulse due to the ALD force directly from its classical expression in \sect{all:radReac}. But in preparation for that comparison there is one step which we must take.
Classical expressions for the force---which involve only the particle's momentum and its derivatives---do not involve any photon phase space. So we must perform the integration over $\df(\wn k)$ which is present in \eqn{eq:radiatedMomentumClassicalLO}. 

To organise the calculation, we integrate over the $\qb_1$ variables in the radiation kernel, \eqn{eq:Rcalculation} using the four-fold delta function, so that we may write the radiated momentum as,
\begin{align}
\Rad^{\mu,(0)}_\class = -\frac{e^6 Q_1^4 Q_2^2}{m_1^2} \int\!\dd^4\qb \dd^4\qb'\;
 e^{-ib\cdot(\qb - \qb')} \del(\ucl_1\cdot(\qb - \qb')) 
 \frac{\del(\ucl_2\cdot\qb)}{\qb^2} \frac{\del(\ucl_2\cdot\qb')}{\qb'^2} \phInt\,,\label{eq:rrmidstage}
\end{align}
where we renamed the remaining variables, $\xferb_2\rightarrow \qb$
and $\xferb'_2\rightarrow \qb'$, in order to match the notation used later
in \sect{ClassicalCalculations}.
After some algebra we find,
\begin{equation}
\begin{aligned}
\phInt = \int \! \df (\kb) \,& \del(\ucl_1\cdot \kb - \wn E) \, \kb^\mu \,
 \left[ 1 + \frac{(\ucl_1\cdot \ucl_2)^2(\qb\cdot \qb')}{\wn E^2} 
 + \frac{(\ucl_2\cdot\kb)^2}{\wn E^2}  \right. \\ 
&\left. - \frac{(\ucl_1\cdot \ucl_2)(\ucl_2\cdot\kb)\kb\cdot(\qb+\qb')}{\wn E^3} + \frac{(\ucl_1\cdot \ucl_2)^2(\kb\cdot\qb)(\kb\cdot\qb')}{\wn E^4}  \right].
\label{eq:phaseSpaceIntegral}
\end{aligned}
\end{equation}
The quantity $\wn E$ is defined to be $\wn E = \ucl_1 \cdot \wn k$; in view of the delta 
function, the integral is constrained so that $\wn E = \ucl_1 \cdot \qb$. This quantity is 
the wavenumber of the photon in the rest frame of particle 1, and is fixed from the point of 
view of the phase space integration. As a result, the integrals are simple: there are two 
delta functions (one explicit, one in the phase space measure) which can be used to perform 
the $\kb^0$ integration and to fix the magnitude of the spatial wavevector. The remaining 
integrals are over angles and are performed in Appendix~\ref{app:integralappendix}.
The radiated momentum takes a remarkably simple form after the phase space integration:

\begin{equation}
\begin{aligned}
\Rad^{\mu,(0)}_\class = -\frac{e^6 Q_1^4 Q_2^2}{3\pi m_1^2} &\int\!\dd^4\qb 
\dd^4\qb'\; e^{-ib\cdot(\qb - \qb')} \del(\ucl_1\cdot(\qb - \qb'))
 \frac{\del(\ucl_2\cdot\qb)}{\qb^2} \frac{\del(\ucl_2\cdot\qb')}{\qb'^2}
  \Theta(\ucl_1\cdot\qb)\\ &\times\left[(\ucl_1\cdot\qb)^2 
  + \qb\cdot\qb'(\ucl_1\cdot \ucl_2)^2\right] \ucl_1^\mu .
\label{eq:radTheta}
\end{aligned}
\end{equation}

The $\Theta$ function is a remnant of the photon phase space volume, so it will be convenient to remove it. The delta functions in the integrand 
in \eqn{eq:radTheta} constrain the components of the vectors $\qb$ and $\qb'$ which lie in 
the two dimensional space spanned by $u_1$ and $u_2$. Let us call the components of $q$ and 
$q'$ in this plane to be $q_\parallel$ and $q'_\parallel$. Then the delta functions set $q_
\parallel = q'_\parallel$. As a result, the integrand (ignoring the $\Theta$ function) is 
symmetric in $q_\parallel \rightarrow - q_\parallel$. Consequently we may symmetrise to find
\begin{equation}
\begin{aligned}
\Rad^{\mu,(0)}_\class = -\frac{e^6 Q_1^4 Q_2^2}{6\pi m_1^2} &\int\!\dd^4\qb \dd^4\qb'\;
 e^{-ib\cdot(\qb - \qb')} \del(\ucl_1\cdot(\qb - \qb')) 
 \frac{\del(\ucl_2\cdot\qb)}{\qb^2} \frac{\del(\ucl_2\cdot\qb')}{\qb'^2} 
 \\ &\hspace*{10mm}\times\left[(\ucl_1\cdot\qb)^2 
                    + \qb\cdot\qb'(\ucl_1\cdot \ucl_2)^2\right]\ucl_1^\mu .
\label{eq:rrResult}
\end{aligned}
\end{equation}
We will see in \sect{all:radReac} that this expression is equal but opposite to the impulse 
obtained from the classical ALD force.

\section{Classical Calculations}
\label{ClassicalCalculations}

In previous sections, we have shown how to understand the electromagnetic scattering of 
point-like particles $Q_i$ using the methods of quantum field theory. We now turn to a 
purely classical approach to the same observables, restricting (once again) to 
electromagnetic scattering for simplicity. 
Our goal in this section is to reassure any skeptical reader that the expressions which we have claimed to be classical, are indeed classical. 
This is straightforward in electrodynamics; it would be harder in
gravity. We expect, however, that an application of our methods in a gravitational context 
will be advantageous.

As usual we have in mind a two-body collision, and we discuss, firstly, the impulse on a particle, secondly the momentum radiated away, 
and lastly the topic of conservation of momentum and the ALD radiation reaction force. 
Our classical approach will be to solve the coupled equations of motion perturbatively. As we will see, these calculations have an iterative structure.
This iteration is straightforward in principle, though it quickly becomes tedious in practice. For this reason we will frequently restrict to the case where 
the mass $m_2$ of particle 2 is much larger than the mass $m_1$ of particle 1, so that particle 2 can be treated as being static. This assumption
was unnecessary using quantum methods\footnote{With the exception of radiation reaction---in that case, we took $m_2$ large to prepare for
a comparison with classical results.}, where the symmetry of Feynman diagrams simplifies matters.

\subsection{The classical electromagnetic impulse}
\label{sec:EDimpulse}

\def\position{x}
Classically, we may take our particles to move along world-lines $\position_i(\tau_i)$ with proper velocities 
$v_i(\tau_i) = d\position_i/d\tau_i$ and momenta $p_i(\tau_i) = m_i v_i(\tau_i)$.
We must solve the Maxwell equation,
\begin{equation}
\partial_\mu F^{\mu\nu}(\position) = J^\nu(\position)
 = e \sum_i Q_i \int \! d \tau_i \; \delta^4(\position - \position_i(\tau_i)) v^\nu_i(\tau_i),
\end{equation}
and the Lorentz force laws\footnote{We will need to include the ALD force law later 
to account for radiation reaction.},
\begin{equation}
\frac{d p^\mu_i}{d \tau_i} = e Q_i \, F^{\mu\nu}(\position_i(\tau_i)) \, v_{i\nu}(\tau_i)\,.
\label{eq:LorentzForce}
\end{equation}
To set up the perturbative method, we assume that the trajectories can be expanded as a power series in the coupling. At zeroth order in the expansion, the trajectories are simply straight lines:
\begin{align}
\position_1(\tau_1) &= b + u_1 \tau_1, \quad \position_2(\tau_2) = u_2 \tau_2,
\end{align}
where $u_1$ and $u_2$ are constant vectors.
Note that, as with our quantum mechanical setup in \eqn{InitialState}, we have aligned the initial trajectory of particle 2 with the spatial origin, and have translated particle 1 through an impact parameter $b$ relative to particle 2. The vectors $u_i$ are the zeroth order terms in the proper velocities: $v_i(\tau_i) = u_i + \mathcal{O}(e^2)$; they will play a prominent role below.

\def\corr#1{\Delta^{(#1)}}
The perturbative expansion proceeds by determining the first order electromagnetic field sourced by the two particles, taken to be moving on straight-line trajectories. Knowledge of the field allows us to compute the first order forces on the particles, and hence the associated (small) first order deviations of the trajectories from straight line motion. Armed with this knowledge of the first order trajectory, we may compute the second order fields, forces and deviations. Iterating this procedure allows us to compute to any desired perturbative order.
More formally, we expand the trajectories as,
\begin{equation}
\begin{aligned}
\position_1(\tau) &= b + u_1 \tau+ \corr1 \position_1(\tau) + \corr2 \position_1(\tau) + \cdots\,, \\
\position_2(\tau) &= u_2 \tau+ \corr1 \position_2(\tau) + 
\corr2 \position_2(\tau) + \cdots\,,
\end{aligned}
\end{equation}
where $\corr{n} \position(\tau)$ is of order $e^{2n}$. We will similarly expand the velocities and forces perturbatively:
\begin{align}
v_i(\tau_i) &= u_i + \corr1 v_i(\tau_i) + \corr2 v_i(\tau_i) ,\\
f_i(\tau_i) &= \corr1 f_i(\tau_i) + \corr2 f_i(\tau_i) ,
\end{align}
where again $\corr{n} v_i(\tau_i)$ and $\corr{n} f_i(\tau_i)$ are of order $e^{2n}$.

\subsubsection{Leading order}
Our first order of business is to determine the leading order electromagnetic fields. Classically, the force on particle 1 is due to the field of particle 2; as the result is symmetric under interchange of the two particles, we will only compute the field due to particle 2. 
Working in Lorenz gauge, the relevant gauge field is,
\begin{equation}
\partial^2 A_2^\mu(x) = Q_2 e \int \! d\tau \, \delta^4(x - u_2 \tau) \, u_2^\mu.
\end{equation}
We find it convenient to work in Fourier space, choosing the conventions
\begin{align}
f(x) &= \int \! \dd^4 \qb \, \tilde f(\qb) \, e^{-i \qb \cdot x}, \\
\tilde f(\qb) &= \int \! d^4 x \, f(x) \, e^{i \qb \cdot x}.
\label{eq:FourierTransform}
\end{align}
\def\Atilde{\widetilde{A}}
The momentum-space gauge field is easily found to be
\begin{equation}
\Atilde_2^\mu(\qb) 
= - \frac{e Q_2}{\qb^2} \; u_2^\mu \; \del(\qb \cdot u_2)\,,
\label{eq:loA2}
\end{equation}
with field strength (in position space)
\begin{equation}
F_2^{\mu\nu} (x) = i e Q_2 \int \! \dd^4 \qb \, \del(\qb \cdot u_2) \, e^{-i \qb \cdot x} \, \frac{\qb^\mu \, u_2^\nu - u_2^\mu \, \qb^\nu}{\qb^2}.
\label{eq:f2}
\end{equation}
The Lorentz force, equation~\eqref{eq:LorentzForce}, on particle 1 requires this field strength evaluated at the position of particle 1. At leading order, we may use the straight line approximation to the trajectory of particle 1, so that the leading order force is
\begin{equation}
\frac{d p^\mu_1}{d \tau_1} =i e^2 Q_1 Q_2 \int \!\dd^4 \qb \, \del(\qb \cdot u_2) \, e^{- i \qb \cdot (b + u_1 \tau_1)} \,  \frac{\qb^\mu \, u_1 \cdot u_2 - u_2^\mu \, \qb \cdot u_1}{\qb^2}.
\label{eq:LOforce}
\end{equation}
The leading order impulse is the total time integral of the force,
\begin{align}
\DeltaPlo &\equiv \int_{-\infty}^\infty d\tau_1 \, \frac{d p^\mu_1}{d \tau_1} \\
&= i e^2 Q_1 Q_2 \int \! \dd^4 \qb \, \del(\qb \cdot u_1) \del(\qb \cdot u_2)  \, e^{-i \qb \cdot b} \, \qb^\mu \, u_1 \cdot u_2  \frac1{\qb^2},
\label{eq:classicalImpulse}
\end{align}
in complete agreement with \eqn{eq:impulseClassicalLO}, which was obtained from a tree-level scattering amplitude. 

\subsubsection{Next-to-leading order}
\label{sec:EDimpulseNLO}

The LO calculation is very simple, and so it is more interesting to look at the next order of perturbation theory. In our quantum-mechanical treatment, this following order was
computed in section~\ref{sec:nloQimpulse} using loop diagrams. In this section, we derive the same expression by iterating the perturbative solution of the classical equations to the next order. The fact that both methods yield the same result is a vivid demonstration that loops involving massive particles are not simply quantum corrections~\cite{Holstein:2004dn}.

In this section we take particle 2 to be static, leaving the more general case as an exercise for the reader. When particle 2 is static, its field strength is given by equation~\eqref{eq:f2} to all orders.

The NLO impulse is the time integral of the NLO Lorentz force. With our assumption of a static particle 2, we know the gauge field acting on particle 1 exactly, and so the origin of this NLO force is simply that particle 1 is not quite moving on a straight line. Thus, to find the correction to the force, we must first determine the motion of particle 1 with NLO accuracy.

We obtain the first perturbative correction to the velocity and to the trajectory of particle 1 by integrating the leading order force, equation~\eqref{eq:LOforce} from time $\tau_1 = -\infty$ to a finite time $\tau_1$. This integral must converge as $\tau_1 \rightarrow -\infty$, so we follow standard practice (see, for example, Jackson~\cite{Jackson:1998nia}  p. 676) and replace $\qb \cdot u_1$ in the argument of the exponential with $\qb \cdot u_1 + i \epsilon$. The correction to the velocity is then
\begin{equation}
\begin{aligned}
m_1 \corr1 v_1^\mu &= i e^2 Q_1 Q_2 \int \!\dd^4 \qb \, \del(\qb \cdot u_2)  \, e^{- i \qb \cdot b} \, \frac{\qb^\mu \, u_1 \cdot u_2 - u_2^\mu \, \qb \cdot u_1}{\qb^2}  \int_{-\infty}^{\tau_1} d \tau_1 \, e^{- i (\qb \cdot u_1 + i \epsilon)\tau_1} \\
&= - e^2 Q_1 Q_2 \int \!\dd^4 \qb \, \del(\qb \cdot u_2) \, e^{-i \qb \cdot (b + u_1 \tau_1)} \, \frac{\qb^\mu \, u_1 \cdot u_2 - u_2^\mu \, \qb \cdot u_1}{\qb^2 (\qb \cdot u_1 + i \epsilon)},
\label{eq:pertVelocity}
\end{aligned}
\end{equation}
where on the second line, we have displayed the $i \epsilon$ convergence factor in the denominator explicitly while leaving it implicit in the argument of the exponential. The leading correction to the position of the particle is given by integrating once more, with the result:
\begin{equation}
\corr1 \position_1^\mu(\tau_1) = -i \frac{e^2 Q_1 Q_2}{m_1} \int \! \dd^4 \qb \, \del(\qb \cdot u_2)  \, e^{-i \qb \cdot (b + u_1 \tau_1)} \, \frac{\qb^\mu \, u_1 \cdot u_2 -  u_2^\mu \, \qb \cdot u_1}{\qb^2 (\qb \cdot u_1 + i \epsilon)^2}.
\label{eq:pertTraj}
\end{equation}
We have now collected the information we need to compute the NLO Lorentz force $\corr1 f^\mu$, and therefore the NLO impulse. Recalling that the field strength involved in the force is given by equation~\eqref{eq:f2}, it is easy to see that the NLO force is
\begin{align}
\corr1 f^\mu = i e^2 Q_1 Q_2 \int \! \dd^4 \wn {\ell} \, \del(\wn {\ell}\cdot u_2) \, & e^{-i \wn {\ell} \cdot (b + u_1 \tau_1)} \, \frac{\wn{\ell}^\mu u_2^\nu - \wn {\ell}^\nu u_2^\mu}{\ellb^2}\nonumber\\&\times\left( \frac{d}{d\tau_1} \corr1 \position_{1\nu}(\tau_1) - i \wn{\ell} \cdot \corr1 \position_1(\tau_1) u_{1\nu}\right),
\label{eq:pertForce}
\end{align}
where we relabelled the variable of integration $\qb \rightarrow \ellb$ for later convenience.
Using our knowledge of the corrected trajectory, we evaluate
\begin{multline}
\frac{d}{d\tau} \corr1 \position_{1\nu}(\tau_1) - i \wn {\ell} \cdot \corr1 \position_1(\tau_1) u_{1\nu} = 
-\frac{e^2 Q_1 Q_2}{m_1} \int \! \dd^4 \wn {q}' \, \del(\wn {q}' \cdot u_2) \, e^{-i \wn {q}' \cdot (b + u_1 \tau_1)} \\
\times \left( \frac{\wn {q}'_\nu \, u_1 \cdot u_2 - u_{2\nu} \, \wn {q}' \cdot u_1}{\wn {q}'^2 (\wn{q}' \cdot u_1 + i \epsilon)} + \frac{\wn {\ell} \cdot \wn{q}' \, u_1 \cdot u_2 \, u_{1\nu}}{\wn {q}'^2 (\wn {q}' \cdot u_1 + i \epsilon)^2} \right) ,
\end{multline}
omitting terms which vanish on the support of the delta functions in equation~\eqref{eq:pertForce}.
The integral $\int_{-\infty}^\infty \, d\tau_1 \, \corr1 f^\mu$ is the next-to-leading order impulse, given explicitly by
\begin{equation}
\begin{aligned}
\DeltaPnlo\big|_{m_2\rightarrow\infty} = \frac{i e^4 Q_1^2 Q_2^2}{m_1} \int \! \dd^4 \qb \, \del(\qb \cdot u_1) \del(&\qb \cdot u_2) e^{-i \qb\cdot b} \! \int \! \dd^4 \ell \frac{\del(\bar{\ell} \cdot u_2)}{\bar{\ell}^2 (\bar{\ell} - \qb)^2} 
\\&\times
\left[ 1+ \frac{\bar{\ell} \cdot (\bar{\ell} -\qb) (u_1 \cdot u_2)^2}{(\bar{\ell} \cdot u_1 - i \epsilon)^2} \right] \bar{\ell}^\mu.
\end{aligned}
\end{equation}
To obtain this result, we shifted one of the variables of integration by setting $\wn{q} = \wn{q}' + \bar{\ell}$. 

This expression for the NLO impulse is not quite in the form we obtained using quantum methods in section~\ref{sec:nloQimpulse}. The necessary rearrangement is as follows. We exploit the change of variable $\bar{\ell}' = \qb - \bar{\ell}$ and define a vector integral
\begin{equation}
I^\mu =  \int \! \dd^4 \bar{\ell} \frac{\del(\bar{\ell} \cdot u_2)}{\bar{\ell}^2 (\bar{\ell} - \qb)^2}\left[ 1+ \frac{\bar{\ell} \cdot (\bar{\ell} -\qb) (u_1 \cdot u_2)^2}{(\bar{\ell} \cdot u_1 - i \epsilon)^2} \right] \bar{\ell}^\mu;
\end{equation}
then we find
\begin{align}
I^\mu &= \int \! \dd^4 \bar{\ell'} \frac{\del(\bar{\ell'} \cdot u_2)}{\bar{\ell'}^2 (\bar{\ell'} - \qb)^2}\left[ 1+ \frac{\bar{\ell'} \cdot (\bar{\ell'} -\qb) (u_1 \cdot u_2)^2}{(\bar{\ell'} \cdot u_1+i \epsilon)^2} \right] \left(\qb^\mu - \bar{\ell'}^\mu\right) \nonumber \\
&=\frac12 \int \! \dd^4 \bar{\ell} \frac{\del(\bar{\ell} \cdot u_2)}{\bar{\ell}^2 (\bar{\ell} - q)^2}\bigg\{\left[ 1+ \frac{\bar{\ell} \cdot (\bar{\ell} -\qb) (u_1 \cdot u_2)^2}{(\bar{\ell} \cdot u_1 + i \epsilon)^2} \right]\qb^\mu \nonumber\\
& \qquad\qquad\qquad\qquad\qquad- i \left[{\bar{\ell} \cdot (\bar{\ell} -\qb) (u_1 \cdot u_2)^2}\del^\prime(\bar{\ell} \cdot u_1) \right] \bar{\ell}^\mu \bigg\}.
\end{align}
Putting the pieces together, we arrive at our final result for the NLO impulse when $m_2\rightarrow\infty$:
\begin{equation}
\begin{aligned}
&\DeltaPnlo\big|_{m_2\rightarrow\infty} =  \int \! \dd^4 \qb \, \del(\qb \cdot u_1) \del(\qb \cdot u_2) e^{-i \qb\cdot b} \, \frac{e^4 Q_1^2 Q_2^2}{2m_1}  \int \! \dd^4 \bar{\ell} \, \frac{\del(\bar{\ell} \cdot u_2)}{\bar{\ell}^2 (\bar{\ell} - \qb)^2} \\
& \quad \times  \bigg[ i \qb^\mu \left( 1 + \frac{\bar{\ell} \cdot (\bar{\ell} -\qb) (u_1 \cdot u_2)^2}{(\bar{\ell} \cdot u_1 + i \epsilon)^2} \right) + 
\bar{\ell}^\mu \, \ellb \cdot (\bar{\ell} -\qb) (u_1 \cdot u_2)^2 \, \del^\prime(\bar{\ell} \cdot u_1) \bigg].\label{eq:classicalNLOimpulseResult}
\end{aligned}
\end{equation}
We have thereby reproduced all terms in \eqn{NLOImpulse} which survive in the limit 
$m_2 \rightarrow \infty$, consistent with our present assumption that $m_2$ is very large.
The full result for the NLO impulse requires accounting for corrections to the field strength of particle 2 due to its motion. The mechanics of the calculation are very much the same as in the discussion above. Taking these additional effects into account, one finds that the NLO impulse is precisely \eqn{NLOImpulse}.

\subsection{Classical Radiated Momentum}

Our next topic is the momentum radiated during a classical collision, using the standard methods of classical field theory. We first discuss a general expression for the momentum radiation which is analogous to the all-order radiation formula, equation~\eqref{eq:radiatedMomentumClassicalAllOrder}, we found using quantum methods. We then apply the formula at leading order, making explicit contact with the LO radiation kernel we computed in section~\ref{sec:LOradiation}.

\subsubsection{General expressions}
\label{sec:genClassicalRadiation}

The electromagnetic stress-energy tensor,
\begin{equation}
T^{\mu\nu}(x) = F^{\mu\alpha}(x) F_\alpha{}^\nu(x) + \frac 14 \eta^{\mu\nu} F^{\alpha\beta}(x) F_{\alpha\beta}(x) \,,
\end{equation}
is the key quantity which describes the distribution and flux of energy and momentum. In particular, the (four-)momentum flux through a three dimensional surface $\partial \Omega$ with surface element $d\Sigma_\nu$ is,
\begin{equation}
K^\mu = \int_{\partial \Omega} d \Sigma_\nu T^{\mu\nu}(x)\,.
\end{equation}
We are interested in the total momentum radiated as two particles scatter. At each time $t$, we therefore surround the two particles with a large sphere. The instantaneous flux of momentum  is measured by integrating over the surface area of the sphere; the total momentum radiated is then the integral of this instantaneous flux over all times. It is straightforward to determine the momentum radiated by direct integration over these spheres using textbook methods, as discussed in Appendix~\ref{app:classicalMomentumAgain}.

A simpler but more indirect method is the following. We wish to use the Gauss theorem to write,
\begin{equation}
K^\mu = \int_{\partial \Omega} d \Sigma_\nu T^{\mu\nu}(x) =  \int \! d^4x \, \partial_\nu T^{\mu\nu}(x)\,.
\end{equation}
However, the spheres surrounding our particle are not the boundary of all spacetime: they do not include the timelike future and past boundaries. To remedy this, we use a trick due to Dirac~\cite{Dirac:1938nz}. 

The radiation we have in mind is causal, so we solve the Maxwell equation with retarded boundary conditions. We denote these fields by $F^{\mu\nu}_\textrm{ret}(x)$.
We could equivalently solve the Maxwell equation using the advanced Green's function. If we wish to determine precisely the same fields $F^{\mu\nu}_\textrm{ret}(x)$ but using the advanced Green's function, we must add a homogeneous solution of the Maxwell equation. Fitting the boundary conditions in this way requires subtracting the incoming radiation field $F^{\mu\nu}_\textrm{in}(x)$ which is present in the advanced solution (but not in the retarded solution) and adding the outgoing radiation field (which is present in the retarded solution, but not the advanced solution.) In other words,
\begin{equation}
F^{\mu\nu}_\textrm{ret}(x) - F^{\mu\nu}_\textrm{adv}(x) = - F^{\mu\nu}_\textrm{in}(x) + F^{\mu\nu}_\textrm{out}(x)\,.
\end{equation}
Now, the radiated momentum $K^\mu$ in which we are interested is described by $F^{\mu\nu}_\textrm{out}(x)$. The field $F^{\mu\nu}_\textrm{in}(x)$ transports the same total amount of momentum in from infinity, ie it transports momentum $-K^\mu$ out. Therefore the difference between the momenta transported out to infinity by the retarded and by the advanced fields is simply $2 K^\mu$. This is useful, because the contributions of the point-particle sources cancel in this difference.

The relationship between the momentum transported by the retarded and advanced field is reflected at the level of the Green's functions themselves. 
The difference in the Green's function takes an instructive form:%
\begin{equation}
\begin{aligned}
\tilde G_\textrm{ret}(\kb) - \tilde G_\textrm{adv}(\kb) &= 
\frac{(-1)}{(\kb^0 + i \epsilon)^2 - \v{\kb}^2} 
- \frac{(-1)}{(\kb^0 - i \epsilon)^2 - \v{\kb}^2} 
\\&=  i     \left( \Theta(\kb^0) - \Theta(-\kb^0) \right) \del(\kb^2)\,.
\end{aligned}
\end{equation}
In this equation, $\v{\kb}$ denotes the spatial components of wavenumber
four-vector $\kb$.
This difference is a homogeneous solution of the wave equation since it is supported 
on $\kb^2 = 0$. The two terms correspond to positive and negative angular frequencies. As 
we will see, the relative sign ensures that the momenta transported to infinity add.

With this in mind, we return to the problem of computing the momentum radiated and write,
\begin{equation}
2 K^\mu = \int_{\partial \Omega} d \Sigma_\nu \left(T^{\mu\nu}_\textrm{ret}(x) -T^{\mu\nu}_\textrm{adv}(x) \right)\,.
\end{equation}
In this difference, the contribution of the sources at timelike infinity cancel, so we may regard the surface $\partial \Omega$ as the boundary of spacetime. Therefore,
\begin{equation}
2K^\mu = \int \! d^4 x \, \partial_\nu \left(T^{\mu\nu}_\textrm{ret}(x) -T^{\mu\nu}_\textrm{adv}(x) \right) =- \int \! d^4 x \, \left( F^{\mu\nu}_\textrm{ret}(x) - F^{\mu\nu}_\textrm{adv}(x)\right) J_\nu(x)\,,
\end{equation}
where the last equality follows from the equations of motion. We now pass to momentum space, noting that,
\begin{equation}
F^{\mu\nu}(x) = -i \int \! \dd^4 \bar k \left( \bar k^\mu \tilde A^\nu(\bar k) - \bar k^\nu \tilde A^\mu(\bar k) \right) e^{-i \bar k \cdot x}\,.
\end{equation}
Using conservation of momentum, the radiated momentum becomes,
\begin{equation}
\begin{aligned}
2K^\mu 
&= i \int \! \dd^4 \bar k \, \bar k^\mu \left( \tilde A^\nu_\textrm{ret}(\bar k) - \tilde A^\nu_\textrm{adv}(\bar k) \right) \tilde J_\nu^*(\bar k), \\
&= -\int \! \dd^4 \bar k \, \bar k^\mu \left(\Theta(\bar k^0) - \Theta(-\bar k^0)\right) \del(\bar k^2) \tilde J^\nu(\bar k) \tilde J_\nu^*(\bar k)\,.
\label{eq:momentumMixed}
\end{aligned}
\end{equation}
The two different $\Theta$ functions arise from the outgoing and incoming radiation fields. Setting $k'^\mu = - k^\mu$ in the second term, and then dropping the prime, it is easy to see that the two terms add as anticipated. We arrive at a simple general result for the momentum radiated:
\begin{equation}
\begin{aligned}
K^\mu &= -\int \! \dd^4 \wn k \,  \Theta(\wn k^0)\del(\wn k^2) \, \wn k^\mu \, \tilde J^\nu(\wn k) \tilde J_\nu^*(\wn k) \\
&= -\int \! \df( \wn k) \, \bar k^\mu \, \tilde J^\nu(\bar k) \tilde J_\nu^*(\bar k) \,.
\label{eq:classicalMomentumRadiated}
\end{aligned}
\end{equation}
It is worth pausing to compare this general classical formula for the radiated momentum to the expression, \eqn{eq:radiatedMomentumClassical} which we derived in 
\sect{sec:classicalradiation}. Evidently the radiation kernel we defined in \eqn{eq:defOfR} is related to the classical current $\tilde J^\mu(\kb)$. This fact was 
anticipated in ref.~\cite{Luna:2017dtq}. Indeed, if we introduce a basis of polarisation vectors $\pol^{\mu\,(h)}(\kb)$ associated with the wavevector $\kb$ with helicity $h$, 
we may write the classical momentum radiated as,
\begin{equation}
K^\mu = \sum_h \int \! \df(k) \, \kb^\mu \, 
\left| \pol^{(h)} \cdot \tilde J(\kb) \right|^2\,,
\label{eq:classicalMomentumRadiated1}
\end{equation}
where here we have written the sum over helicities explicitly. In the next subsection, we will take this observation a step further by demonstrating that the leading order radiation kernel is proportional to $\pol^{(h)} \cdot \tilde J(\bar k)$ on the support of the phase space integral in \eqn{eq:classicalMomentumRadiated1}.

\subsubsection{Application at leading order}
\label{sec:classicalLOradiation}

We have established a convenient, general formula for the momentum radiated in the classical theory. It is now rather straightforward to use this expression, as well as our perturbative knowledge of the trajectories, in order to determine the momentum radiated order by order in perturbation theory. We will again simplify our classical discussion by assuming that $m_2$ is so large that we can treat particle 2 as static. Our goal in this section will be to compare the classical momentum radiated with our previous quantum calculation for the expectation of this quantity, in the limit where this expectation is very sharply peaked. As we have seen, the general expressions for the radiation have the same structure so it will be enough for us to compare the radiation kernel to the classical current.

Static particles do not radiate, so we can ignore the contribution of particle 2 while evaluating the momentum radiated. To see that this is indeed the case, note that the part of the current involving particle 2 is
\begin{equation}
J_2^\mu = e Q_2 \int \! d\tau_2 \, u_2^\mu \, \delta^{(4)}(x - u_2 \tau_2) \Rightarrow
 \tilde J_2(\kb) = e Q_2 \; u_2^\mu \; \del(\kb \cdot u_2)\,.
\end{equation}
Working in the rest frame of particle 2, the delta function in $\tilde J_2(\wn k)$ forces $\wn k^0 = 0$. But on the other hand the integral over $\wn k$ in the radiated momentum, \eqn{eq:classicalMomentumRadiated}, requires that $\kb^0 = |\v {\kb}|$. 
The only solution is $\kb^\mu = 0$, so the integral vanishes.

Therefore we may replace the current $\tilde J^\mu(\kb)$ in the radiated momentum with the contribution $\tilde J_1^\mu(\wn k)$ from particle 1. This current is,
\begin{equation}
\begin{aligned}
	\tilde J_1^\mu(\wn k) &= e Q_1 \int \! d \tau_1 \, v_1^\mu(\tau_1) e^{i \wn k \cdot \position_1 (\tau_1) } \\
	&= e Q_1 u_1^\mu \del(\bar k \cdot u_1) e^{i \wn k \cdot b} 
\\&\hphantom{=}
	+ e Q_1 \int \! d \tau_1 e^{i \wn k \cdot ( b + u_1 \tau_1)} 
	\left( \corr1 v_1^\mu(\tau_1) 
	   + i u_1^\mu \wn{k} \cdot \corr1 \position_1(\tau_1) \right)
	   +\Ord(e^5)\,.
\end{aligned}
\end{equation}
We can ignore the leading-order static part of this current just as we ignored the static current of particle 2. The dynamical part involves the perturbative correction to the trajectory of particle 1, which we computed in section~\ref{sec:EDimpulseNLO} in order to determine the NLO impulse on particle 1. 
With the help of these results, we can write down the NLO current of particle 1. Comparison to the quantum calculation is facilitated by resolving this current onto a basis of polarisation vectors and using \eqn{eq:classicalMomentumRadiated} for the momentum radiated. We find,
\begin{equation}
\begin{aligned}
	\pol^{(h)} \cdot \tilde J_1(\wn k) &\rightarrow \frac{e^3 Q_1^2 Q_2}{m_1}  \int \! \dd^4 \qb_1 \dd^4 \qb_2 \,  \del(\qb_1 \cdot u_1)  \del(\qb_2 \cdot u_2)\del^{(4)}(\wn k - \qb_1 - \qb_2) \, e^{i \qb_1 \cdot b} \frac{1}{\qb_2^2} 
\\& \hspace*{25mm} \times \biggl[ u_2 \cdot \pol^{(h)} 
	- \frac{u_1 \cdot u_2 \, \qb_2 \cdot \pol^{(h)}}{\wn k \cdot u_1} 
	-\frac{\wn k \cdot u_2 \, u_1 \cdot \pol^{(h)}}{\wn k \cdot u_1} 
\\& \hspace*{31mm} 
	+ \frac{\wn k \cdot \qb_2 \, u_1 \cdot u_2 \, u_1 \cdot \pol^{(h)}}
	       {(\wn k \cdot u_1)^2} \biggr]+\Ord(e^5)\,.
\end{aligned}
\end{equation}
We are now ready to compare this classical result to equation~\eqref{eq:RcalculationAfterWFs} for the radiation kernel we encountered in the quantum case. 
The expressions are proportional, with proportionality constant $e^3 \hbar^2$ (and an irrelevant sign). This factor is precisely provided by the definition of the radiated momentum in terms of the radiation kernel, \eqn{eq:radiatedMomentumClassicalLO}. Thus, we see once again that the quantum and classical calculations match---as they must in the classical limit.

The calculation we have described in this subsection is an Abelian version of the non-Abelian calculation of Goldberger and Ridgway~\cite{Goldberger:2016iau}, which was motivated by the double-copy relation between gauge theory and gravity. The fact that the classical current can be reproduced by taking a limit of a scattering amplitude was pointed out 
in ref.~\cite{Luna:2017dtq}. Recently, the classical computation was pushed to one higher 
order in perturbation theory by Shen~\cite{Shen:2018ebu}. It would be interesting to 
reproduce Shen's result using the methods of scattering amplitudes.

\subsection{Momentum conservation and the radiation reaction force}
\label{all:radReac}

Finally we turn to conservation of momentum in the classical theory. This is a celebrated 
problem in classical field theory, where the point particle
approximation leads to well-known issues. Problems arise because of the singular nature of 
the point particle source. In particular, the electromagnetic field at the position of a 
point charge is infinite, so to make sense of the Lorentz force acting on the particle the 
traditional route is to subtract the particle's own field from the full electromagnetic field in 
the force law. The result is a well-defined force, but conservation of momentum is lost.

Conservation of momentum is restored by including another force, the Abraham-Lorentz-Dirac (ALD) force~\cite{Lorentz,Abraham,Dirac:1938nz}, acting on the particles. This gives rise to an impulse on particle 1 in addition to the impulse due to the Lorentz force; the change in the momentum due to the ALD force balances the momentum lost to radiation. The ALD impulse is, 
\begin{equation}
\Delta p^\mu_1 = \frac{e^2 Q_1^2}{6\pi m_1}\int_{-\infty}^\infty d\tau\left(\frac{d^2p_1^\mu}{d\tau^2} + \frac{p_1^\mu}{m_1^2}\frac{dp_1}{d\tau}\cdot\frac{dp_1}{d\tau}\right).
\label{eq:ALDclass}
\end{equation}
To keep our discussion as simple as possible, we again take particle 2 to be
static in this section.

Working in perturbation theory, the lowest order contribution to $dp_1 / d\tau$ is of order 
$e^2$, due to the LO Lorentz force. Therefore $\Delta p^\mu_1$ is at least of order $e^4$. 
However, this potential contribution to the ALD impulse vanishes. To see this, recall that 
the LO force on particle 1 is given by~\eqn{eq:LOforce}. The acceleration due to this 
leading order Lorentz force in turn gives rise to an ALD impulse of,
\begin{equation}
\Delta p^\mu_1 = \frac{e^4 Q_1^3 Q_2}{6\pi m_1}\int\!\dd^4\qb\, \del(\qb\cdot u_1) \del(\qb\cdot u_2) \, e^{-i\wn{q}\cdot b} \, \qb\cdot u_1 \, \frac{\qb^\mu \, u_1\cdot u_2 - u_2^\mu \, \qb\cdot u_1}{\qb^2} = 0\,.
\end{equation}
An alternative point of view on the same result is to perform the time integral in equation~\eqref{eq:ALDclass}, noting that the second term in the ALD force is higher order. The impulse is then proportional to $f^\mu(+\infty) - f^\mu(-\infty)$, the difference in the asymptotic Lorentz forces on particle 1. But at asymptotically large times the two particles are infinitely far away, so the Lorentz forces must vanish. Since this second argument does not rely on perturbation theory we may ignore the first term in the ALD force law in the remainder of the section.

Thus, the first non-vanishing impulse due to radiation reaction is of order $e^6$. Since we only need the leading order Lorentz force,~\eqn{eq:LOforce}, to evaluate the ALD impulse, we can anticipate that the result will be very simple.
Indeed, integrating the ALD force, we find that the impulse on particle 1 due to radiation reaction is
\begin{multline}
\Delta {p^\mu_1} = \frac{e^6 Q_1^4 Q_2^2}{6\pi m_1^2} u_1^\mu \int\! \dd^4\qb\,\dd^4\qb'\, \del(\qb\cdot u_2) \del(\qb'\cdot u_2)  \del(u_1\cdot(\qb-\qb')) \, e^{-ib\cdot(\qb-\qb')} \frac{1}{\qb^2 \qb'^2} \\ \times\left[(\qb\cdot u_1)^2 + \qb\cdot\qb'(u_1\cdot u_2)^2 \right]\,.
\label{eq:classicalRadiationImpulse}
\end{multline}
This is precisely the expression~\eqref{eq:rrResult} we found using our quantum mechanical approach in~\sect{sec:ALD}. In that case, the simple final result arose after integrating the square of a more complicated five-point tree amplitude over the phase space of the photon.

\section{Discussion and conclusions}
\label{sec:discussion}

In order to apply on-shell scattering amplitudes to the calculation of classically
observable quantities, one needs a definition of the observables in the
quantum theory.  One also needs a set of rules and a path for taking the classical
limit of the quantum observables.  Ideally the rules will be straightforward,
simpler to apply than the full quantum calculation, and the path will lead gently
downhill.

In this article, we have shown how to construct suitable expressions for two
example observables, the momentum transfer or impulse~(\ref{eq:defl1})
on a particle and the momentum emitted as radiation~(\ref{eq:radiationTform})
during the scattering of two spinless
point particles.  We have shown how to restore $\hbar$s and classify
momenta in \sect{RestoringHBar}; in \sect{sec:PointParticleLimit}, how to choose suitable wavefunctions for the 
incoming particles and under what conditions the classical limit is simple.  
In \sect{sec:PointParticleLimit}, we further gave simplified 
leading- and next-to-leading-order expressions 
in terms of on-shell scattering amplitudes for
the impulse in \eqns{eq:impulseGeneralTerm1classicalLO}{eq:classicalLimitNLO}, 
and for the radiated momentum in \eqn{eq:radiatedMomentumClassical}.  These expressions
apply directly to both electrodynamics and gravity.
In \sect{sec:examples}, we used explicit expressions for the amplitudes in quantum electrodynamics
to obtain results in the classical theory. We have been careful throughout to
ensure that our methods correctly incorporate conservation of momentum,
without the need to introduce an analogue of the Abrahams--Lorentz--Dirac 
radiation reaction.

Other observables should be readily accessible by similar derivations: the same
two observables, but for the scattering of spinning particles; the change in spin
during scattering; the polarization of emitted radiation; the radiation flux as
a function of spherical angle; and more.
Higher-order corrections, to the extent they are unambiguously defined in the
classical theory, require the harder work of computing two- and higher-loop
amplitudes, but the formalism of this article will continue to apply.

Our setup has features in common with two related, but somewhat separate, areas
of current interest. One area is the study of the potential between two massive
bodies~\cite{Donoghue:1993eb,Donoghue:1994dn,Donoghue:2001qc,BjerrumBohr:2002ks,%
BjerrumBohr:2002kt,Neill:2013wsa,Bjerrum-Bohr:2013bxa,Bjerrum-Bohr:2014lea,%
Bjerrum-Bohr:2014zsa,Bjerrum-Bohr:2016hpa,Cachazo:2017jef,Guevara:2017csg,%
Bjerrum-Bohr:2018xdl}. The second is the study of particle scattering in the eikonal 
regime~\cite{Eikonal1,Amati:1990xe,%
Eikonal2,DAppollonio:2010krb,%
Eikonal3,Luna:2016idw,Collado:2018isu}. 
Diagrammatically, the study of the potential is evidently
closely related to the impulse of the present article. To some extent this is by design:
we wished to construct an on-shell observable related to the potential. But we have also
been able to construct an additional observable, the radiated momentum, which is related
to the gravitational flux.

It is interesting that classical physics emerges in the study of the high-energy limit of quantum 
scattering~\cite{Eikonal1}, 
see also ref.~\cite{Damour:2016gwp,Damour:2017zjx}. 
Indeed the classical center-of-momentum scattering angle can be obtained from the eikonal function
(see, for example ref.~\cite{DAppollonio:2010krb}). 
This latter function must therefore be related as well to the impulse, 
even though we have not taken any high-energy limit.
Indeed, the impulse and the scattering angle are equivalent at LO and NLO, 
because no momentum is radiated at these orders. Therefore
the scattering angle completely determines the
change in momentum of the particles (and vice versa). 
The connection to the eikonal function should be interesting to explore.

At NNLO, on the other hand, the equivalence between the angle and the impulse fails. This is
because of radiation: knowledge of the angle tells you where the particles went,
but not how fast. In this respect the impulse is more informative than the angle.
Eikonal methods are still applicable in the radiative case~\cite{Amati:1990xe},
so they should reproduce the high-energy limit of the expectation value of the radiated momentum. 
Meanwhile at low energies, methods based on soft theorems could provide a bridge between the impulse
and the radiated momentum~\cite{Laddha:2018rle}.

The NLO scattering angle is, in fact, somewhat simpler 
than the impulse: see ref.~\cite{Luna:2016idw} for example. 
Thanks to the exponentiation at play
in the eikonal limit, it is the triangle diagram which is responsible for the NLO correction. But the impulse contains
additional contributions, as we discussed in \sect{sec:nloQimpulse}.
Perhaps this is because the impulse must satisfy an on-shell constraint, unlike the angle. 

We restricted attention to spinless scattering in this article. In this context,
the impulse (or equivalently, the angle)
is the only physical observable at LO and NLO, and completely determines the interaction
Hamiltonian between the two particles~\cite{Damour:2016gwp,Damour:2017zjx}. The situation is richer in the case of arbitrarily aligned
spins: then the change in spins of the particles is an observable which is not 
determined by the scattering angle. We expect that this observable can also be extracted
from scattering amplitudes using our methods.

As in any application of traditional scattering amplitudes, however,
time-dependent phenomena are not readily accessible. This reflects the fact that 
amplitudes are the matrix elements of a time evolution operator from the far past to 
the far future. For a direct application of our
methods to the time-dependent gravitational waveform, we must overcome this limitation. 
One possible path
for future investigation would start from the fact that the two observables we have
discussed are essentially
expectation values. They are therefore most naturally discussed using the time-dependent in-in formalism, which has a well-known
Schwinger--Keldysh diagrammatic formulation. Whether the double copy applies in this 
context, offering an avenue to simpler calculations in gravity, remains to be
explored.

An advantage of our methods is that they naturally incorporate the actual radiated 
flux, which is obviously a key physical quantity for gravitational wave observatories. 
The formalism presented in this paper opens the door to applying the many tools and
recent advances in the study of scattering amplitudes to computing a variety of
observables at higher orders in massless classical field theories.  The application
of one of these insights, the double copy, to observables in gravity should prove
particularly fruitful.

\acknowledgments
We thank Richard Ball, Arjun Berera, Zvi Bern, Lucile Cangemi, John Joseph Carrasco, Clifford Cheung, Maria Derda, Einan Gardi, 
Grisha Korchemsky, Andr\'es Luna, 
Rafael Porto, Justin Vines, and Chris White for helpful discussions and comments.
DAK also thanks the Galileo Galilei Institute 
where this work was completed during the 2018
\textsl{Amplitudes in the LHC era\/} program.
 This work was supported in part by the French Agence Nationale pour la
 Recherche, under grant ANR--17--CE31--0001--01. BM is supported by STFC studentship ST/R504737/1. DOC is an IPPP associate, and thanks the IPPP for on-going support as well as for hospitality during this work. He is supported in part by the Marie Curie FP7 grant 631370 and by the STFC consolidated grant ÒParticle Physics at the Higgs CentreÓ.

\appendix
\section{Conventions}
\label{app:conventions}

We have chosen a mostly-minus signature metric. To deal with a proliferation of factors of $2 \pi$ appearing in measures, we use a short-hand notation first defined in equations~\eqref{delDefinition} and~\eqref{eq:ddxDefinition}
\begin{align}
\del^{(n)}(u) &\equiv (2\pi)^n \delta^{(n)}(u) = \int \! d^n x \, e^{i u \cdot x} \, ,\\
\dd^n q &\equiv \frac{d^nq}{(2\pi)^n} \, .
\end{align}
The measure for integrals over Lorentz-invariant phase space, \eqn{eq:dfDefinition}, is
\begin{equation}
\df(k_i) \equiv \dd^4 k_i\delp(k_i^2-m_i^2)\,.
\end{equation}

We occasionally find it convenient to separate a Lorentz vector $x^\mu$ into its time component $x^0$ and its three spatial components $\v x$ so that $x^\mu = (x^0, x^i) = (x^0, \v x)$ (where $i = 1, 2, 3$.)

Our convention for Fourier transforms, \eqn{eq:FourierTransform}, is
\begin{align}
f(x) &= \int \! \dd^4 \qb \, \tilde f(\qb) \, e^{-i \qb \cdot x}, \\
\tilde f(\qb) &= \int \! d^4 x \, f(x) \, e^{i \qb \cdot x}.
\end{align}

\section{Linear Wavefunction Integrals}
\label{app:wavefunctions}

In order to compute the wavefunction normalization in \eqn{LinearExponential},
we must compute the following integral,
\begin{equation}
m^{-2}\int \df(p)\;\exp\biggl[-\frac{2 p\cdot u}{m\xi}\biggr]\,.
\end{equation}
Let us introduce the following parametrization for the on-shell phase space,
\begin{equation}
p^\mu = E\bigl(\cosh\zeta,\,\sinh\zeta\sin\theta\cos\phi,\,\sinh\zeta\sin\theta\sin\phi,
               \sinh\zeta\cos\theta\bigr)\,,
\label{Parametrization}               
\end{equation}
so that,
\begin{equation}
\begin{aligned}
\df(p) &= (2\pi)^{-3}\dd E d\zeta d\Omega_2\,\del(E^2-m^2)\Theta(E)\,
  E^3 \sinh^2\zeta
  \\&=
(2\pi)^{-3}\dd E d\zeta d\theta d\phi \,\del(E^2-m^2)\Theta(E)\,
  E^3 \sinh^2\zeta\sin\theta\,,
\end{aligned}
\end{equation}
with $\zeta$ running over $[0,\infty]$, $\theta$ over $[0,\pi]$,
and $\phi$ over $[0,2\pi]$.
Performing the $E$ integration, we obtain,
\begin{equation}
\df(p)  \rightarrow \frac{m^2}{2(2\pi)^{3}} 
d\zeta d\theta d\phi\,\sinh^2\!\zeta\sin\theta\,,
\label{Measure}
\end{equation}
along with $E=m$ in the integrand.  The integral must be a Lorentz-invariant
function of $u$; as the only available Lorentz invariant is $u^2=1$, we conclude
that the result must be a function of $\xi$ alone.  We can compute it in the rest
frame of $u$, where our desired integral is,
\begin{equation}
\frac1{2(2\pi)^3} \int_0^\infty d\zeta\;\sinh^2\!\zeta\,\int_0^\pi d\theta\;\sin\theta\,
\int_0^{2\pi} d\phi\; \exp\biggl[-2\frac{\cosh\zeta}{\xi}\biggr]
=\frac1{2(2\pi)^2} \xi K_1(2/\xi)\,,
\end{equation}
where $K_1$ is a modified Bessel function of the second kind.  The normalization
condition~(\ref{WavefunctionNormalization}) then yields,
\begin{equation}
\frac{2\sqrt2\pi}{\xi^{1/2} K_1^{1/2}(2/\xi)}\,,
\end{equation}
for the wavefunction's normalization.

Next, we compute $\langle p^\mu\rangle$.  Lorentz invariance implies that the expectation
value must be proportional to $u^\mu$;
again computing in the rest frame, we find that,
\begin{equation}
\langle p^\mu\rangle = m u^\mu \,\frac{K_2(2/\xi)}{K_1(2/\xi)}\,.
\end{equation}
The phase-space measure fixes $\langle p^2\rangle = m^2$, so we conclude that,
\begin{equation}
\begin{aligned}
\frac{\langle(\Delta p)^2\rangle}{\langle p^2\rangle} &=
1-\frac{\langle p\rangle^2}{\langle p^2\rangle} 
\\&= 1-\frac{K_2^2(2/\xi)}{K_1^2(2/\xi)}
\\&=-\frac32\xi+\Ord(\xi^2)\,.
\end{aligned}
\end{equation}

We must further evaluate an integral with an on-shell delta function,
\begin{equation}
T(\qb) = \frac{\Norm^2}{\hbar m^3}\exp\biggl[-\frac{\hbar \qb\cdot u}{m\xi}\biggr]\int \df(p)\;
\del(2 p\cdot\qb/m+\hbar\qb^2/m)\,\exp\biggl[-\frac{2 p\cdot u}{m\xi}\biggr]\,.
\end{equation}
This integral is dimensionless, and can depend only on two Lorentz invariants,
$\qb\cdot u$ and $\qb^2$, along with $\xi$.  
It is convenient to write it as a function of two
dimensionless variables built out of these invariants,
\begin{equation}
\begin{aligned}
\omega &\equiv \frac{\qb\cdot u}{\sqrt{-\qb^2}}\,,
\\ \tau &\equiv \frac{\hbar\sqrt{-\qb^2}}{2m}\,.
\end{aligned}
\end{equation}
We again work in the rest frame of $u^\mu$, and without loss of generality,
choose the $z$-axis of the $p$ integration to lie along the direction of
$\v{\qb}$.  The only components that appear in the integral are then
$\qb^0$ and $\qb^z$; after integration, we can obtain the dependence on
$\tau$ and $\omega$ via the replacements,
\begin{equation}
\begin{aligned}
\qb^0 &\rightarrow \frac{2 m\omega\tau}{\hbar}\,,
\\ \qb^z &\rightarrow \frac{2m\sqrt{1+\omega^2}\tau}{\hbar}\,;
\end{aligned}
\end{equation}
and hence,
\begin{equation}
-\!\qb^2 \rightarrow \frac{4m^2\tau^2}{\hbar^2}\,.
\end{equation}
Using the measure~(\ref{Measure}), we find,
\begin{equation}
\begin{aligned}
\frac1{2\pi\,\hbar m\,\xi K_1(2/\xi)} &\exp\biggl[-\frac{\hbar \qb^0}{m\xi}\biggr]
\\&\times \int_0^\infty d\zeta\;\sinh^2\!\zeta\,\int_0^\pi d\theta\;\sin\theta\,
\int_0^{2\pi} d\phi\; \exp\biggl[-2\frac{\cosh\zeta}{\xi}\biggr]
\\& \hspace*{15mm}\times \del(2 \qb^0 \cosh\zeta - 2\qb^z \sinh\zeta \cos\theta
+\hbar\qb^2/m)\,.
\end{aligned}
\end{equation}
The $\phi$ integral is trivial, and we can use the delta function to do the
$\theta$ integral,
\begin{equation}
\begin{aligned}
\frac1{2\hbar m\,\qb^z\,\xi K_1(2/\xi)} \exp\biggl[-\frac{\hbar \qb^0}{m\xi}\biggr]
\int_0^\infty &d\zeta\;\sinh\zeta\;
\exp\biggl[-2\frac{\cosh\zeta}{\xi}\biggr]
\\& \times 
\Theta\bigl(1+ \qb^0 \coth\zeta/\qb^z +\hbar\qb^2\csch\zeta/(2 m\qb^z)\bigr)\,
\\& \times 
\Theta\bigl(1-\qb^0 \coth\zeta/\qb^z -\hbar\qb^2\csch\zeta/(2 m\qb^z)\bigr)\,.
\end{aligned}
\end{equation}
In the $\hbar\rightarrow 0$ limit, the first theta function will have no effect,
even with $\qb^2<0$.  Changing variables to $w=\cosh\zeta$, 
the second theta function will impose the constraint,
\begin{equation}
w \ge \frac{\qb^z\sqrt{1-\hbar^2\qb^2/(4m^2)}}{\sqrt{-\qb^2}} -\frac{\hbar\qb^0}{2m}\,.
\end{equation}
In terms of $\omega$ and $\tau$,
this constraint is,
\begin{equation}
w \ge \sqrt{1+\omega^2}\sqrt{1+\tau^2}-\omega\tau\,.
\end{equation}
Up to corrections of $\Ord(\hbar)$, the right-hand side is greater than 1,
and so becomes the lower limit of integration.  The result for the integral is then,
\begin{equation}
\begin{aligned}
T(\qb) &=\frac1{8 m^2\sqrt{1+\omega^2}\tau\,K_1(2/\xi)} 
\exp\biggl[-2\frac{\omega\tau}{\xi}\biggr]
 \exp\biggl[-\frac2{\xi}\Bigl(\sqrt{1+\omega^2}\sqrt{1+\tau^2}-\omega\tau\Bigr)\biggr]
\\&=\hspace{5mm}
\frac1{4 \hbar m\sqrt{(\qb\cdot u)^2-\qb^2}\,K_1(2/\xi)} 
 \exp\biggl[-\frac2{\xi}\frac{\sqrt{(\qb\cdot u)^2-\qb^2}}{\sqrt{-\qb^2}}
                               \sqrt{1-\hbar^2\qb^2/(4m^2)}\biggr]
\,.
\end{aligned}
\label{TIntegral}
\end{equation}

\def\Eqb{E_{\qb}}
How does this function behave in the $\hbar,\xi\rightarrow 0$ limit (with
$\qb$ fixed)?
The Bessel function simplifies,
\begin{equation}
\frac1{K_1(2/\xi)} \sim \frac2{\sqrt{\pi}\sqrt{\xi}} \exp\biggl[\frac2{\xi}\biggr]\,,
\label{BesselLimit}
\end{equation}
so we must take into account a modification of the exponent.
  In the limit, $\sqrt{\xi}\sim \hbar$, so
that $\hbar\sqrt{\xi}\sim \xi$, and $T$ has the form shown in \eqn{LimitForm},
\begin{equation}
\frac1{\xi}\exp\biggl[-\frac{f(\qb)}{\xi}\biggr]\,,
\end{equation}  
which will yield a delta function so long as $f(\qb)$ is positive.
  To figure out its argument, recall that
$\qb^2<0$, and use a parametrization analogous that in \eqn{Parametrization},
\begin{equation}
\qb^\mu = \Eqb\bigl(\sinh\zeta,\,\cosh\zeta\sin\theta\cos\phi,
                        \,\cosh\zeta\sin\theta\sin\phi,
                        \,\cosh\zeta\cos\theta\bigr)\,,
\end{equation}
so that again working in the rest frame of $u^\mu$,
the exponent in \eqn{TIntegral} (including the term from \eqn{BesselLimit}) is,
\begin{equation}
-\frac2{\xi}\Bigl(\cosh\zeta\sqrt{1+\hbar^2 \Eqb^2/(4m^2)}-1\Bigr)\,,
\end{equation}
so that the delta function will ultimately localize 
\begin{equation}
\cosh\zeta \rightarrow \frac1{\sqrt{1+\hbar^2 \Eqb^2/(4m^2)}} = 
1-\frac{\hbar^2 \Eqb^2}{8m^2}+\Ord(\hbar^4)\,,
\end{equation}
or translating back to Lorentz-invariant expressions, localize
\begin{equation}
\qb\cdot u + \frac{\hbar^2}{8m^2}\qb\cdot u\, \qb^2
\end{equation}
to zero.

\section{Angular integrals}
\label{app:integralappendix}

\newcommand{\littleInt}{y}
In \sect{sec:ALD} we encountered an integral over the on-shell phase space of a photon, namely \eqn{eq:phaseSpaceIntegral} which we reproduce here for ease of discussion:
\begin{equation}
\begin{aligned}
\phInt = \int \! \df (\wn k) \,& \del(u_1\cdot \bar{k} - \wn E) \, \bar{k}^\mu \, \left[ 1 + \frac{(u_1\cdot u_2)^2(\qb\cdot \qb')}{\wn E^2} + \frac{(u_2\cdot\bar{k})^2}{\wn E^2}  \right. \\ 
&\left. - \frac{(u_1\cdot u_2)(u_2\cdot\bar{k})\bar{k}\cdot(\qb+\qb')}{\wn E^3} + \frac{(u_1\cdot u_2)^2(\bar{k}\cdot\qb)(\bar{k}\cdot\qb')}{\wn E^4}  \right].
\end{aligned}
\end{equation}
In this appendix we will perform the integral over $\wn{k}$. We begin by defining
\begin{align}
\littleInt_0 &\equiv \int\! \df(\wn k) \; \del(u_1\cdot\bar{k}- \wn E), \\
\littleInt_1^\mu &\equiv \int\! \df(\wn k) \; \del(u_1\cdot\bar{k}- \wn E ) \; \wn k^\mu, \\
\littleInt_2^{\mu\nu} &\equiv \int\! \df(\wn k) \; \del(u_1\cdot\bar{k}- \wn E ) \; \wn k^\mu \wn k^\nu, \\
\littleInt_3^{\mu\nu\rho} &\equiv \int\! \df(\wn k) \; \del(u_1\cdot\bar{k}- \wn E ) \; \wn k^\mu \wn k^\nu \wn k^\rho.
\end{align}
We may then write
\begin{equation}
\begin{aligned}
\phInt =  &\left( 1 + \frac{(u_1\cdot u_2)^2(\qb\cdot \qb')}{\wn E^2} \right) \littleInt_1^\mu   \\
&\quad + \left( \frac{u_{2\nu} u_{2\rho}}{\wn E^2} - \frac{(u_1\cdot u_2) \, u_{2\nu} \, (\qb_\rho+\qb'_\rho)}{\wn E^3} + \frac{(u_1\cdot u_2)^2 \, \qb_\nu \qb'_\rho)}{\wn E^4} \right) \littleInt_3^{\mu\nu\rho}\,.
\end{aligned}
\end{equation}
Thus we need to know the integrals $\littleInt_1^\mu$ and $\littleInt_3^{\mu\nu\rho}$. It is also convenient to compute $\littleInt_0$, but we will omit the calculation of $\littleInt_2$.

We begin by computing $\littleInt_0$: as we will see, this integral appears in the evaluation of the rest. It is convenient to work in the frame where particle 1 is stationary, so $u_1^\mu = (1,\v 0)$. Then we can write
\begin{equation}
\littleInt_0 = \frac{1}{(2 \pi)^2} \int_{-\infty}^\infty d \wn k^0 \int_0^\infty d \wn \lambda \, \wn \lambda^2 \int \! d \Omega_2 \, \Theta(\wn k^0) \, \delta(\wn k^0 - \wn E) \delta((\wn k^0)^2 - \wn \lambda^2),
\end{equation}
where $\lambda = | \v k|$.
Each integral is straightforward. The integral over solid angle $d \Omega_2$ evaluates to $4\pi$; we perform the $\wn k^0$ integral using the first delta function to discover a factor $\Theta(\wn E)$, and finally we use the second delta function to perform the $\wn k$ integral. The result is
\begin{equation}
\littleInt_0 = \frac{\wn E}{2\pi} \Theta(\wn E).
\end{equation}

To evaluate $y_1^\mu$, we exploit Lorentz covariance which dictates that
\begin{equation}
\littleInt_1^\mu = \int\! \df(\wn k) \; \del(u_1\cdot\wn{k} -\wn E) \; \bar{k}^\mu = a_1 u_1^\mu\,,
\label{eq:evali1}
\end{equation}
where $a_1$ is a scalar factor to be determined. In fact it is trivial to compute $a_1$ by dotting this expression into $u_1$; then the delta function enforces $a_1=\littleInt_1\cdot u_1 = \wn{E} \, \littleInt_0$, so that 
\begin{equation}
\littleInt_1^\mu = \frac{\wn E^2}{2\pi}u_1^\mu\; \Theta(\wn E).
\end{equation}

In the same manner, Lorentz covariance and the symmetries of the integral require that,
\begin{equation}
\littleInt_3^{\mu\nu\rho} = a_3 u_1^\mu u_1^\nu u_1^\rho + b_3 \left(u_1^\mu\eta^{\nu\rho} + u_1^\nu\eta^{\mu\rho} + u_1^\rho\eta^{\mu\nu}\right),
\end{equation}
in terms of scalar factors $a_3$ and $b_3$ which we must determine. To do so, we contract $\littleInt_3^{\mu\nu\rho}$ with $u_{1\mu} u_{1\nu} u_{1\rho}$ and with $\eta_{\mu\nu} u_1^\rho$ to develop simultaneous equations for $a_3$ and $b_3$:
\begin{align}
u_{1\mu} u_{1\nu} u_{1\rho} \littleInt_3^{\mu\nu\rho} &= a_3 + 3b_3 = \wn E^3 \, \littleInt_0\,,\\
\eta_{\mu\nu} u_{1\rho} \littleInt_3^{\mu \nu\rho} &= a_3 + 6b_3 = 0\,.
\end{align}
As this system has the solution $a=\frac{\wn E^4}{\pi} \Theta(\wn E)$ and $b=-\frac{\wn E^4}{6\pi} \Theta(\wn E)$, we learn that
\begin{equation}
\littleInt_3^{\mu\nu\rho} = \frac{\wn E^4}{\pi}\left(u_1^\mu u_1^\nu u_1^\rho - \frac16 \left(u_1^\mu\eta^{\nu\rho} + u_1^\nu\eta^{\mu\rho} + u_1^\rho\eta^{\mu\nu}\right)\right) \Theta(\wn E).
\end{equation}

\section{An Alternative Classical Point of View on Momentum Radiation}
\label{app:classicalMomentumAgain}

In the main text, we provided a classical derivation of a formula for the momentum radiated in electrodynamics. The result, \eqn{eq:classicalMomentumRadiated}, was
\begin{equation}
K^\mu = -\int \! \df(\kb) \; \kb^\mu \tilde J^\nu(\kb) \tilde J_\nu^*(\kb)\,.
\end{equation}
The derivation maintained covariance, but involved a rather indirect extraction of the radiation field. In this appendix, we will derive the same formula from a somewhat simpler point of view, using textbook methods (see, for example,~\cite{schwingerBook}).

We begin as in section~\ref{sec:genClassicalRadiation} with the observation that the momentum radiated can be obtained by integrating the stress-energy tensor over the surfaces of large two-dimensional spheres, and over all time:
\begin{equation}
K^\mu = \lim_{|\v x| \rightarrow \infty} 
 \int_{-\infty}^\infty dt \int \! d\Omega_2 \, |\v x|^2 \, n_\nu T^{\mu\nu}(x)\,,
\end{equation}
where $n \cdot e_r = 1$ and $e_r$ is the unit radial outgoing vector on the spheres. To evaluate this integral, we need an expression for the gauge field $A^\mu(x)$. Working in Lorenz gauge, this can be obtained by solving the Maxwell equation using a Green's function $G(x)$:
\begin{equation}
A^\mu (x) = \int \! d^4 x' \, G(x-x') J^\mu(x')\,.
\end{equation}
In our situation, we assume that there is no incoming radiation, so the appropriate choice is the retarded Green's function $G_\mathrm{ret}(x)$ which is explicitly
\begin{equation}
G_\textrm{ret}(x) = \frac{1}{2\pi} \Theta(x^0) \delta(x^2) = \frac{1}{4\pi | \v x |} \delta(x^0 - |\v x|)\,,
\end{equation}
where $x^\mu = (t, \v x)$. As we will see, it is useful to Fourier transform $A^\mu(x)$ in the time dimension alone. We write,
\begin{equation}
\begin{aligned}
	A^\mu (\omega, \v x) &\equiv \int \, dt \, e^{i \omega t} A^\mu(t, \v x) \\
	&= \int \! d t' d^3 x' \, \frac{e^{i \omega t'} \, e^{i \omega |\v x - \v x'|}}{4\pi |\v x - \v x'|} \, J^\mu(t', \v x') \\
	&= \frac{1}{4\pi} \int \! d^3 x' \, \frac{e^{i \omega |\v x - \v x'|}}{|\v x -\v x'|} \, J^\mu(\omega, \v x')\,.
\label{eq:gaugeFieldExact}
\end{aligned}
\end{equation}
This is an exact expression for the gauge field. However, we only need to know the gauge field on the surface of spheres surrounding our interacting particles, with radii very large compared to the separation of the particles. 
So we can expand the fields at large $|\v x|$. In particular, the $\v x'$ integral in~\eqref{eq:gaugeFieldExact} extends only over the spatial support of the source $J^\mu(t', \v x')$. We assume that radii of the spheres is always very large compared to this spatial size (which in our application will be of order $b$). We can therefore assume that $|\v x'|/|\v x| \ll 1$.

Some care needs to be taken in the expansion of equation~\eqref{eq:gaugeFieldExact}, however, since another length scale appears in the problem. This is the wavelength $\lambda$ of the radiation, of order $1/\omega$. This wavelength need not be of order $b$, and in particular the quantity $\omega |\v x'| \sim |\v x'| / \lambda$ need not be small. We therefore expand the field as
\begin{equation}
	A^\mu (\omega, \v x) \simeq \frac{e^{i \omega |\v x|}}{4\pi |\v x|} \int \! d^3 x' \, e^{-i \omega \, \v{\hat x} \cdot \v x'} \, J^\mu(\omega, \v x')\,,
\label{eq:gaugeFieldStep}
\end{equation}
where $\v{\hat x} = \v x / |\v x|$ is a unit three-dimensional vector. Defining the wavevector $\wn{\v k} = \omega \v {\hat x}$, can recognise the spatial Fourier transform present in \eqn{eq:gaugeFieldStep}, and learn that
\begin{equation}
	A^\mu (\omega, \v x) \simeq \frac{e^{i \omega |\v x|}}{4\pi |\v x|}  \, 
	J^\mu(\omega, \v{\kb}) = \frac{e^{i \omega |\v x|}}{4\pi |\v x|}  \, \tilde J^\mu(\kb)\,,
\end{equation}
where $\wn{k}^\mu = (\omega, \v k)$. Note that $\wn k \cdot \wn k = 0$.
For later use, we remark that
\begin{equation}
	A^\mu(-\omega, \v x) = \frac{e^{-i \omega |\v x|}}{4\pi |\v x|}  \,
	  \tilde J^\mu(-\kb)\,.
\end{equation}
It is straightforward to obtain the field strength $F^{\mu \nu}$ from our gauge field. Neglecting terms which are subdominant at large distances, we find that
\begin{equation}
\begin{aligned}
	F^{\mu \nu}(x) &\equiv \int \frac{d\omega}{2 \pi} \, e^{-i \omega t} F^{\mu \nu}(\omega, \v x) \\
	&= -i \int \frac{d\omega}{2 \pi} \, e^{-i \omega t}  \left(\wn k^\mu A^\nu(\omega, \v x) - \wn k^\nu A^\mu(\omega, \v x) \right)\,.
\end{aligned}
\end{equation}
It is also worth noting that $\wn k_\mu A^\mu(\omega, \v x) = 0$ as a consequence of current conservation. 

We now return to the momentum radiated. We first trade the time integral of the stress-energy tensor for a frequency integral:
\begin{equation}
\begin{aligned}
	\int_{-\infty}^\infty dt  &\left(F^{\mu \alpha}(x) F_\alpha{}^\nu(x) + \frac 14 \eta^{\mu \nu} F^{\alpha \beta}(x) F_{\alpha \beta}(x) \right) \\
	&=  \int \frac{d\omega}{2 \pi} \int_{-\infty}^\infty dt \, e^{-i \omega t} \left(F^{\mu \alpha}(\omega, \v x) F_\alpha{}^\nu(t, \v x) + \frac 14 \eta^{\mu \nu} F^{\alpha \beta}(\omega, \v x) F_{\alpha \beta}(t, \v x) \right) \\
	&=  \int \frac{d\omega}{2 \pi} \left(F^{\mu \alpha}(\omega, \v x) F_\alpha{}^\nu(-\omega, \v x) + \frac 14 \eta^{\mu \nu} F^{\alpha \beta}(\omega, \v x) F_{\alpha \beta}(-\omega, \v x) \right) \\
	&= \int \frac{d\omega}{2 \pi} \, \wn{k}^\mu \wn{k}^\nu A^{\alpha}(\omega, \v x) A_\alpha(-\omega, \v x) \,.
\end{aligned}
\end{equation}
Armed with these results, it is an easy matter to complete the derivation. The momentum radiated is
\begin{equation}
\begin{aligned}
	K^\mu &= \lim_{|\v x| \rightarrow \infty} \int \frac{d\omega}{2 \pi} \int \! d\Omega_2 \, |\v x|^2 \, n_\nu \wn k^\mu \wn k^\nu A^{\alpha}(\omega, \v x) A_\alpha(-\omega, \v x) \\
	&= - \frac{1}{(2 \pi)^3} \int_{0}^\infty d\omega \, \omega^2 \int \! d\Omega_2 \, \frac{1}{2\omega} \wn k^\mu\, \tilde J^{\alpha}(\wn k) \tilde J_\alpha(-\wn k) \\
	&= - \frac{1}{(2 \pi)^3} \int \! d^3 \wn{k} d \wn k^0 \, \frac{\delta (\wn k^0 - |\v {\wn k}|)}{2 |\v {\wn k}|}  \, \wn k^\mu\, \tilde J^{\alpha}(\wn k) \tilde J_\alpha(-\wn k) \\
&= - \int \! \df(\wn k) \, \wn k^\mu \tilde J^{\alpha}(\wn k) \tilde J^*_\alpha(\wn k)\,,
\end{aligned}
\end{equation}
as expected. In this derivation, we lost manifest Lorentz invariance during the calculation, but we were able to restore it at the end because the observable of interest is Lorentz invariant.


\begin{thebibliography}{99}

\bibitem{LIGO}
B.~P.~Abbott {\it et al.} [LIGO Scientific and Virgo Collaborations],
Phys.\ Rev.\ Lett.\  {\bf 116}, no. 6, 061102 (2016)
doi:10.1103/PhysRevLett.116.061102
[arXiv:1602.03837 [gr-qc]];
Phys.\ Rev.\ Lett.\  {\bf 116}, no. 24, 241103 (2016)
doi:10.1103/PhysRevLett.116.241103
[arXiv:1606.04855 [gr-qc]];
Phys.\ Rev.\ Lett.\  {\bf 118}, no. 22, 221101 (2017)
doi:10.1103/PhysRevLett.118.221101
[arXiv:1706.01812 [gr-qc]];
Phys.\ Rev.\ Lett.\  {\bf 119}, no. 14, 141101 (2017)
doi:10.1103/PhysRevLett.119.141101
[arXiv:1709.09660 [gr-qc]];
Phys.\ Rev.\ Lett.\  {\bf 119}, no. 16, 161101 (2017)
doi:10.1103/PhysRevLett.119.161101
[arXiv:1710.05832 [gr-qc]].
    
\bibitem{Buonanno:2014aza}
  A.~Buonanno and B.~S.~Sathyaprakash,
  arXiv:1410.7832 [gr-qc].

\bibitem{ADM}
  R.~Arnowitt and S.~Deser,
  Phys.\ Rev.\  {\bf 113}, 745 (1959)
  doi:10.1103/PhysRev.113.745;
%
  R.~L.~Arnowitt, S.~Deser and C.~W.~Misner,
  Phys.\ Rev.\  {\bf 116}, 1322 (1959)
  doi:10.1103/PhysRev.116.1322;
%
  R.~L.~Arnowitt, S.~Deser and C.~W.~Misner,
  Phys.\ Rev.\  {\bf 117}, 1595 (1960)
  doi:10.1103/PhysRev.117.1595;
%
  R.~L.~Arnowitt, S.~Deser and C.~W.~Misner,
  Gen.\ Rel.\ Grav.\  {\bf 40}, 1997 (2008)
  doi:10.1007/s10714-008-0661-1
  [gr-qc/0405109].

\bibitem{Schafer:2018kuf}
  G.~Sch{\"a}fer and P.~Jaranowski,
  Living Rev.\ Rel.\  {\bf 21}, no. 1, 7 (2018)
  doi:10.1007/s41114-018-0016-5
  [arXiv:1805.07240 [gr-qc]].

\bibitem{Blanchet:2013haa}
  L.~Blanchet,
  Living Rev.\ Rel.\  {\bf 17}, 2 (2014)
  doi:10.12942/lrr-2014-2
  [arXiv:1310.1528 [gr-qc]].

\bibitem{EOB1}
  A.~Buonanno and T.~Damour,
  Phys.\ Rev.\ D {\bf 59}, 084006 (1999)
  doi:10.1103/PhysRevD.59.084006
  [gr-qc/9811091];
  A.~Buonanno and T.~Damour,
  Phys.\ Rev.\ D {\bf 62}, 064015 (2000)
  doi:10.1103/PhysRevD.62.064015
  [gr-qc/0001013].

\bibitem{EOB2}
  T.~Damour,
  Phys.\ Rev.\ D {\bf 64}, 124013 (2001)
  doi:10.1103/PhysRevD.64.124013
  [gr-qc/0103018];
%
  T.~Damour, P.~Jaranowski and G.~Schaefer,
  Phys.\ Rev.\ D {\bf 62}, 084011 (2000)
  doi:10.1103/PhysRevD.62.084011
  [gr-qc/0005034].

\bibitem{NumericalRelativity}
  F.~Pretorius,
  Phys.\ Rev.\ Lett.\  {\bf 95}, 121101 (2005)
  doi:10.1103/PhysRevLett.95.121101
  [gr-qc/0507014];
%
  F.~Pretorius,
  arXiv:0710.1338 [gr-qc].

\bibitem{EFT1}
  W.~D.~Goldberger and I.~Z.~Rothstein,
  Phys.\ Rev.\ D {\bf 73}, 104029 (2006)
  doi:10.1103/PhysRevD.73.104029
  [hep-th/0409156].

\bibitem{EFT2}
  R.~A.~Porto,
  Phys.\ Rept.\  {\bf 633}, 1 (2016)
  doi:10.1016/j.physrep.2016.04.003
  [arXiv:1601.04914 [hep-th]];
%
  M.~Levi,
  arXiv:1807.01699 [hep-th].

\bibitem{EFT3}
  W.~D.~Goldberger and I.~Z.~Rothstein,
  Phys.\ Rev.\ D {\bf 73}, 104030 (2006)
  doi:10.1103/PhysRevD.73.104030
  [hep-th/0511133];
%
  W.~D.~Goldberger and I.~Z.~Rothstein,
  Gen.\ Rel.\ Grav.\  {\bf 38}, 1537 (2006)
  [Int.\ J.\ Mod.\ Phys.\ D {\bf 15}, 2293 (2006)]
  doi:10.1007/s10714-006-0345-7, 10.1142/S0218271806009698
  [hep-th/0605238];
%
  R.~A.~Porto,
  Phys.\ Rev.\ D {\bf 73}, 104031 (2006)
  doi:10.1103/PhysRevD.73.104031
  [gr-qc/0511061];
%
  R.~A.~Porto and I.~Z.~Rothstein,
  Phys.\ Rev.\ Lett.\  {\bf 97}, 021101 (2006)
  doi:10.1103/PhysRevLett.97.021101
  [gr-qc/0604099];
%
  R.~A.~Porto,
  Phys.\ Rev.\ D {\bf 77}, 064026 (2008)
  doi:10.1103/PhysRevD.77.064026
  [arXiv:0710.5150 [hep-th]];
%
  R.~A.~Porto and I.~Z.~Rothstein,
  arXiv:0712.2032 [gr-qc];
%
  B.~Kol and M.~Smolkin,
  Class.\ Quant.\ Grav.\  {\bf 25}, 145011 (2008)
  doi:10.1088/0264-9381/25/14/145011
  [arXiv:0712.4116 [hep-th]];
%
  C.~R.~Galley and B.~L.~Hu,
  Phys.\ Rev.\ D {\bf 79}, 064002 (2009)
  doi:10.1103/PhysRevD.79.064002
  [arXiv:0801.0900 [gr-qc]];
%
  R.~A.~Porto and I.~Z.~Rothstein,
  Phys.\ Rev.\ D {\bf 78}, 044012 (2008)
  Erratum: [Phys.\ Rev.\ D {\bf 81}, 029904 (2010)]
  doi:10.1103/PhysRevD.78.044012, 10.1103/PhysRevD.81.029904
  [arXiv:0802.0720 [gr-qc]];
%
  M.~Levi,
  Phys.\ Rev.\ D {\bf 82}, 064029 (2010)
  doi:10.1103/PhysRevD.82.064029
  [arXiv:0802.1508 [gr-qc]].
%
  B.~Kol,
  Gen.\ Rel.\ Grav.\  {\bf 40}, 2061 (2008)
  [Int.\ J.\ Mod.\ Phys.\ D {\bf 17}, 2617 (2009)]
  doi:10.1007/s10714-008-0673-x, 10.1142/S0218271808014151
  [arXiv:0804.0187 [hep-th]].

\bibitem{EFT4}
  R.~A.~Porto and I.~Z.~Rothstein,
  Phys.\ Rev.\ D {\bf 78}, 044013 (2008)
  Erratum: [Phys.\ Rev.\ D {\bf 81}, 029905 (2010)]
  doi:10.1103/PhysRevD.81.029905, 10.1103/PhysRevD.78.044013
  [arXiv:0804.0260 [gr-qc]];
%
  J.~B.~Gilmore and A.~Ross,
  Phys.\ Rev.\ D {\bf 78}, 124021 (2008)
  doi:10.1103/PhysRevD.78.124021
  [arXiv:0810.1328 [gr-qc]];
%
  Y.~Z.~Chu,
  Phys.\ Rev.\ D {\bf 79}, 044031 (2009)
  doi:10.1103/PhysRevD.79.044031
  [arXiv:0812.0012 [gr-qc]];
%
  W.~D.~Goldberger and A.~Ross,
  Phys.\ Rev.\ D {\bf 81}, 124015 (2010)
  doi:10.1103/PhysRevD.81.124015
  [arXiv:0912.4254 [gr-qc]];
%
  C.~R.~Galley and M.~Tiglio,
  Phys.\ Rev.\ D {\bf 79}, 124027 (2009)
  doi:10.1103/PhysRevD.79.124027
  [arXiv:0903.1122 [gr-qc]];
%
  B.~Kol and M.~Smolkin,
  Phys.\ Rev.\ D {\bf 80}, 124044 (2009)
  doi:10.1103/PhysRevD.80.124044
  [arXiv:0910.5222 [hep-th]];
%
  R.~A.~Porto,
  Class.\ Quant.\ Grav.\  {\bf 27}, 205001 (2010)
  doi:10.1088/0264-9381/27/20/205001
  [arXiv:1005.5730 [gr-qc]];
%
  M.~Levi,
  Phys.\ Rev.\ D {\bf 82}, 104004 (2010)
  doi:10.1103/PhysRevD.82.104004
  [arXiv:1006.4139 [gr-qc]];
%
  R.~A.~Porto, A.~Ross and I.~Z.~Rothstein,
  JCAP {\bf 1103}, 009 (2011)
  doi:10.1088/1475-7516/2011/03/009
  [arXiv:1007.1312 [gr-qc]];
%
  B.~Kol, M.~Levi and M.~Smolkin,
  Class.\ Quant.\ Grav.\  {\bf 28}, 145021 (2011)
  doi:10.1088/0264-9381/28/14/145021
  [arXiv:1011.6024 [gr-qc]];
%
  S.~Foffa and R.~Sturani,
  Phys.\ Rev.\ D {\bf 84}, 044031 (2011)
  doi:10.1103/PhysRevD.84.044031
  [arXiv:1104.1122 [gr-qc]];
%
  M.~Levi,
  Phys.\ Rev.\ D {\bf 85}, 064043 (2012)
  doi:10.1103/PhysRevD.85.064043
  [arXiv:1107.4322 [gr-qc]].

\bibitem{EFT5}
  S.~Hergt, J.~Steinhoff and G.~Schaefer,
  Annals Phys.\  {\bf 327}, 1494 (2012)
  doi:10.1016/j.aop.2012.02.006
  [arXiv:1110.2094 [gr-qc]];
%
  S.~Foffa and R.~Sturani,
  Phys.\ Rev.\ D {\bf 87}, no. 4, 044056 (2013)
  doi:10.1103/PhysRevD.87.044056
  [arXiv:1111.5488 [gr-qc]];
%
  A.~Ross,
  Phys.\ Rev.\ D {\bf 85}, 125033 (2012)
  doi:10.1103/PhysRevD.85.125033
  [arXiv:1202.4750 [gr-qc]];
%
  R.~A.~Porto, A.~Ross and I.~Z.~Rothstein,
  JCAP {\bf 1209}, 028 (2012)
  doi:10.1088/1475-7516/2012/09/028
  [arXiv:1203.2962 [gr-qc]];
%
  C.~R.~Galley and A.~K.~Leibovich,
  Phys.\ Rev.\ D {\bf 86}, 044029 (2012)
  doi:10.1103/PhysRevD.86.044029
  [arXiv:1205.3842 [gr-qc]];
%
  W.~D.~Goldberger, A.~Ross and I.~Z.~Rothstein,
  Phys.\ Rev.\ D {\bf 89}, no. 12, 124033 (2014)
  doi:10.1103/PhysRevD.89.124033
  [arXiv:1211.6095 [hep-th]];
%
  C.~R.~Galley and R.~A.~Porto,
  JHEP {\bf 1311}, 096 (2013)
  doi:10.1007/JHEP11(2013)096
  [arXiv:1302.4486 [gr-qc]];
%
  O.~Birnholtz, S.~Hadar and B.~Kol,
  Phys.\ Rev.\ D {\bf 88}, no. 10, 104037 (2013)
  doi:10.1103/PhysRevD.88.104037
  [arXiv:1305.6930 [hep-th]];
%
  S.~Chakrabarti, T.~Delsate and J.~Steinhoff,
  Phys.\ Rev.\ D {\bf 88}, 084038 (2013)
  doi:10.1103/PhysRevD.88.084038
  [arXiv:1306.5820 [gr-qc]];
%
  S.~Foffa and R.~Sturani,
  Class.\ Quant.\ Grav.\  {\bf 31}, no. 4, 043001 (2014)
  doi:10.1088/0264-9381/31/4/043001
  [arXiv:1309.3474 [gr-qc]];
%
  S.~Foffa,
  Phys.\ Rev.\ D {\bf 89}, no. 2, 024019 (2014)
  doi:10.1103/PhysRevD.89.024019
  [arXiv:1309.3956 [gr-qc]];
%
  M.~Levi and J.~Steinhoff,
  JCAP {\bf 1412}, no. 12, 003 (2014)
  doi:10.1088/1475-7516/2014/12/003
  [arXiv:1408.5762 [gr-qc]].

\bibitem{EFT6}
  M.~Levi and J.~Steinhoff,
  JHEP {\bf 1506}, 059 (2015)
  doi:10.1007/JHEP06(2015)059
  [arXiv:1410.2601 [gr-qc]];
%
  M.~Levi and J.~Steinhoff,
  JHEP {\bf 1509}, 219 (2015)
  doi:10.1007/JHEP09(2015)219
  [arXiv:1501.04956 [gr-qc]];
%
  M.~Levi and J.~Steinhoff,
  JCAP {\bf 1601}, 011 (2016)
  doi:10.1088/1475-7516/2016/01/011
  [arXiv:1506.05056 [gr-qc]];
%
  M.~Levi and J.~Steinhoff,
  JCAP {\bf 1601}, 008 (2016)
  doi:10.1088/1475-7516/2016/01/008
  [arXiv:1506.05794 [gr-qc]];
%
  C.~R.~Galley, A.~K.~Leibovich, R.~A.~Porto and A.~Ross,
  Phys.\ Rev.\ D {\bf 93}, 124010 (2016)
  doi:10.1103/PhysRevD.93.124010
  [arXiv:1511.07379 [gr-qc]];
%
  M.~Levi and J.~Steinhoff,
  arXiv:1607.04252 [gr-qc];
%
  S.~Foffa, P.~Mastrolia, R.~Sturani and C.~Sturm,
  Phys.\ Rev.\ D {\bf 95}, no. 10, 104009 (2017)
  doi:10.1103/PhysRevD.95.104009
  [arXiv:1612.00482 [gr-qc]];
%
  R.~A.~Porto and I.~Z.~Rothstein,
  Phys.\ Rev.\ D {\bf 96}, no. 2, 024062 (2017)
  doi:10.1103/PhysRevD.96.024062
  [arXiv:1703.06433 [gr-qc]];
%
  R.~A.~Porto,
  Phys.\ Rev.\ D {\bf 96}, no. 2, 024063 (2017)
  doi:10.1103/PhysRevD.96.024063
  [arXiv:1703.06434 [gr-qc]];
%
  M.~Levi and J.~Steinhoff,
  Class.\ Quant.\ Grav.\  {\bf 34}, no. 24, 244001 (2017)
  doi:10.1088/1361-6382/aa941e
  [arXiv:1705.06309 [gr-qc]];
%
  N.~T.~Maia, C.~R.~Galley, A.~K.~Leibovich and R.~A.~Porto,
  Phys.\ Rev.\ D {\bf 96}, no. 8, 084064 (2017)
  doi:10.1103/PhysRevD.96.084064
  [arXiv:1705.07934 [gr-qc]];
%
  N.~T.~Maia, C.~R.~Galley, A.~K.~Leibovich and R.~A.~Porto,
  Phys.\ Rev.\ D {\bf 96}, no. 8, 084065 (2017)
  doi:10.1103/PhysRevD.96.084065
  [arXiv:1705.07938 [gr-qc]].

\bibitem{Neill:2013wsa}
  D.~Neill and I.~Z.~Rothstein,
  Nucl.\ Phys.\ B {\bf 877}, 177 (2013)
  doi:10.1016/j.nuclphysb.2013.09.007
  [arXiv:1304.7263 [hep-th]].

\bibitem{Bjerrum-Bohr:2013bxa}
N.~E.~J.~Bjerrum-Bohr, J.~F.~Donoghue and P.~Vanhove,
JHEP {\bf 1402}, 111 (2014)
doi:10.1007/JHEP02(2014)111
[arXiv:1309.0804 [hep-th]].

\bibitem{Bjerrum-Bohr:2014lea}
N.~E.~J.~Bjerrum-Bohr, B.~R.~Holstein, L.~Plant\'e and P.~Vanhove,
Phys.\ Rev.\ D {\bf 91}, no. 6, 064008 (2015)
doi:10.1103/PhysRevD.91.064008
[arXiv:1410.4148 [gr-qc]].

\bibitem{Bjerrum-Bohr:2014zsa}
N.~E.~J.~Bjerrum-Bohr, J.~F.~Donoghue, B.~R.~Holstein, L.~Plant\'e and P.~Vanhove,
Phys.\ Rev.\ Lett.\  {\bf 114}, no. 6, 061301 (2015)
doi:10.1103/PhysRevLett.114.061301
[arXiv:1410.7590 [hep-th]].
  
\bibitem{Bjerrum-Bohr:2016hpa}
N.~E.~J.~Bjerrum-Bohr, J.~F.~Donoghue, B.~R.~Holstein, L.~Plant\'e and P.~Vanhove,
JHEP {\bf 1611}, 117 (2016)
doi:10.1007/JHEP11(2016)117
[arXiv:1609.07477 [hep-th]].
      
\bibitem{Cachazo:2017jef}
  F.~Cachazo and A.~Guevara,
  arXiv:1705.10262 [hep-th].

\bibitem{Guevara:2017csg}
  A.~Guevara,
  arXiv:1706.02314 [hep-th].
  
\bibitem{Bjerrum-Bohr:2018xdl}
N.~E.~J.~Bjerrum-Bohr, P.~H.~Damgaard, G.~Festuccia, L.~Plant\'e and P.~Vanhove,
arXiv:1806.04920 [hep-th].

\bibitem{Donoghue:1993eb}
  J.~F.~Donoghue,
  Phys.\ Rev.\ Lett.\  {\bf 72}, 2996 (1994)
  doi:10.1103/PhysRevLett.72.2996
  [gr-qc/9310024].
  
\bibitem{Donoghue:1994dn}
  J.~F.~Donoghue,
  Phys.\ Rev.\ D {\bf 50}, 3874 (1994)
  doi:10.1103/PhysRevD.50.3874
  [gr-qc/9405057].

\bibitem{Donoghue:1996mt}
  J.~F.~Donoghue and T.~Torma,
  Phys.\ Rev.\ D {\bf 54}, 4963 (1996)
  doi:10.1103/PhysRevD.54.4963
  [hep-th/9602121].
  
\bibitem{Donoghue:2001qc}
  J.~F.~Donoghue, B.~R.~Holstein, B.~Garbrecht and T.~Konstandin,
  Phys.\ Lett.\ B {\bf 529}, 132 (2002)
  Erratum: [Phys.\ Lett.\ B {\bf 612}, 311 (2005)]
  doi:10.1016/S0370-2693(02)01246-7, 10.1016/j.physletb.2005.03.018
  [hep-th/0112237].
  
\bibitem{BjerrumBohr:2002ks}
  N.~E.~J.~Bjerrum-Bohr, J.~F.~Donoghue and B.~R.~Holstein,
  Phys.\ Rev.\ D {\bf 68}, 084005 (2003)
  Erratum: [Phys.\ Rev.\ D {\bf 71}, 069904 (2005)]
  doi:10.1103/PhysRevD.68.084005, 10.1103/PhysRevD.71.069904
  [hep-th/0211071].
  
\bibitem{BjerrumBohr:2002kt}
  N.~E.~J.~Bjerrum-Bohr, J.~F.~Donoghue and B.~R.~Holstein,
  Phys.\ Rev.\ D {\bf 67}, 084033 (2003)
  Erratum: [Phys.\ Rev.\ D {\bf 71}, 069903 (2005)]
  doi:10.1103/PhysRevD.71.069903, 10.1103/PhysRevD.67.084033
  [hep-th/0211072].
 
\bibitem{Holstein:2004dn}
  B.~R.~Holstein and J.~F.~Donoghue,
  Phys.\ Rev.\ Lett.\  {\bf 93}, 201602 (2004)
  doi:10.1103/PhysRevLett.93.201602
  [hep-th/0405239].
  
\bibitem{Damour:2016gwp}
  T.~Damour,
  Phys.\ Rev.\ D {\bf 94}, no. 10, 104015 (2016)
  doi:10.1103/PhysRevD.94.104015
  [arXiv:1609.00354 [gr-qc]].

\bibitem{Damour:2017zjx}
  T.~Damour,
  Phys.\ Rev.\ D {\bf 97}, no. 4, 044038 (2018)
  doi:10.1103/PhysRevD.97.044038
  [arXiv:1710.10599 [gr-qc]].

\bibitem{Kawai:1985xq}
  H.~Kawai, D.~C.~Lewellen and S.~H.~H.~Tye,
  Nucl.\ Phys.\ B {\bf 269}, 1 (1986)
  doi:10.1016/0550-3213(86)90362-7.
 
\bibitem{DoubleCopy}
  Z.~Bern, J.~J.~M.~Carrasco and H.~Johansson,
  Phys.\ Rev.\ D {\bf 78}, 085011 (2008)
  doi:10.1103/PhysRevD.78.085011
  [arXiv:0805.3993 [hep-ph]]; 
  Z.~Bern, J.~J.~M.~Carrasco and H.~Johansson,
  Phys.\ Rev.\ Lett.\  {\bf 105}, 061602 (2010)
  doi:10.1103/PhysRevLett.105.061602
  [arXiv:1004.0476 [hep-th]].

\bibitem{Monteiro:2014cda}
  R.~Monteiro, D.~O'Connell and C.~D.~White,
  JHEP {\bf 1412}, 056 (2014)
  doi:10.1007/JHEP12(2014)056
  [arXiv:1410.0239 [hep-th]].
  
\bibitem{Luna:2016due}
  A.~Luna, R.~Monteiro, I.~Nicholson, D.~O'Connell and C.~D.~White,
  JHEP {\bf 1606}, 023 (2016)
  doi:10.1007/JHEP06(2016)023
  [arXiv:1603.05737 [hep-th]].

\bibitem{Goldberger:2016iau}
W.~D.~Goldberger and A.~K.~Ridgway,
Phys.\ Rev.\ D {\bf 95} (2017) 125010
doi:10.1103/PhysRevD.95.125010
[arXiv:1611.03493 [hep-th]].
  
\bibitem{Luna:2016hge}
  A.~Luna, R.~Monteiro, I.~Nicholson, A.~Ochirov, D.~O'Connell, N.~Westerberg and C.~D.~White,
  JHEP {\bf 1704}, 069 (2017)
  doi:10.1007/JHEP04(2017)069
  [arXiv:1611.07508 [hep-th]].

\bibitem{Luna:2018dpt}
  A.~Luna, R.~Monteiro, I.~Nicholson and D.~O'Connell,
  arXiv:1810.08183 [hep-th].

\bibitem{Goldberger:2017frp}
  W.~D.~Goldberger, S.~G.~Prabhu and J.~O.~Thompson,
  Phys.\ Rev.\ D {\bf 96}, no. 6, 065009 (2017)
  doi:10.1103/PhysRevD.96.065009
  [arXiv:1705.09263 [hep-th]].
  
\bibitem{Shen:2018ebu}
  C.~H.~Shen,
  arXiv:1806.07388 [hep-th].

\bibitem{Goldberger:2017vcg}
  W.~D.~Goldberger and A.~K.~Ridgway,
  Phys.\ Rev.\ D {\bf 97}, no. 8, 085019 (2018)
  doi:10.1103/PhysRevD.97.085019
  [arXiv:1711.09493 [hep-th]].

\bibitem{Plefka:2018dpa}
  J.~Plefka, J.~Steinhoff and W.~Wormsbecher,
  arXiv:1807.09859 [hep-th].

\bibitem{Goldberger:2017ogt}
  W.~D.~Goldberger, J.~Li and S.~G.~Prabhu,
  Phys.\ Rev.\ D {\bf 97}, no. 10, 105018 (2018)
  doi:10.1103/PhysRevD.97.105018
  [arXiv:1712.09250 [hep-th]].

\bibitem{Li:2018qap}
  J.~Li and S.~G.~Prabhu,
  Phys.\ Rev.\ D {\bf 97}, no. 10, 105019 (2018)
  doi:10.1103/PhysRevD.97.105019
  [arXiv:1803.02405 [hep-th]].

\bibitem{Holstein:2008sx}
  B.~R.~Holstein and A.~Ross,
  arXiv:0802.0716 [hep-ph].

\bibitem{Cheung:2018wkq}
  C.~Cheung, I.~Z.~Rothstein and M.~P.~Solon,
  arXiv:1808.02489 [hep-th].

\bibitem{Laddha:2018rle}
  A.~Laddha and A.~Sen,
  JHEP {\bf 1809}, 105 (2018)
  doi:10.1007/JHEP09(2018)105
  [arXiv:1801.07719 [hep-th]].

\bibitem{Lorentz}
H.~A.~Lorentz, 
Arch.\ N\'eerl.\ Sci.\ Exactes Nat.\ 25, 363 (1892).

\bibitem{Abraham}
M.~Abraham, 
Ann.\ d.\ Phys.\ 10, 105 (1903);
``Theorie der Elektrizit{\"a}t'', vol.~II (Teubner, Leipzig, 1905);
Ann.\ d.\ Phys.\ 14, 236 (1904).

\bibitem{Dirac:1938nz}
  P.~A.~M.~Dirac,
  Proc.\ Roy.\ Soc.\ Lond.\ A {\bf 167}, 148 (1938)
  doi:10.1098/rspa.1938.0124.

\bibitem{LandauLifshitz}
L.~D.~Landau and E.~M.~Lifshitz,
\textit{The Classical Theory of Fields\/}, 4${}^{\textrm{th}}$ ed.,
ISBN 978-0750627689 (Butterworth-Heinemann, Oxford, 1975).

\bibitem{Higuchi:2002qc}
  A.~Higuchi,
  Phys.\ Rev.\ D {\bf 66}, 105004 (2002)
  Erratum: [Phys.\ Rev.\ D {\bf 69}, 129903 (2004)]
  doi:10.1103/PhysRevD.66.105004, 10.1103/PhysRevD.69.129903
  [quant-ph/0208017].
 
\bibitem{Galley:2006gs}
  C.~R.~Galley, B.~L.~Hu and S.~Y.~Lin,
  Phys.\ Rev.\ D {\bf 74}, 024017 (2006)
  doi:10.1103/PhysRevD.74.024017
  [gr-qc/0603099].

\bibitem{Galley:2010es}
  C.~R.~Galley, A.~K.~Leibovich and I.~Z.~Rothstein,
  Phys.\ Rev.\ Lett.\  {\bf 105}, 094802 (2010)
  doi:10.1103/PhysRevLett.105.094802
  [arXiv:1005.2617 [gr-qc]].

\bibitem{Birnholtz:2013nta}
  O.~Birnholtz, S.~Hadar and B.~Kol,
  Phys.\ Rev.\ D {\bf 88}, no. 10, 104037 (2013)
  doi:10.1103/PhysRevD.88.104037
  [arXiv:1305.6930 [hep-th]].
  
\bibitem{Birnholtz:2014fwa}
  O.~Birnholtz, S.~Hadar and B.~Kol,
  Int.\ J.\ Mod.\ Phys.\ A {\bf 29}, no. 24, 1450132 (2014)
  doi:10.1142/S0217751X14501322
  [arXiv:1402.2610 [hep-th]].

\bibitem{Birnholtz:2014gna}
  O.~Birnholtz,
  Int.\ J.\ Mod.\ Phys.\ A {\bf 30}, no. 02, 1550011 (2015)
  doi:10.1142/S0217751X15500116
  [arXiv:1410.5871 [physics.class-ph]].

\bibitem{RelativisticWavefunctions}
  M.~H.~Al-Hashimi and U.-J.~Wiese,
  Annals Phys.\  {\bf 324}, 2599 (2009)
  doi:10.1016/j.aop.2009.09.001
  [arXiv:0907.5178 [quant-ph]].

\bibitem{Eikonal1}
  D.~Amati, M.~Ciafaloni and G.~Veneziano,
  Phys.\ Lett.\ B {\bf 197}, 81 (1987)
  doi:10.1016/0370-2693(87)90346-7;
%
  G.~'t Hooft,
  Phys.\ Lett.\ B {\bf 198}, 61 (1987)
  doi:10.1016/0370-2693(87)90159-6;
%
  I.~J.~Muzinich and M.~Soldate,
  Phys.\ Rev.\ D {\bf 37}, 359 (1988)
  doi:10.1103/PhysRevD.37.359;
%
  D.~Amati, M.~Ciafaloni and G.~Veneziano,
  Int.\ J.\ Mod.\ Phys.\ A {\bf 3}, 1615 (1988)
  doi:10.1142/S0217751X88000710.

\bibitem{Amati:1990xe}
  D.~Amati, M.~Ciafaloni and G.~Veneziano,
  Nucl.\ Phys.\ B {\bf 347}, 550 (1990)
  doi:10.1016/0550-3213(90)90375-N.

\bibitem{Eikonal2}
  D.~Amati, M.~Ciafaloni and G.~Veneziano,
  Phys.\ Lett.\ B {\bf 289}, 87 (1992)
  doi:10.1016/0370-2693(92)91366-H;
%
  D.~N.~Kabat and M.~Ortiz,
  Nucl.\ Phys.\ B {\bf 388}, 570 (1992)
  doi:10.1016/0550-3213(92)90627-N
  [hep-th/9203082];
%
  D.~Amati, M.~Ciafaloni and G.~Veneziano,
  Nucl.\ Phys.\ B {\bf 403}, 707 (1993)
  doi:10.1016/0550-3213(93)90367-X;
%
  I.~J.~Muzinich and S.~Vokos,
  Phys.\ Rev.\ D {\bf 52}, 3472 (1995)
  doi:10.1103/PhysRevD.52.3472
  [hep-th/9501083].

\bibitem{DAppollonio:2010krb}
  G.~D'Appollonio, P.~Di Vecchia, R.~Russo and G.~Veneziano,
  JHEP {\bf 1011}, 100 (2010)
  doi:10.1007/JHEP11(2010)100
  [arXiv:1008.4773 [hep-th]].

\bibitem{Eikonal3}
  S.~Melville, S.~G.~Naculich, H.~J.~Schnitzer and C.~D.~White,
  Phys.\ Rev.\ D {\bf 89}, no. 2, 025009 (2014)
  doi:10.1103/PhysRevD.89.025009
  [arXiv:1306.6019 [hep-th]];
%
  R.~Akhoury, R.~Saotome and G.~Sterman,
  arXiv:1308.5204 [hep-th];
%
  G.~D'Appollonio, P.~Di Vecchia, R.~Russo and G.~Veneziano,
  JHEP {\bf 1505}, 144 (2015)
  doi:10.1007/JHEP05(2015)144
  [arXiv:1502.01254 [hep-th]];
%
  M.~Ciafaloni, D.~Colferai and G.~Veneziano,
  Phys.\ Rev.\ Lett.\  {\bf 115}, no. 17, 171301 (2015)
  doi:10.1103/PhysRevLett.115.171301
  [arXiv:1505.06619 [hep-th]];
%
  G.~D'Appollonio, P.~Di Vecchia, R.~Russo and G.~Veneziano,
  JHEP {\bf 1603}, 030 (2016)
  doi:10.1007/JHEP03(2016)030
  [arXiv:1510.03837 [hep-th]];
%
  M.~Ciafaloni, D.~Colferai, F.~Coradeschi and G.~Veneziano,
  Phys.\ Rev.\ D {\bf 93}, no. 4, 044052 (2016)
  doi:10.1103/PhysRevD.93.044052
  [arXiv:1512.00281 [hep-th]].

\bibitem{Luna:2016idw}
  A.~Luna, S.~Melville, S.~G.~Naculich and C.~D.~White,
  JHEP {\bf 1701}, 052 (2017)
  doi:10.1007/JHEP01(2017)052
  [arXiv:1611.02172 [hep-th]].

\bibitem{Collado:2018isu}
  A.~K.~Collado, P.~Di Vecchia, R.~Russo and S.~Thomas,
  JHEP {\bf 1810}, 038 (2018)
  doi:10.1007/JHEP10(2018)038
  [arXiv:1807.04588 [hep-th]].

\bibitem{Sucher:1994qe}
  J.~Sucher,
  Phys.\ Rev.\ D {\bf 49}, 4284 (1994).
  doi:10.1103/PhysRevD.49.4284

\bibitem{Luna:2017dtq}
  A.~Luna, I.~Nicholson, D.~O'Connell and C.~D.~White,
  JHEP {\bf 1803}, 044 (2018)
  doi:10.1007/JHEP03(2018)044
  [arXiv:1711.03901 [hep-th]].

\bibitem{Jackson:1998nia}
  J.~D.~Jackson,
  \textit{Classical Electrodynamics\/}, 
  ISBN 978-0471309321 (3${}^{\textrm{rd}}$ ed.) (Wiley, New York, 1998).

\bibitem{schwingerBook}
  J.~Schwinger, L.~L.~DeRaad, Jr., K.~A.~Milton and W.~Y.~Tsai,
  \textit{Classical Electrodynamics\/}, 
  ISBN 978-0738200569 (Perseus, 1998).


\end{thebibliography}
\end{document}